\documentclass[a4paper,fleqn]{cas-dc}
\usepackage {xcolor}
\usepackage[numbers]{natbib}
\usepackage{eurosym,makecell,comment}
\usepackage{multirow}
\usepackage{threeparttable}
\usepackage{framed}
\usepackage{hyperref}
\usepackage{url}
\usepackage{verbatim}
\usepackage{graphicx}
 \usepackage{grffile} 
\usepackage{subcaption}
\usepackage{float}
\usepackage{microtype}
\usepackage{colortbl}
\usepackage{amsmath,amssymb,amsfonts}
\usepackage{tikz}
\usepackage{pgfplots}
\usepackage{listings}
\usepackage{array}
\usepackage{ragged2e}

\usepackage{pifont}
\usepackage{algorithm}
\usepackage{algpseudocode}
\usepackage{balance}

\usepackage{soul}
\usetikzlibrary{calc}

\pgfplotsset{compat=1.18}

\newcommand{\ie}{\textit{i.e., }}  % from the Latin id est = that is
\newcommand{\eg}{\textit{e.g., }}  % from the Latin exempli gratia = for example
\def \1{\textit{(i)}}
\def \2{\textit{(ii)}}
\def \3{\textit{(iii)}}
\def \4{\textit{(iv)}}
\def \5{\textit{(v)}}

\usepackage{etoolbox}
\newtoggle{finalPaper}

\setstcolor{blue}

\toggletrue{finalPaper} % final version

\iftoggle{finalPaper} {

	\newcommand{\rmvtxt}[1]{}}
{

	\newcommand{\rmvtxt}[1]{\st{#1}}}

\newtoggle{removeItalic}
\togglefalse{removeItalic} % regular command
\iftoggle{removeItalic} {
    \renewcommand{\textit}[1]{#1}
}

\newcommand{\solution}{\textit{DART}}
\newcommand{\voyager}{\textit{Voyager}}
\newcommand{\sentinel}{\textit{Sentinel}}

\begin{document}
\let\WriteBookmarks\relax
\def\floatpagepagefraction{1}
\def\textpagefraction{.001}
\shorttitle{Decentralized Federated Learning Model Robustness Analysis}
\shortauthors{Feng et~al.}

\title[mode = title]{\solution{}: A Solution for Decentralized Federated Learning Model Robustness Analysis}

\author[1]{Chao Feng*}[orcid=0000-0002-0672-1090]
\author[1]{Alberto {Huertas Celdrán}}[orcid=0000-0001-7125-1710]
\author[1]{Jan {von der Assen}}[orcid=0000-0002-0591-8887]
\author[2]{Enrique Tom\'as {Mart\'inez Beltr\'an}}[orcid=0000-0002-5169-2815]
\author[3]{Gérôme Bovet}[orcid=0000-0002-4534-3483]
\author[1]{Burkhard Stiller}[orcid=0000-0002-7461-7463]

\address[1]{Communication Systems Group CSG, Department of Informatics IfI, University of Zurich UZH, 8050 Zürich, Switzerland}
\address[2]{Department of Information and Communications Engineering, University of Murcia, 30100 Murcia, Spain}
\address[3]{Cyber-Defence Campus within Armasuisse Science \& Technology, 3602 Thun, Switzerland}

\cortext[cor1]{Corresponding author.
Email address: cfeng@ifi.uzh.ch (C. Feng)}


\maketitle

\begin{abstract}
Federated Learning (FL) has emerged as a promising approach to address privacy concerns inherent in Machine Learning (ML) practices. However, conventional FL methods, particularly those following the Centralized FL (CFL) paradigm, utilize a central server for global aggregation, which exhibits limitations such as bottleneck and single point of failure. To address these issues, the Decentralized FL (DFL) paradigm has been proposed, which removes the client-server boundary and enables all participants to engage in model training and aggregation tasks. Nevertheless, as CFL, DFL remains vulnerable to adversarial attacks, notably poisoning attacks that undermine model performance. While existing research on model robustness has predominantly focused on CFL, there is a noteworthy gap in understanding the model robustness of the DFL paradigm. In this paper, a thorough review of poisoning attacks targeting the model robustness in DFL systems, as well as their corresponding countermeasures, are presented. Additionally, a solution called \solution{} is proposed to evaluate the robustness of DFL models, which is implemented and integrated into a DFL platform. Through extensive experiments, this paper compares the behavior of CFL and DFL under diverse poisoning attacks, pinpointing key factors affecting attack spread and effectiveness within the DFL. It also evaluates the performance of different defense mechanisms and investigates whether defense mechanisms designed for CFL are compatible with DFL. The empirical results provide insights into research challenges and suggest ways to improve the robustness of DFL models for future research.

\end{abstract}

\section{Introduction}
\label{sec:intro}

% ML and why FL
The rapid expansion of the Internet-of-Things (IoT) has led to the interconnection of more than 15 billion devices, resulting in the generation of a staggering 330 Exabytes of data on a daily basis~\cite{Duarte2023}. Machine Learning (ML) has become a crucial tool for efficiently handling and analyzing such immense datasets~\cite{silva_FederatedLearning_2023}. The ML pipeline involves various stages, such as data collection, centralization, preprocessing, training, and inference. However, data aggregation into a single server or data center presents challenges, particularly in IoT environments where data is collected and stored in a distributed manner~\cite{beltran_DecentralizedFederated_2022}. Centralizing data is challenging due to network and storage limitations, and sharing data online raises security and privacy concerns, users are wary of sharing sensitive information from IoT devices~\cite{rodriguez-barroso_SurveyFederated_2023}. Consequently, there is a growing research interest in designing an innovative ML paradigm that effectively tackles the challenges related to data processing and privacy preservation. 

\begin{figure}[ht!]
\centering
\includegraphics[width=1\linewidth]{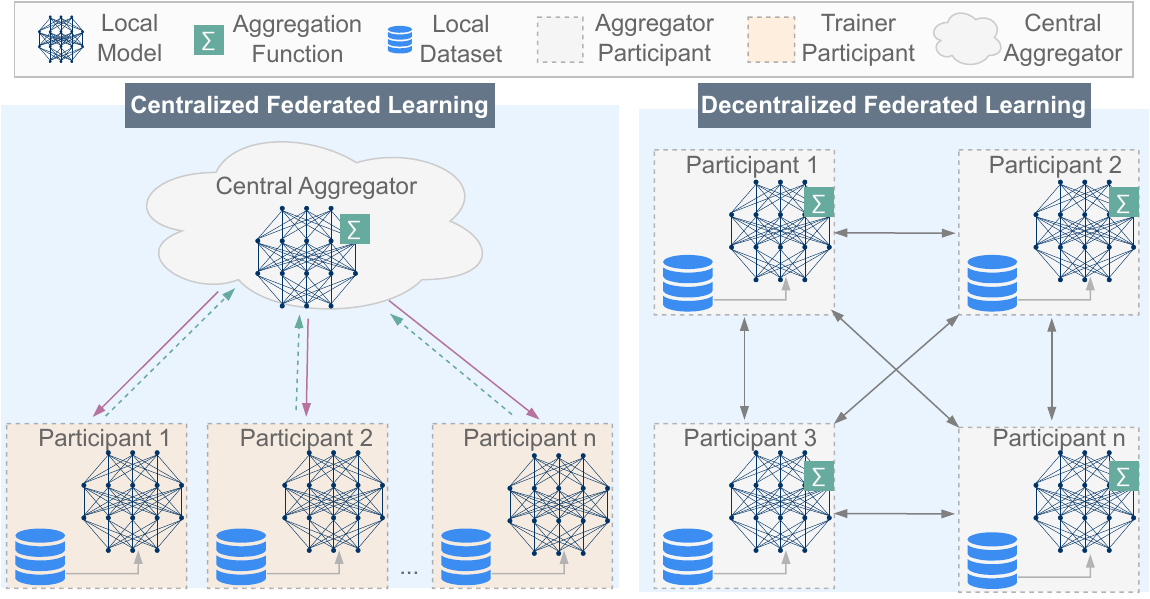}
\caption{Federation Aggregation Architecture of FL}
\label{fig:fl_architectures}
\end{figure}

% CFL and DFL
Federated Learning (FL) is a privacy-preserving ML paradigm where data is distributed and retained locally across clients (\eg IoT devices) instead of centralized in a single location~\cite{googleresearch_FederatedLearning_2017}.  As illustrated in the left side of \figurename~\ref{fig:fl_architectures}, the local dataset in each client is employed to feed and train the local model algorithm and evaluate the model outcomes. These local models are then sent to a central aggregation server, where a global model is generated and distributed back to the clients for further training. This iterative process continues until the global model reaches convergence or a predefined number of aggregation rounds is reached. Throughout this federated procedure, clients prioritize privacy by only transmitting the models over the network while keeping the underlying data securely stored on their devices. Additionally, the smaller size of the models compared to the raw data helps overcome bandwidth and storage limitations. Thus, FL is well-suited for implementation in distributed networks that prioritize privacy preservation, such as healthcare and autonomous driving sections~\cite{huertas2022iotj}. Nevertheless, the incorporation of the central aggregation server, referred to as Centralized FL (CFL), introduces new risks to the system. The central aggregator becomes a single-point-of-failure risk of the CFL paradigm, leading to network congestion and processing bottleneck~\cite{beltran_DecentralizedFederated_2022}.

To cope with the single-point-of-failure risk of the CFL paradigm, a decentralized approach known as Decentralized FL (DFL) has been proposed~\cite{beltran_DecentralizedFederated_2022}. The DFL paradigm eliminates the distinction between servers and clients, allowing all participants to serve as trainers and aggregators, as shown on the right side of \figurename~\ref{fig:fl_architectures}. In this approach, participants train models using their own local data and then exchange these models with other participants through the network. Local model aggregation occurs within each participant, incorporating knowledge shared by others while ensuring that data privacy is preserved as only the models are shared. This decentralization allows the FL process to be fully autonomous and more resistant to network failures. Moreover, the DFL system offers increased stability by avoiding the risks associated with a single aggregator, distinguishing it from the CFL paradigm. Besides, the DFL paradigm allows for flexible interconnection using various network topologies, providing greater adaptability.

% threats to FL
However, in addition to the benefits that CFL and DFL offer, the inherently distributed nature of this FL system makes it vulnerable to poisoning attacks~\cite{tian_ComprehensiveSurvey_2023}. These attacks involve malicious participants manipulating the training data or inserting malicious elements into their local models, resulting in inaccurate outputs~\cite{xia_PoisoningAttacks_2023}. These compromised models are distributed across the FL system and aggregated with benign models. This aggregation results in declining performance for benign models, reducing their robustness. These poisoning attacks can target both CFL and DFL systems. Therefore, there is a growing body of research dedicated to protecting the model robustness of FL systems and mitigating the detrimental effects of poisoning attacks~\cite{lyu_PrivacyRobustness_2022}. 

% overview and contributions of the paper
Given the significance of poisoning attacks on DFL, it is crucial to prioritize the strength and security of the models. However, existing studies predominantly focus on the impact of poisoning attacks on CFL, indicating the need for greater attention to safeguarding DFL systems against such attacks. Moreover, there is a scarcity of practical tools available for effectively analyzing the robustness of DFL. Therefore, this work proposes and implements a DFL model robustness analysis solution. The contributions of this work can be summarized as follows:

%todo change here
\begin{itemize}
    \item An analysis of the poisoning attacks targeting DFL. This paper offers a comprehensive overview of poisoning attacks, classifying and examining these attacks from the perspectives of both attack intention and attack strategy.
    
    \item An overview of existing defenses against poisoning attacks. This paper comprehensively reviews defense mechanisms to protect FL (both CFL and DFL) from poisoning attacks. The existing defense mechanisms are thoroughly examined, categorized, and analyzed from various perspectives.
    
    \item Design and prototype a model robustness analysis solution for DFL. This paper proposes a model robustness analysis solution for DFL, called \solution{}. \solution{} comprises an attack component that enables the execution of untargeted and targeted poisoning attacks, as well as a defense component that incorporates multiple defense mechanisms to protect DFL models from poisoning attacks. Meanwhile, \solution{} is prototypically implemented and integrated into a real DFL platform to benchmark the performance of different attacks and the effectiveness of defense mechanisms.
    
    \item Experimental comparison of the model robustness between CFL and DFL and benchmark of the effectiveness of defense mechanisms. This paper conducts a series of experiments to compare the behavior and evaluate the robustness of CFL and DFL paradigms when subjected to targeted and untargeted attacks in MNIST, FashionMNIST, and Cifar10 datasets. Besides, this study conducts extensive experiments to benchmark the effectiveness of various defense mechanisms and offers recommendations for selecting the optimal defense mechanism.
        
    \item Analysis of the challenges and opportunities in the field of mode robustness. The paper delves into the learning lessons and highlights the research challenges from the literature review and experimental analysis while pinpointing the potential avenues for exploration in the DFL model robustness field.
    
\end{itemize}

% innovation
In summary, the primary innovation of this work lies in its comprehensive review of existing literature on model robustness in DFL, as well as its empirical assessment of defense mechanisms against poisoning attacks. Then, this paper argues that DFL overlay network topology plays a crucial role in model robustness, underscoring the importance of considering topology when designing novel attacks or defenses. In addition, it analyzes the potential issues and challenges for improving the security of the DFL model against poisoning attacks in real-world scenarios, including both offensive and defensive considerations. Finally, this work offers recommendations for improving the robustness of the model in DFL and proposes practical roadmaps for implementation.

% structure
The subsequent sections of this paper are structured as follows: Section~\ref{sec:related} presents an introduction to the relevant literature, while Section~\ref{sec:background} provides a comprehensive overview of poisoning attacks and the corresponding defense mechanisms. Afterwards, Section~\ref{sec:framework} introduces the framework proposed in the paper, encompassing the overarching architecture, conception, and implementation, as well as the integration with a DFL platform. The experimental study is detailed in Section~\ref{sec:experiments}. Lastly, Section~\ref{sec:challenges} discusses lessons learned, challenges encountered, and potential research opportunities, while Section~\ref{sec:conclusion} offers concluding remarks.

\section{Related Work}

\label{sec:related}

\begin{table}[H]
\centering
\caption{Comparison of Attributes among Existing Robustness Analysis of FL Research }
\label{tab:relatedwork}
\resizebox{\columnwidth}{!}{%
\begin{tabular}{lccccc}
\toprule
Research &
  CFL &
  DFL &
  \begin{tabular}[c]{@{}l@{}}Defense \\ Included\end{tabular} &
  \begin{tabular}[c]{@{}l@{}}Overview \\ of Attacks\end{tabular} &
  \begin{tabular}[c]{@{}l@{}}Experimental \\ Analysis\end{tabular} \\ \midrule
  \cite{benmalek:hal-03620400} 2022   & \checkmark & \checkmark & \checkmark & \checkmark & x \\
  \cite{BLANCOJUSTICIA2021104468} 2021 & \checkmark & x & \checkmark & \checkmark & \checkmark \\
  \cite{Chen10020431} 2022             & \checkmark & x & \checkmark & \checkmark & x \\
  \cite{Jere9308910} 2021              & \checkmark & x & \checkmark & \checkmark & x \\
  \cite{Kumar10274102} 2023            & \checkmark & x & \checkmark & \checkmark & x \\
  \cite{liu2022threats} 2022           & \checkmark & x & \checkmark & \checkmark & x \\
  \cite{lyu_PrivacyRobustness_2022} 2022               & \checkmark & x & \checkmark & \checkmark & x \\
\cite{MOTHUKURI2021619} 2021         & \checkmark & \checkmark & \checkmark & \checkmark & x \\
\cite{NAIR2023103723} 2023           & \checkmark & x & x & \checkmark & x \\
\cite{qammar2022federated} 2022      & \checkmark & x & \checkmark & \checkmark & x \\
\cite{rodriguez-barroso_SurveyFederated_2023} 2023  & \checkmark & x & \checkmark & \checkmark & \checkmark \\
\cite{Wang9771619} 2022              & \checkmark & x & \checkmark & x & x \\
\cite{xia_PoisoningAttacks_2023} 2023              & \checkmark & x & \checkmark & \checkmark & x \\
\cite{Zhang9634881} 2021             & \checkmark & x & \checkmark & \checkmark & x \\
This Work                            & \checkmark & \checkmark & \checkmark & \checkmark & \checkmark \\\midrule
\end{tabular}%

}
\end{table}

There have been a few overview and survey articles for FL model robustness studies, discussing both theoretical and practical considerations related to adversarial attacks and defense mechanisms. This section conducts a comparative analysis of these articles and presents an overview in \tablename~\ref{tab:relatedwork}.

\cite{MOTHUKURI2021619} analyzed the security and privacy risks faced by FL from the perspectives of the sources of vulnerabilities and the types of adversarial attacks. A comparison was made between FL and other distributed ML methods regarding their vulnerability to adversarial attacks. The authors analyzed both CFL and fully-connected DFL architectures within the FL framework. However, the analysis was predominantly qualitative, lacking quantitative assessment and neglecting to compare the distinctions between CFL and DFL. Similarly, \cite{lyu_PrivacyRobustness_2022} presented a comprehensive assessment of the risks and threats confronting FL, outlining the impact of potential attacks. Furthermore, this survey presented an overview of defensive measures and assessed their susceptibility to compromise.

\cite{Kumar10274102} summarized the security risks and impacts of adversarial attacks on FL systems and proposed a multidimensional analysis framework from the source of the attack, the attack's impact, the attack's budget, and the attack's visibility. A taxonomy of adversarial attacks is also proposed. In terms of defense mechanisms, the effects of different defense mechanisms are compared through quantitative literature analysis. Nevertheless, this survey lacks empirical analysis to validate the legitimacy of these attacks and defense measures.

\cite{BLANCOJUSTICIA2021104468} and \cite{rodriguez-barroso_SurveyFederated_2023} employed a combination of literature analysis and empirical analysis to examine and summarize the security and privacy risks associated with FL systems. The study presented an overview of attacks and defenses regarding security and privacy and utilized experimental methods to validate various types of attacks empirically. However, it is important to note that the survey solely focused on CFL and did not address the emerging challenges faced by DFL systems.

\cite{Zhang9634881}, \cite{Jere9308910}, \cite{qammar2022federated}, \cite{benmalek:hal-03620400}, and \cite{NAIR2023103723} conducted extensive research on the vulnerabilities of FL systems. They developed taxonomies of adversarial attacks. While \cite{ liu2022threats} focused on analyzing the threats throughout the entire life cycle of FL, considering various stages of the FL pipeline. They classified and summarized the attacks, examined their potential impact, and analyzed the available defense strategies concurrently. \cite{Chen10020431} and \cite{Wang9771619} evaluated the efficacy and viability of various defense strategies, classifying and categorizing defenses according to which attacks can be effectively mitigated. \cite{xia_PoisoningAttacks_2023} centered on examining poisoning attacks, where various types of such attacks were categorized, and diverse strategies and methods for carrying out these attacks were elucidated. Additionally, the research entailed an analysis of the countermeasures employed against different attacks, their effectiveness, and the vulnerability of data distribution.

In conclusion, the current research examining and analyzing the security of FL mainly focuses on CFL systems, with only a limited number of articles and surveys exploring DFL. Additionally, these works rely on qualitative analysis and theoretical taxonomies, which are lacking in quantitative and experimental analysis. Thus, there is a need for a practical-oriented module that can quantitatively evaluate and analyze the robustness of DFL models using experiments while also appraising the efficacy of different defense mechanisms.

\section{Poisoning Attacks and Defense Mechanisms}
\label{sec:background}
Malicious attacks pose a significant obstacle in FL, especially poisoning attacks, which are designed to undermine the robustness of the FL model and diminish the usability of its output. Given the distributed nature of FL, detecting and mitigating poisoning attacks presents a substantial challenge. This Section provides an overview of poisoning attacks and reviews existing defense mechanisms on both CFL and DFL.

\subsection{Poisoning Attacks on FL}
\label{subsec:poisoningattacks}
In FL systems, where nodes are distributed across various entities, adversaries can tamper with data or models to disrupt the effectiveness and robustness of benign models. In contrast to CFL, detecting malicious activity in DFL poses greater challenges due to the absence of a central entity controlling the entire process. Poisoning attacks targeting DFL can be categorized based on two viewpoints: the attack intention and attack strategy~\cite{xia_PoisoningAttacks_2023}. The attack intention indicates the desired outcome of the adversary, whether it is to misclassify a particular target or to undermine the overall performance. The attack strategy entails the creation of a poisoning attack through the manipulation of either the data or the model.

% attack intention
Poisoning attacks can be categorized into three groups based on the intentions of the adversary, namely untargeted, targeted, and backdoor attacks~\cite{tian_ComprehensiveSurvey_2023,xia_PoisoningAttacks_2023}. The goal of untargeted attacks is to reduce the overall performance of the model, hindering its convergence. Conversely, targeted attacks involve intentional manipulation of the training data by the adversary to cause incorrect or biased predictions for specific target inputs.  The objective of the targeted attacks is to infiltrate the model in such a way that only a particular target set or class is misclassified, while the rest of the set is correctly classified~\cite{xia_PoisoningAttacks_2023}. Another type of attack is the backdoor attack, where the adversary inserts one or more triggers into the model during training, which can then be exploited during the inference process~\cite{bagdasaryan_HowBackdoor_2019}. Essentially, the model behaves normally without the trigger, but an attacker can elicit a desired prediction or classification by presenting the trigger during the inference stage.

In the CFL paradigm, targeted attacks are data-driven processes and thus tend to be launched at the client side. Untargeted attacks in the CFL paradigm can be executed without reliance on data, enabling them to be launched from either the client or server side. In contrast, in the DFL paradigm, the traditional differentiation between client and server is eliminated, allowing any node to launch both targeted and untargeted attacks. Consequently, the attack surfaces in the DFL paradigm are increased.

\begin{table*}[t!]\centering
\caption{Overview of Defense Mechanisms Against Poisoning Attacks in FL}
\scriptsize
\renewcommand{\arraystretch}{1.5}
% \newcolumntype{L}[1]{>{\raggedright\let\newline\\\arraybackslash\hspace{0pt}}m{#1}}
\setlength\tabcolsep{3pt}

\begin{tabular}{llllccc}
\toprule
\multirow{2}{*}{\textbf{Category}} &\multirow{2}{*}{\textbf{Type}} &\multirow{2}{*}{\textbf{Method}} &\multirow{2}{*}{\textbf{Technique}} &\multirow{2}{*}{\textbf{Paradigm}} &\multicolumn{2}{c}{\textbf{Objective}} \\\cmidrule(lr){6-7}
\textbf{} &\textbf{} &\textbf{} &\textbf{} & & U & T \\\midrule
%\cmidrule{5-7}
\multirow{11}{*}{Byzantine-Robust}  & \multirow{6}{*}{Geometry}  & \textit{COMED} \cite{yin_ReputationBasedResilient_2017} &Coordinate-wise median &CFL &\checkmark &x \\%\cmidrule[0.1pt]{3-7}
 & &\textit{TrimmedMean} \cite{yin_ByzantineRobustDistributed_2018} &Filtered mean &CFL &\checkmark &x \\%\cmidrule{3-7}
& &\textit{RFA} \cite{pillutla_RobustAggregation_2022} &Geometric median &CFL &\checkmark &x \\%\cmidrule{3-7}
& &\textit{Krum} \cite{blanchard_MachineLearning_2017} &Euclidean distance &CFL &\checkmark &x \\%\cmidrule{3-7}
& &\textit{Multi-Krum} \cite{blanchard_MachineLearning_2017} &Euclidean distance &CFL &\checkmark &x \\%\midrule
& &\textit{Bulyan} \cite{mhamdi_HiddenVulnerability_2018} &Krum and TrimmedMean &CFL &\checkmark &x \\ \cmidrule(l){2-7}
Aggregation& \multirow{3}{*}{Regularization} &\textit{Zeno++} \cite{xie_ZenoRobust_2021} &Approximated gradient descent score &CFL &\checkmark &x \\
& &\textit{AFA} \cite{munoz-gonzalez_ByzantineRobustFederated_2019} &Gradient similarity, Hidden Markov model &CFL &\checkmark &x \\
& &\textit{RSA} \cite{li_RSAByzantineRobust_2019} &Norm regularization &CFL &\checkmark &x \\ \cmidrule(l){2-7}
&Decomposition &\textit{DnC} \cite{shejwalkar_ManipulatingByzantine_2021} &Dimensionality Reduction (PCA) and SVD &CFL &\checkmark &x \\
& &\textit{RLR} \cite{ozdayi_DefendingBackdoors_2021} &Learning rate decomposition &CFL &- &\checkmark \\\cmidrule{1-7}
\multirow{6}{*}{Anomaly Detection} &Validation &\textit{ERR, LFR} \cite{fang_LocalModel_2021} &Global validation &CFL &\checkmark &\checkmark \\
% & &Baffle &Loss feed-back &CFL &x &\checkmark \\
& &\textit{PDG}\textit{A}N \cite{zhao_PDGANNovel_2020} &Model accuracy auditing &CFL &\checkmark &\checkmark \\ \cmidrule(l){2-7}
& \multirow{4}{*}{Gradient-based} &\textit{FoolsGold}~\cite{fung_MitigatingSybils_2020} &Gradient similarity (cosine) &CFL &\checkmark &\checkmark \\
& &\textit{FLDetector} \cite{zhang_FLDetectorDefending_2022} &Hessian-based gradient consistency &CFL &\checkmark &\checkmark \\
& &Li et al. \cite{li_LearningDetect_2020} &Spectral anomaly detection &CFL &\checkmark &\checkmark \\
& &\textit{Sniper} \cite{zhao_PDGANNovel_2020} &Graph clustering &CFL &- &\checkmark \\\cmidrule{1-7}
MTD & & \voyager{} \cite{feng2023voyager} & Manipulating of the network topology & DFL & \checkmark & x \\ \cmidrule{1-7}
\multirow{7}{*}{Hybrid Mechanism} & &\textit{FLTrust} \cite{cao_FLTrustByzantinerobust_2021} &ReLU-clipped cosine similarity, norm thresholding &CFL &\checkmark &\checkmark \\
& &\textit{Trusted DFL} \cite{gholami_TrustedDecentralized_2022} &Trusted aggregation &DFL &\checkmark &- \\
& &\textit{FedInv} \cite{zhao_FedInvByzantineRobust_2022} &Gradient-based clustering distribution divergences &CFL &\checkmark &\checkmark \\
& &\textit{Norm Clipping }\cite{ozdayi_DefendingBackdoors_2021} &Clipping and worker momentum &DFL &\checkmark &\checkmark \\
& &\textit{DeepSight} \cite{rieger_DeepSightMitigating_2022} &Classification and clustering with update fingerprinting &CFL &\checkmark &\checkmark \\
& &\textit{FLAME} \cite{nguyen_FLAMETaming_2021} &Clustering (cosine similarity), adaptive clipping, noising &CFL &\checkmark &\checkmark \\
& & \sentinel{} \cite{feng2023sentinel} & Model similarity, bootstrap validation and normalization & DFL & \checkmark & \checkmark \\ \cmidrule{1-7}
Post-Aggregation  & &Wu et al. \cite{wu_MitigatingBackdoor_2021} &Neuron Pruning &CFL &x &\checkmark \\
& &Sun et al. \cite{sun_CanYou_2019} &Weak Differential Privacy &CFL &x &\checkmark \\
\bottomrule
\end{tabular}
\label{tab:rw}

Untargeted Attacks (U), Targeted Attacks (T)
\end{table*}

% attack strategy
From the attack strategy perspective, poisoning attacks can be divided into data poisoning and model poisoning~\cite{xia_PoisoningAttacks_2023}. Data poisoning involves manipulating training data to indirectly affect the model's training process, such as changing labels, \ie label flipping, or inserting malicious samples, \ie sample poisoning~\cite{rodriguez-barroso_SurveyFederated_2023}. Both label flipping and sample poisoning can be used as untargeted attacks or targeted attacks. In the context of backdoor attacks, the attacker can choose to incorporate either semantic or artificial backdoors. Semantic backdoors use naturally existing cues, like a specific colored striped background, as triggers. Artificial backdoors are intentionally created by injecting particular triggers, such as adding certain symbols to a sample~\cite{nguyen_BackdoorAttacks_2023}. The effectiveness of data poisoning attacks is limited by the need for a large number of poisoned samples~\cite{rodriguez-barroso_SurveyFederated_2023}. In model poisoning attacks, malicious participants can manipulate the locally trained model directly by modifying shared weights or gradients~\cite{xia_PoisoningAttacks_2023}. These attacks pose a greater threat as they have a higher success rate and are more difficult to detect. Model poisoning can be divided into two categories: random weights generation, which involves adding random noise to the transmitted model, and optimized weights generation, which involves optimizing the distribution of the added noise to decrease the chances of detection~\cite{rodriguez-barroso_SurveyFederated_2023, tian_ComprehensiveSurvey_2023}. 

In DFL, participants possess knowledge not only of their own local data and models but also of the models belonging to their neighboring nodes. Therefore, malicious nodes can strategically optimize their attacks and execute more threatening attacks while avoiding detection, thereby heightening the complexity of designing effective defense mechanisms.

Theoretically, while DFL improves system fault tolerance and mitigates single-point-of-failure, it also expands the attack surface and potentially exposes the system to a greater array of attack vectors. Thereby, DFL does not offer superiority over CFL in terms of model robustness.

\subsection{Defense Mechanisms}
\label{sec:defenses}

Most defense methods are designed for the centralized setting, which involves a less complex model aggregation process by a single aggregation server. However, in DFL, the lack of a centralized control entity and the increased array of attack vectors mentioned in Section~\ref{subsec:poisoningattacks} pose challenges when developing defense strategies for DFL. This paper categorizes the approaches for mitigating model poisoning attacks into five categories: Byzantine-robust aggregation, anomaly detection, Moving Target Defense (MTD), hybrid, and post-aggregation. 

As shown in \tablename~\ref{tab:rw}, these articles have been reviewed, compared, and analyzed across various dimensions, such as the techniques employed, the paradigms in which they are implemented (CFL or DFL), and their effectiveness in addressing both targeted and untargeted attacks.

%Byzantine-Robust Aggregation
\subsubsection{Byzantine-Robust Aggregation}\label{sec:related_work:robust_aggregation}
The main objective of Byzantine-robust defense techniques is to establish a robust (referred to as Byzantine resilience) aggregation rule that prevents the model's performance from being compromised by malicious updates~\cite{mhamdi_HiddenVulnerability_2018, rieger_DeepSightMitigating_2022}. These defense methods can be categorized into geometric measures, regularization, and decomposition.

%Geometric, coordinate-wise

There are two types of geometric approaches: coordinate-wise filtering and vector-wise filtering. The Coordinate-wise Median (\textit{COMED}) is a defensive strategy that uses dimension-wise filtering to protect against attacks~\cite{yin_ReputationBasedResilient_2017}. \textit{COMED} is like the median concept but applied in high-dimensional spaces to remove outliers. This approach is effective against model replacement attacks. Another approach called \textit{TrimmedMean} method~\cite{yin_ByzantineRobustDistributed_2018} also uses dimension-wise filtering. It eliminates the lowest and highest values in each dimension using a trimming parameter. The underlying assumption of this approach is that malicious updates can be identified as outliers compared to benign updates. 

Vector-wise filtering is a commonly used geometric measure. \textit{RFA}~\cite{pillutla_RobustAggregation_2022} utilizes the geometric median as an aggregation function, which helps defend against untargeted poisoning attacks. However, this defense mechanism can be compromised if an attacker uses a singular-flipping attack to manipulate the geometric center~\cite{li_LearningDetect_2020}. \textit{Krum}~\cite{blanchard_MachineLearning_2017} is a widely used robust aggregation function that assigns scores to client updates based on their similarity to other updates and selects the update with the lowest score to update the global model. \textit{Multi-Krum} is a modified version that selects multiple client updates for the next global model. \textit{Bulyan}~\cite{mhamdi_HiddenVulnerability_2018} is proposed as a two-stage aggregation protocol that overcomes the limitations of \textit{Krum}, but experiments have shown that it performs worse than\textit{ Multi-Krum} by potentially reducing benign updates~\cite{sun_CanYou_2019}.

% regularization-based
The second category of Byzantine-robust methods is the regularization-based approach. \textit{Zeno++}~\cite{xie_ZenoRobust_2021} is an asynchronous protocol for robust CFL. This approach proposes an estimated score for gradient descent that considers the loss and update magnitude. This helps reduce the impact of malicious gradient updates. However, a drawback is that the server needs a validation dataset, which is a constraint in a federated setting. Additionally, relying on performance rather than parameters adds extra computational burden\cite{guo_ByzantineresilientDecentralized_2021}. Adaptive Federated Averaging (\textit{AFA})~\cite{munoz-gonzalez_ByzantineRobustFederated_2019} not only filters out potentially malicious updates but also completely blocks adversaries from participating. \textit{AFA} employs a Hidden Markov Model (HMM) based on gradient cosine similarity to predict a client's behavior, thereby incorporating a learning component into the defense system. This predictive score is then utilized as a weight factor in the aggregation process, serving as a regularization term. Another method, \textit{RSA}~\cite{li_RSAByzantineRobust_2019}, integrates $\ell_1$-norm regularization into Stochastic Gradient Descent (SGD) to enhance robustness against Byzantine attacks.

% decomposition
The last class of Byzantine-robust is the decomposition-based approach. Divide-and-Conquer (\textit{DnC})~\cite{shejwalkar_ManipulatingByzantine_2021} uses spectral analysis, \ie singular value decomposition, to identify and filter out poisoned updates.  The evaluations show that these techniques perform well in scenarios where the data is Independent and Identically Distributed (IID). However, when dealing with non-IID datasets, \textit{DnC} is not effective in preventing attacks if the adversary has knowledge from benign clients. Robust Learning Rate (\textit{RLR})~\cite{ozdayi_DefendingBackdoors_2021} tackles this problem by implementing a mechanism where clients vote for the direction of the global model update based on the algebraic sign of their update vector. Each dimension of the update vector requires a certain learning threshold to be met by the total number of votes. However, a challenge in \textit{RLR} is determining an appropriate learning threshold that the number of malicious clients cannot exceed.

% Anomaly Detection
\subsubsection{Anomaly Detection}\label{sec:related_work:anomaly_detection}
Defense mechanisms that employ anomaly detection, also known as Byzantine detection, aim to identify and eliminate potentially harmful updates \cite{zhang_SurveyTrustworthy_2023}. Unlike Byzantine robustness, these anomaly detection schemes do not incorporate the defense strategy into the aggregation rules. 

Error Rate-based Rejection (\textit{ERR}) and Loss Function-based Rejection (\textit{LFR}) are designed in \cite{fang_LocalModel_2021} to enhance the robustness of CFL. These methods involve evaluating the performance of client models using a validation dataset on the server side. Before combining the received models, a certain number of updates that have the most impact on either the loss or validation error are discarded. However, a drawback of these methods is that the server needs to collect a clean dataset, which goes against the privacy-preserving principles of FL. To address this limitation, \textit{PDGAN}~\cite{zhao_PDGANNovel_2020} utilizes a generative adversarial network (GAN) to generate a synthetic dataset for outlier model detection. However, it should be noted that the GAN needs to be trained before the task starts, and the server must have sufficient computing power to train an ML model in a reasonable time while performing aggregation.

\begin{figure*}[h]
\centering
\includegraphics[width=1\linewidth]{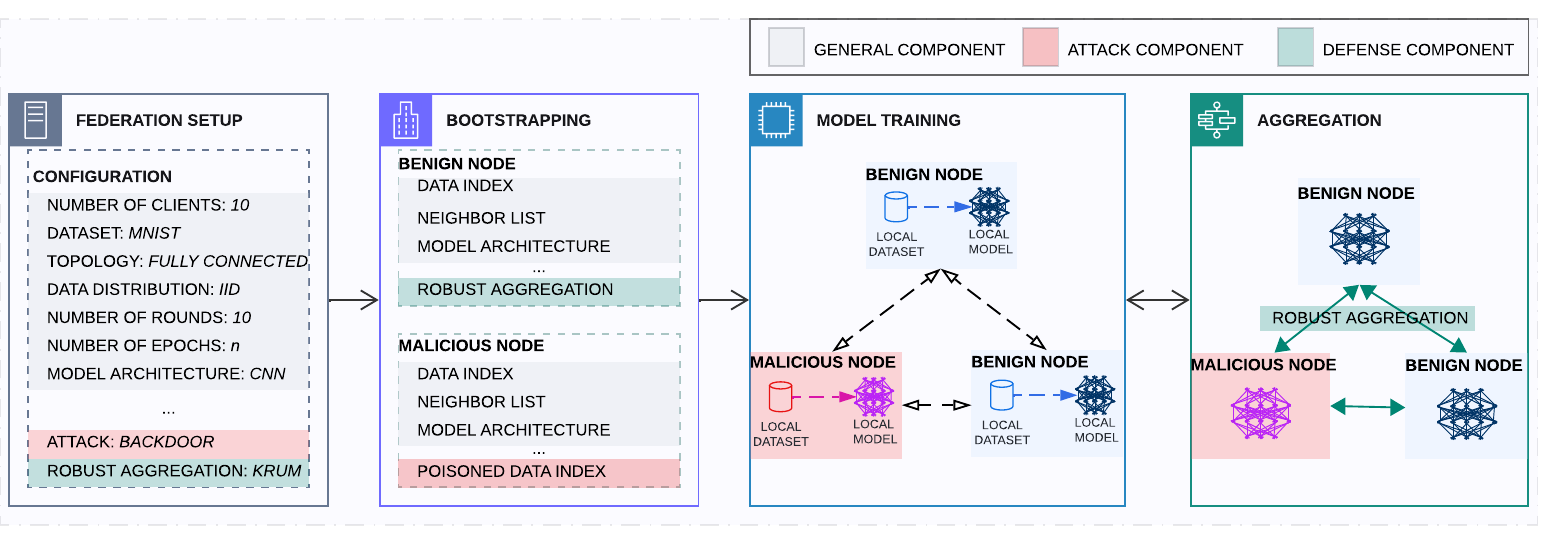}
\caption{\solution{} within the DFL Procedure}
\label{fig:dart}
\end{figure*}

\textit{FoolsGold}~\cite{fung_MitigatingSybils_2020} is a gradient similarity-based anomaly detection method that aims to identify grouped adversaries. It utilizes the calculation of gradient cosine similarity to detect malicious attacks that exhibit high similarity. However, \textit{FoolsGold} is ineffective against individual adversaries. \textit{FLDetector}~\cite{zhang_FLDetectorDefending_2022} predicts a client's update by analyzing its past contributions. This defense mechanism has shown effectiveness against various adaptive attacks, such as scaling attacks, distributed backdoors, and untargeted model poisoning. However, the computational cost of predicting consistency is high. Another proposed defense mechanism suggested in \cite{li_LearningDetect_2020} involves training a spectral anomaly detection model to identify malicious updates in a low-dimensional latent space. This approach trains an autoencoder model on benign model updates to detect anomalous models. However, selecting appropriate benign model updates for training the autoencoder model is crucial. \textit{Sniper}~\cite{cao_UnderstandingDistributed_2019}maps client updates into a graph-cluster using gradient similarity and effectively defends against poisoning attacks, but its performance is reduced in a more heterogeneous setting.

\subsubsection{MTD-based Techniques}\label{sec:related_work:mtd}

MTD is an innovative security strategy that seeks to minimize the consequences of cyberattacks by actively or passively modifying specific system elements~\cite{cai2016moving}. Proposed in \cite{feng2023voyager}, \voyager{} is an MTD-based aggregation protocol that reactively alters the network topology to enhance the resilience of the DFL system. The \voyager{} protocol comprises three stages: an anomaly detector, a network topology explorer, and a connection deployer. The anomaly detector assesses the received models against shared models from other participants to identify any abnormal models, thereby triggering the MTD policy. The network topology explorer aims to identify more reliable participants. Lastly, the connection deployer establishes connections with these candidate participants and exchanges their respective models to create an aggregated model.

\subsubsection{Hybrid Defense Techniques}\label{sec:related_work:hybrid}
Hybrid defenses combine robust aggregation and anomaly detection techniques. In the \textit{FLTrust} approach ~\cite{cao_FLTrustByzantinerobust_2021}, the server uses a clean dataset to train a reference model and compares it to received model updates using cosine similarity. These updates are then aggregated based on their trust scores. However, obtaining a root dataset may not be possible in a CFL architecture. \textit{Trusted DFL}~\cite{gholami_TrustedDecentralized_2022} introduces trusted aggregation in a decentralized architecture by evaluating nodes based on their behavior and update consistency. Each node shares its local trust score, enabling other nodes to compute a global trust score for aggregation. \textit{FedInv}~\cite{zhao_FedInvByzantineRobust_2022} addresses the limitations of \textit{FLTrust} and \textit{PDGAN} by using a privacy-preserving model inversion strategy to generate dummy data from clients' gradient updates. These updates are scored based on distribution divergence and aggregated using majority clustering. However, the computational complexity of \textit{FedInv} is currently being discussed.

\begin{figure}[ht!]
\centering
\includegraphics[width=1\linewidth]{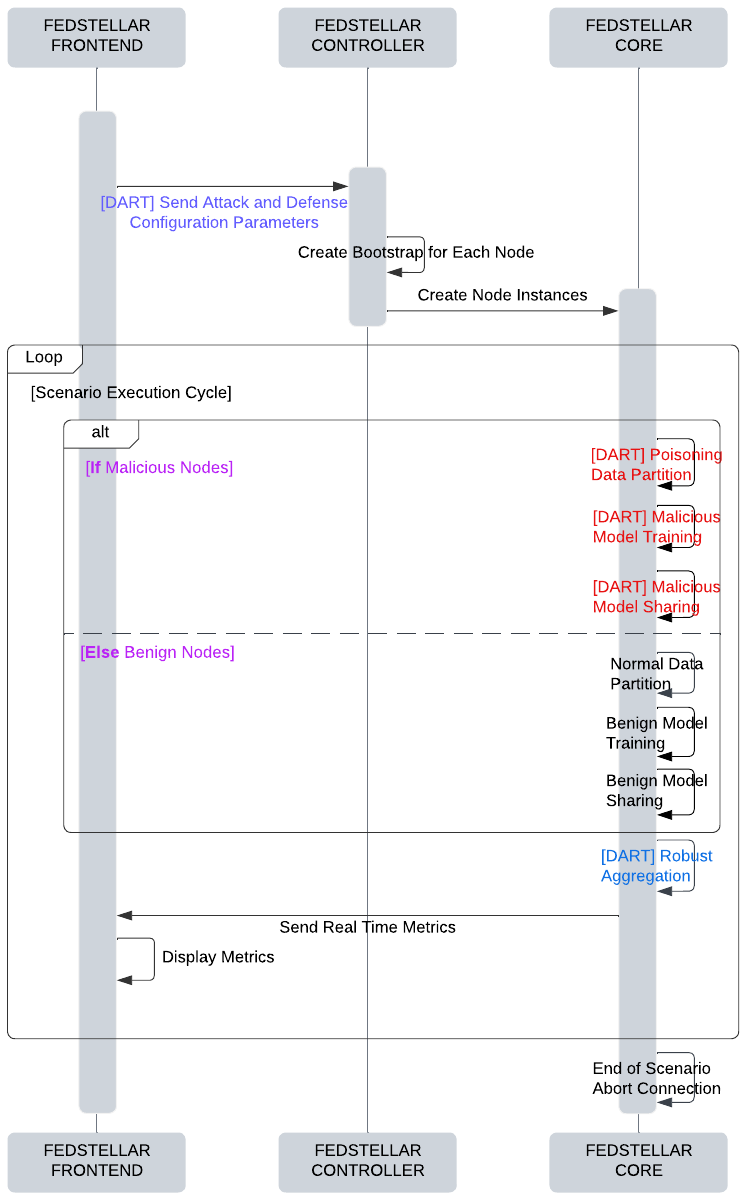}
\caption{{Sequence Diagram Showing the Procedure of {\textit{DART}}}}
\label{fig:dart_sequence}
\end{figure}

\textit{Norm Clipping}~\cite{ozdayi_DefendingBackdoors_2021} is a defense mechanism that limits the scale of model weights to protect against semantic backdoor attacks in CFL. However, it only considers the magnitude of an update and not its direction, making it difficult to determine an appropriate threshold. \textit{DeepSight}~\cite{rieger_DeepSightMitigating_2022} combines filtering and clipping to significantly reduce an attacker's possibilities by analyzing the neural network structure and creating fingerprints. \textit{DeepSight} is effective but computationally complex. \textit{FLAME}~\cite{nguyen_FLAMETaming_2021} proposes a hybrid approach using dynamic clustering, adaptive clipping, and noising. Client updates are clustered based on similarity, filtered for outliers, and then clipped. The aggregated model is smoothed using adaptive noising. However, \textit{FLAME} fails when the number of attackers exceeds half of the cluster.

\sentinel{}~\cite{feng2023sentinel} is a hybrid defense strategy to enhance the resilience of DFL against poisoning attacks. It introduces a three-phase aggregation protocol to strengthen security measures. In the first phase, a layer-wise average cosine similarity metric is used to identify and filter out suspicious model updates. The remaining updates are then aggregated based on local bootstrap validation loss, with aggregation weights determined through adaptive loss distance mapping. In the final phase, trusted models are normalized using the local model norm as a threshold to minimize the impact of potential stealth attacks.
\subsubsection{Post-Aggregation Techniques}\label{sec:related_work:post_aggregation}
Post-aggregation techniques refer to approaches that do not disrupt the process of aggregation. Instead, modifications to model updates are made after the aggregation has taken place. \cite{wu_MitigatingBackdoor_2021} proposed a method to defend against backdoor attacks in the CFL setting. They use neuron pruning to identify and deactivate maliciously activated neurons during an attack. Clients submit voting sequences based on their averaged activation values, and the server aggregates these votes to determine which neurons to prune. To enhance the defense, \cite{sun_CanYou_2019} suggested incorporating weak differential privacy using Gaussian noise. Adding even small amounts of noise can protect against attacks, but it may decrease the accuracy of the main task. However, determining the right amount of noise to add is challenging.

Overall, a significant amount of ongoing research is focused on enhancing the robustness of the FL systems model. However, most of the research focuses on CFL and does not consider DFL's unique characteristics. These studies also have limitations, such as being computationally complex and sensitive to data distribution.

\section{\solution{} Solution}
\label{sec:framework}

This section bridges the research gap highlighted in the previous section by designing and implementing a robustness analysis module for DFL systems. This module enables analysis of the robustness of DFL models when confronted with poisoning attacks and evaluation of the effectiveness of various defense mechanisms.

\subsection{Design of the \solution{} module}
In general, the training procedure of a DFL model can be divided into four stages, as shown in \figurename~\ref{fig:dart}. Initially, on the DFL platform, the user establishes the fundamental configuration of the federation, including the desired number of participants, the dataset to be used for training, the client interconnection topology, or the number of rounds of aggregation. After that, the user-defined configuration is transformed into specific bootstrapping for each individual node, encompassing details such as the actual training samples for each node, the neighbors to which they are connected, and the specific model architecture. With this configuration information, nodes start local model training while establishing communication channels with their neighboring nodes. Once the local model training is finished, each node exchanges and aggregates its respective models. These local training and model aggregation processes are repeated until the model reaches convergence or reaches a predetermined number of aggregation rounds. 

% \begin{figure}[ht]
%     \centering
%     \includegraphics[width=1\linewidth]{figures/architecutre.pdf}
%     \caption{\solution{} Module in \textit{Fedstellar} Platform}
%     \label{fig:arch_in_fed}
% \end{figure}

However, this procedure lacks consideration for the robustness analysis of the DFL model, meaning users cannot customize and execute poisoning attacks or select and implement defense mechanisms. Therefore, this study proposes the \solution{} module, consisting of two components: the attack component (highlighted red in \figurename~\ref{fig:dart}) and the defense component (highlighted green in \figurename~\ref{fig:dart}). When defining the desired federation, users can choose the type of poisoning attack they wish to deploy. Depending on whether the user wants to evaluate only the robustness of the DFL model or to assess the effectiveness of the defenses against poisoning attacks, the user can choose to deploy a defense mechanism.  During the bootstrapping phase, the benign node selects the defense mechanism while the malicious node chooses the desired poisoning attack. In the model training node, the malicious node trains the poisoned model and the benign node trains the uninfected model. During the aggregation phase, the benign node implements defenses to filter or mitigate attacks from the malicious node.

\begin{figure}[t]
\centering
\includegraphics[width=1\linewidth]{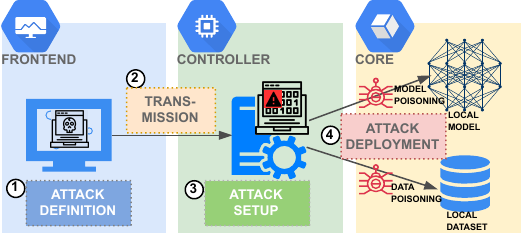}
\caption{Attack Component of \solution{}}
\label{fig:attackcomponent}
\end{figure}

Specifically, as depicted in the sequence diagram shown in Figure {\ref{fig:dart_sequence}}, upon selecting the DFL scenario in the frontend interface, the user could also choose the desired attack and defense settings to be executed. These configuration parameters are then communicated to the controller through the REST API. Subsequently, the controller writes these configuration parameters to the bootstrap items of each node, specifying which nodes are functioning as benign nodes and which are operating as malicious nodes. The core creates instances of the nodes based on their respective configuration parameters and starts executing the scenario. A normal dataset is provided for benign nodes to train the local model, which is then shared with neighboring nodes. In contrast, malicious nodes are supplied with poisoned data and either train a malicious model through data poisoning or directly tamper with the trained local model through model poisoning, transmitting the malicious model to neighboring nodes. Next, the node follows the robust aggregation algorithm chosen in the configuration for model aggregation, proceeding to the next round of the scenario execution loop.

\subsection{\solution{} Module in \textit{Fedstellar} Platform}
\textit{Fedstellar}~\cite{beltran_FedstellarPlatform_2023} is a DFL platform that provides a user-friendly interface to easily deploy and train DFL models. Therefore, this paper adopts \textit{Fedstellar} as the infrastructure and incorporates a model robustness analysis module, called \solution{}. As shown in \figurename~\ref{fig:arch_in_fed}, \textit{Fedstellar} consists of three layers, the frontend, the controller, and the core, corresponding to the four steps of the DFL model training process.

\begin{itemize}
	\item \textbf{\textsc{Frontend}}. This layer corresponds to the federation setup step and enables the user and system interaction, allowing the user to personalize their own FL scenario through diverse configurations. Moreover, the system provides real-time monitoring functionalities, including the tracking of resource usage and model performance during the training process. Building upon this layer, \solution{} provides the option to choose attacks and defense mechanisms.
 
	\item \textbf{\textsc{Controller}}. This layer corresponds to the bootstrapping step and functions as the central hub of the platform. It receives user configurations from the frontend, manages the federated scenario, selects appropriate learning algorithms and datasets, and establishes network connections to enable a FL process. The attack and defense are specified within this layer. 
	\item \textbf{\textsc{Core}}. This layer corresponds to the training and aggregation steps. Its operations entail data preparation, model training, ensuring secure communication among participants, and storing the federated models. Therefore, this layer encompasses the primary functionality of \solution{}, including the deployment and execution of attacks and defenses.
\end{itemize}

 \begin{figure}[t]
    \centering
    \includegraphics[width=1\linewidth]{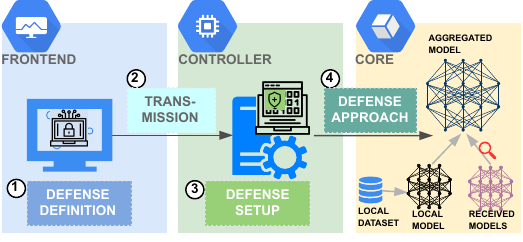}
    \caption{Defense Component of \solution{}}
    \label{fig:defensecomponent}
\end{figure}

\subsubsection{\solution{}: Attack Component}
The \solution{} attack component consists of three parts, as illustrated in \figurename~\ref{fig:attackcomponent}. The process includes four steps; first, when the user selects the DFL configuration in the Frontend, the user can also select the attack definitions, such as what type of attack is performed, the percentage of nodes to be attacked, and the percentage of noise injection. These configurations are transmitted to the Controller as JSON files. The Controller then sets up specific attacks in the nodes based on these attack configurations, such as which nodes are configured as malicious nodes and what attack strategies these nodes use. 

The Controller utilizes this attack information along with other configurations to bootstrap the node's operation. In the fourth step, the Core starts the execution of these attacks. These attacks encompass two parts: one involves manipulating the data, such as altering the training data of the model to facilitate a data poisoning attack, while the other entails directly modifying the model, known as a model poisoning attack. The attack component of \solution{} integrates the following poisoning attacks: Untargeted Label Flipping, Untargeted Sample Poisoning, Random Model Poisoning, Targeted Label Flipping, and Backdoor Attack.

\subsubsection{\solution{}: Defense Component}
Similarly to the previous component, the \solution{} defense component is incorporated within the three layers of \textit{Fedstellar} and operates based on the same underlying principles, as shown in \figurename~\ref{fig:defensecomponent}. Firstly, the user operating the Frontend chooses the specific defense mechanism that they wish to evaluate. In the second step, these configurations are transmitted to the backend Controller as JSON files. In the third step, the Controller converts these configurations into bootstrap for the node, including which defense mechanisms to employ. In the fourth step, the Core starts to train the local model, as well as run the defense mechanisms to protect its model from malicious models. The defense component incorporates a range of defense mechanisms, namely \textit{Krum}, \textit{Median}, \textit{TrimmedMean}, \textit{FLTrust}, \sentinel{}, and the MTD-based \voyager{}.

\subsection{{Implementation of \solution{}}}

\begin{figure*}[t]
    \centering
    \includegraphics[width=1\linewidth]{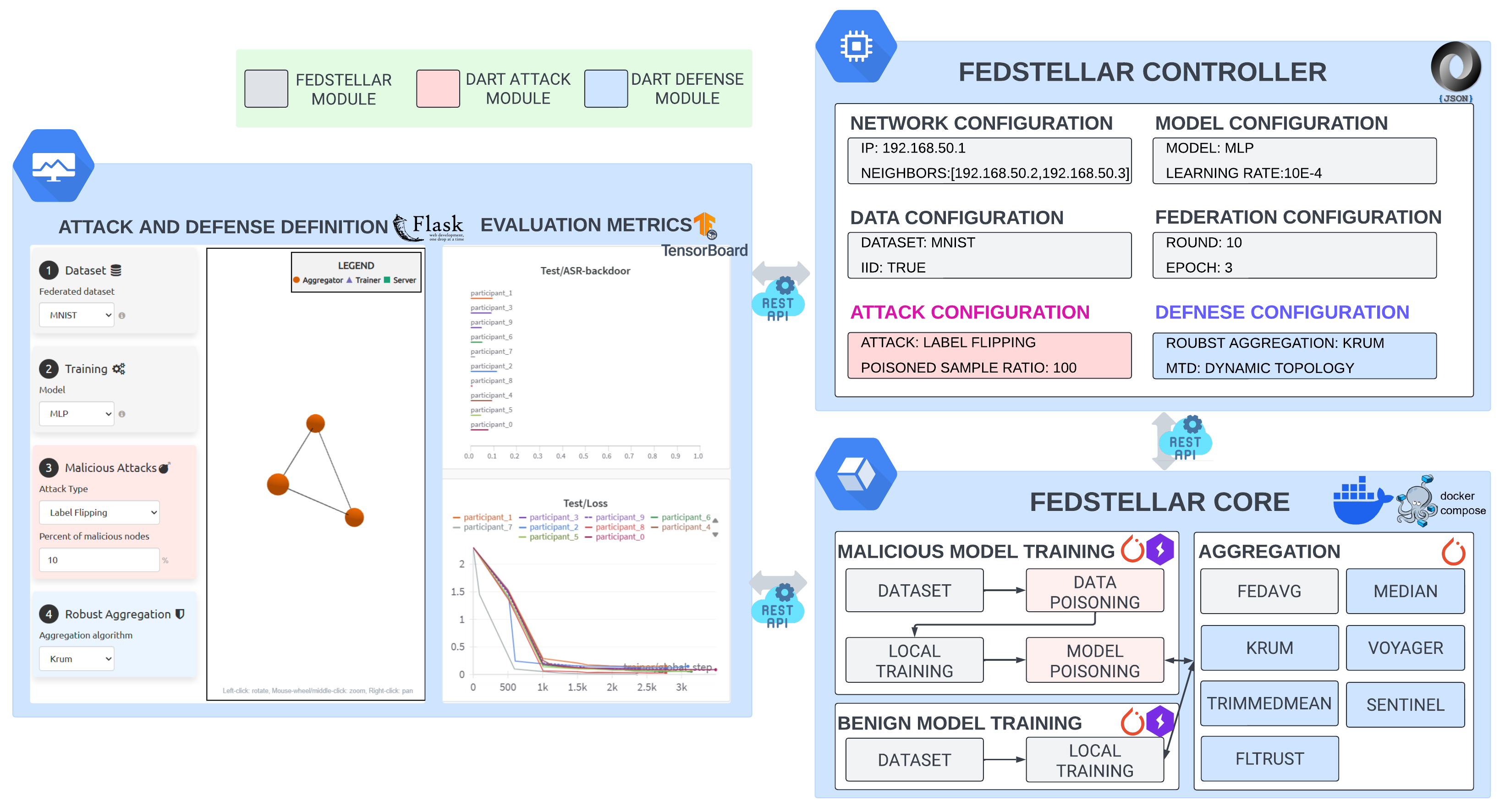}
    \caption{{Integration of \textit{DART} Module in \textit{Fedstellar} Platform}}
    \label{fig:arch_in_fed}
\end{figure*}

Figure {\ref{fig:arch_in_fed}} illustrates the implementation and integration of DART within the \textit{Fedstellar} platform. \textit{Fedstellar} provides the fundamental DFL scenario configuration elements on the frontend, while \textit{DART} enhances this by introducing the option for malicious attacks and defense mechanisms. Users are able to select which attacks to execute and determine the ratio of malicious nodes within the federation. Additionally, users can choose from various robust aggregation functions for defense, such as \textit{Krum} and \textit{TrimmedMean}. The frontend of \textit{DART} is developed using the Flask framework {\cite{Flask}}, with user selections communicated to the backend controller through the REST API.

The controller generates bootstrapping details for each node based on the user's selections, encompassing network settings, neighbor connections, and datasets. \textit{DART} adds configuration specifics for attacks and defenses. Malicious nodes are configured with attack settings, while benign nodes are designated as "no attack". The chosen robust aggregation function for defense is incorporated into the node's bootstrap as the model aggregation algorithm. Configuration details for each node are stored in JSON format on the file system to facilitate node bootstrap and are also transmitted to the core system through REST API.

\begin{algorithm}[h]
\caption{{\textit{DART} PROCESS IN EACH NODE}}
\label{alg:dart}
\begin{algorithmic}[1]
\small
\Require $D_{i}$: Local Dataset, $\mathcal{F}$: Model Training, $\mathcal{A}$: Aggregation Function, $P$: Neighbor List, $R$: Total Rounds, $\mathcal{T}$: Model Transmission, $\mathcal{DP}$: Data Poisoning, $\mathcal{MP}$: Model Poisoning. 
\State $r \gets$ 0
\State $m_i \gets$ Local Model
\State $M_i \gets$ Neighbor Model List

\If{Data Poisoning}   \Comment{Poison Local Dataset}
    \State $D_{i} \gets \mathcal{DP}(D_{i})$  
\EndIf

\For{$r \leq R$}
    \State $M_i \gets [\ ]$ 
    \State $m_i \gets \mathcal{F}(D_{i}, m_i)$ \Comment{Local Model Training}

    \If{Model Poisoning}   \Comment{Poison Local Model}
        \State $M_{i} \gets \mathcal{MP}(M_{i})$  
    \EndIf

    \State $\mathcal{T} (M_i, P)$ \Comment{Transmit Local Model to Neighbors}
    \For{$j$ in $P$}          \Comment{Receive Models from Neighbors}
    \State $M_i \gets M_i \cup \{e\}$
    \EndFor
    \State $m_i \gets \mathcal{A}(m_i, M_i)$ \Comment{Model Aggregation}
\EndFor
\end{algorithmic}
\end{algorithm}

The core system creates instances of individual nodes through the bootstrap information. Docker containers {\cite{docker}} are used to virtualize the nodes. When the docker instance is started, the nodes follow the configuration information and get the local data. The \textit{DART} process in each node is shown in Algorithm {\ref{alg:dart}}. If a node is benign, it will use the regular model training pipeline to train the model locally, transmit the local model to its neighbors, and wait for the model from its neighbors. If a node is malicious, then it will manipulate the data according to attack configuration, get the poisoned local dataset, train the malicious model, and then share it with the neighbors; or the node modifies the local model, gets the poisoned model, and shares it with the neighbors. The local model is developed using the Pytorch framework {\cite{paszke2019pytorchimperativestylehighperformance}}, with Pytorch Lightning {\cite{pytorch_lightning}} serving as the model trainer for model optimization.

Following the acquisition of models shared by neighboring nodes, the node utilizes the selected algorithm outlined in the bootstrap configuration to aggregate models. After the aggregation is completed, the next round of local model training and model aggregation is performed. Throughout the model training and testing phases, metrics, including model performance, attack, and defense effectiveness, are transmitted from core to frontend utilizing a REST API. The frontend then visualizes these metrics via TensorBoard {\cite{tensorboard}}.

\section{Experimental Analysis}
\label{sec:experiments}

This section compares and analyzes the impact of poisoning attacks on the model robustness of CFL and DFL paradigms and identifies the factors influencing their efficacy. Besides, this section benchmarks various defense mechanisms in diverse attack scenarios.

\subsection{Datasets and Federation Setups} 
\label{subsection:setups}
The following datasets and deep learning model are chosen to evaluate the aggregation algorithms implemented for the Fedstellar framework:

\begin{itemize}
    \item \textbf{MNIST}~\cite{lecun_MNISTHandwritten_2010} consists of handwritten digits represented by 28×28 grayscale images. It compromises $60\,000$ training samples and $10\,000$ test samples. A three layers multilayer perceptron (MLP) with a linear input layer of size $784 \times 256$, a linear hidden layer of size $256 \times 128$, and a linear output layer of size $128 \times 10$ is used for classification.
   
    \item \textbf{FashionMNIST}~\cite{xiao_FashionMNISTNovel_2017} consists of $60\,000$ training samples and $10\,000$ test samples, which are 28×28 grayscale images with 10 classes.For the FashionMNIST dataset, the same MLP as for MNIST is used.
    
    \item \textbf{Cifar10}~\cite{krizhevsky_LearningMultiple_2009} consists of $50\,000$ training and $10\,000$ test images with 10 classes of 32×32 color images. A small convolutional neural network (CNN) designed for mobile applications is used for this task \cite{howard_MobileNetsEfficient_2017}. 
    
\end{itemize}

\begin{table}[h]
\centering
\caption{Configuration of Experiments}
\label{tab:config}
\resizebox{\columnwidth}{!}{%
\begin{tabular}{lll}
\toprule
\multicolumn{2}{l}{\textbf{CONFIGURATION}} & \textbf{VALUE} \\ \midrule
\multirow{5}{*}{\textbf{Federation}} & Number of Clients & 10 \\ \cmidrule(l){2-3}
 & Total Rounds & 10 \\ \cmidrule(l){2-3}
 & Epochs in Each Round & 3 \\ \cmidrule(l){2-3}
 & Paradigm & CFL, DFL \\ \cmidrule(l){2-3}
 & Data Distribution & IID \\ \hline
\multirow{3}{*}{\textbf{Robustness}} & Attacks & \begin{tabular}[c]{@{}l@{}}Untargeted Label Flipping\\ Untargeted Sample Poisoning\\ Random Model Poisoning\\ Targeted Label Flipping \\ Backdoor Attack\end{tabular} \\ \cmidrule(l){2-3}
 & Poisoned Node Ratios (PNR) & 0\%, 10\%, 30\%, 50\%, 70\%, 90\% \\ \cmidrule(l){2-3}
 & Aggregation Function & \begin{tabular}[c]{@{}l@{}} \textit{FedAvg}, \textit{Krum},   \textit{Median}, \\ \textit{TrimmedMean}, \textit{FLTrust}, \\ \textit{Sentinal},   \textit{Voyager}\end{tabular} \\ \hline 
\multirow{3}{*}{\textbf{Network}} & Bandwidth & 1 Mbps \\ \cmidrule(l){2-3}
 & Delay & 0 ms \\ \cmidrule(l){2-3}
 & Topology & \begin{tabular}[c]{@{}l@{}}Fully connected, Ring, \\ Star,   Random\end{tabular} \\\midrule
\end{tabular} 
}
\end{table}

\label{subsection:untargetedattack}
\begin{figure*}[h]
\centering
\includegraphics[width=0.70\linewidth]{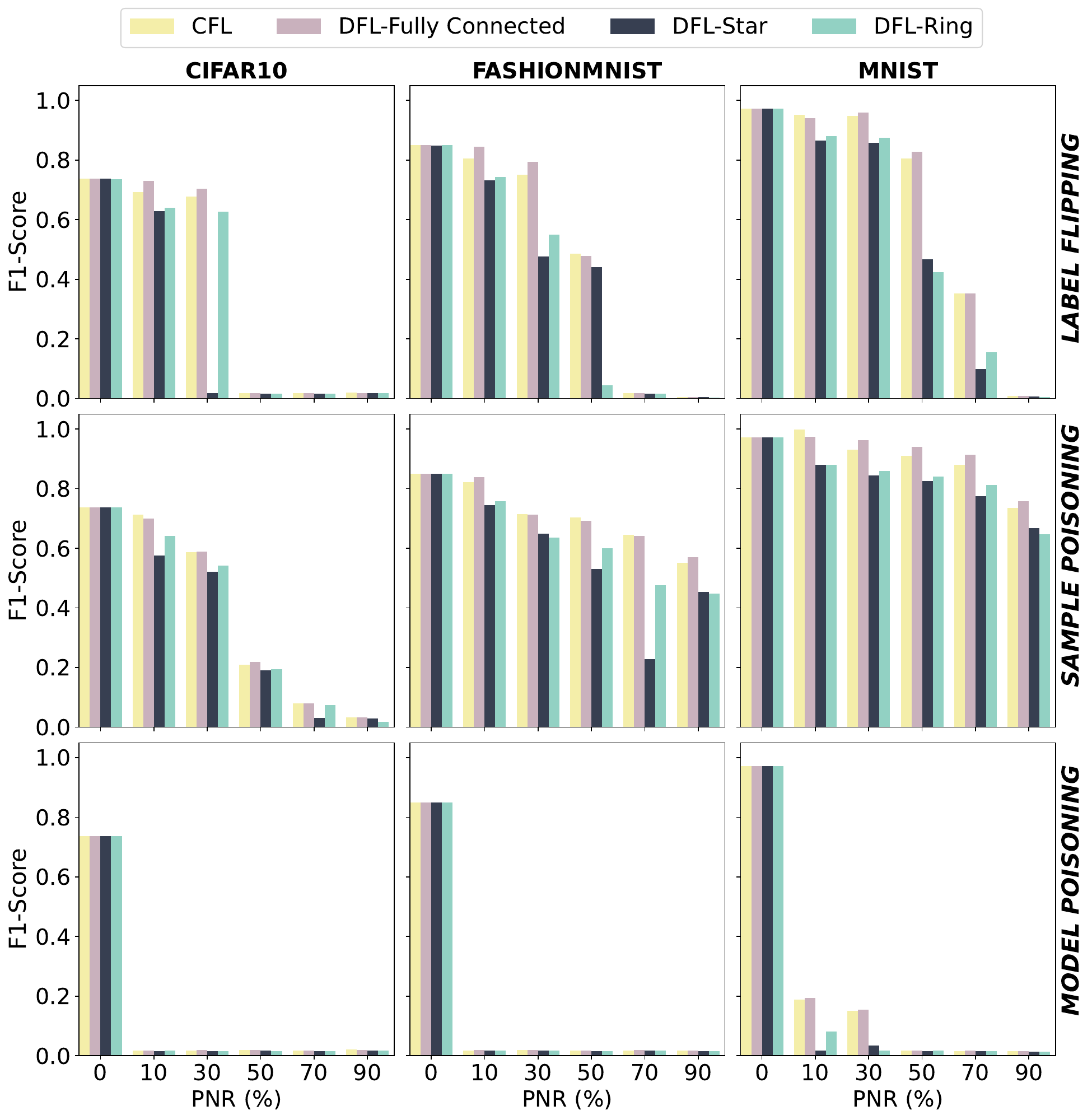}
\caption{Average F1-Score Results for Untargeted Poisoning Attacks}
\label{fig:untargetedattacks}
\end{figure*}

The experiment configurations are listed in \tablename~\ref{tab:config}. The learning process is designed to run for a total of 10 rounds, with each round comprising three epochs. All experiments are performed in the federation composed of 10 clients, and the data are evenly distributed across the clients in an IID manner. All experiments are carried out on the \textit{Fedstellar} platform in synchronous mode. The federation network has a bandwidth of 1 Mbps without loss and delay. For the experiments with different attacks, the Poisoned Node Ratios (PNR) are increased from 0\% to 90\%. The defense mechanisms \textit{Krum}, \textit{Median}, \textit{TrimmedMean}, \textit{FLTrust}, \textit{Sentinal}, and \voyager{} are implemented, assessed, and compared. Furthermore, the experiments employ four different participants interconnection topologies, \ie ring, star, random, and fully connected.

\subsection{Poisoning Attack on CFL and DFL}

This experiment first analyzes the performance of CFL and DFL in the presence of different types of poisoning attacks. When executing the attacks, it assumes that all the attacks occur during the model training process and recur in each round. These attacks encompass both data poisoning and model poisoning strategies with untargeted and targeted intentions. Experiments with malicious client proportions ranging from 0\% to 90\% are executed to observe the effect of different degrees of maliciousness on the model robustness. 

\subsubsection{Experimental Analysis of Untargeted Attacks}
\begin{figure*}[h]
\centering
\includegraphics[width=0.70\linewidth]{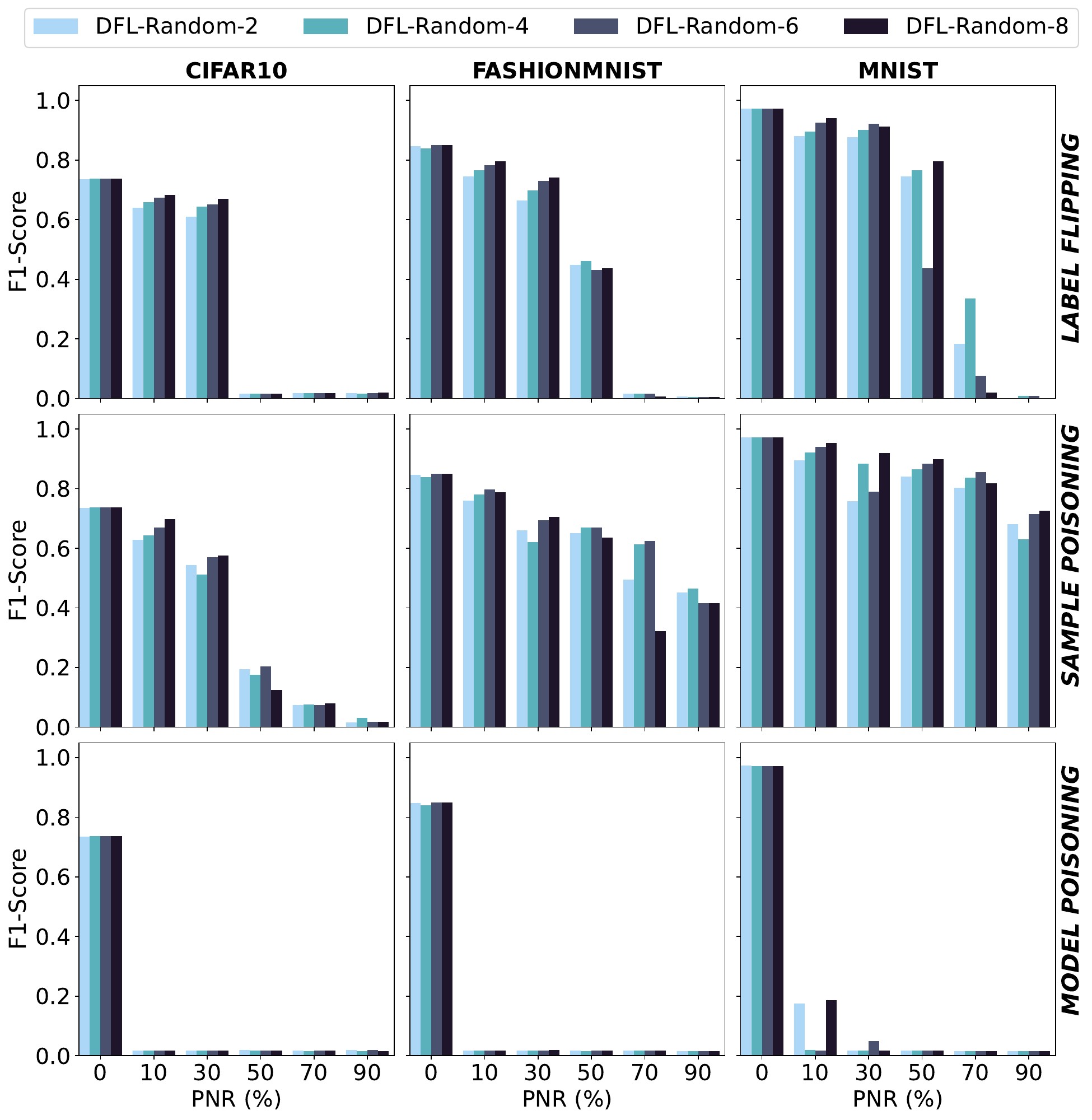}
\caption{Average F1-Score Results for Untargeted Poisoning Attacks with Random Topologies}
\label{fig:untargeted-random}
\end{figure*}

Three untargeted attacks have been implemented to compare the attack performance in CFL and DFL, including Untargeted Label Flipping,  Untargeted Sample Poisoning, and Random Model Poisoning attacks. The detailed attack setups are listed below:
\begin{itemize}
 \item  \textbf{Untargeted Label Flipping}: Randomly replacing all the labels corresponding to the samples in the training set such that the model is unable to map the data patterns of the samples to the correct labels. Thus, the loss function cannot correctly guide the optimization direction of the model.
 
 \item  \textbf{Untargeted Sample Poisoning}: 30\% of Gaussian noise is added to all of the samples in the training set to mask the correct features in the data. As a result, the model fails to recognize effective patterns in the data and cannot classify the data correctly.
 
 \item \textbf{Random Model Poisoning}: Before sending the model for aggregation, the malicious node directly adds 30\% of Gaussian noise to the parameters of the trained local model to devastate the model's effectiveness.

\end{itemize}

\paragraph{\textbf{\1 Evaluation Metrics for Untargeted Attacks}} \mbox{}\\

The purpose of an untargeted attack is to spread malice in the FL system and reduce the effectiveness of the models of all nodes. Therefore, the average F1-Score in the model's benign nodes is calculated as the metric to evaluate the model's robustness and the effectiveness of the untargeted attack. A lower F1-Score indicates that the untargeted attack is more effective, but the FL models have lower robustness. This experiment is carried out in CFL and DFL utilizing fully connected, ring, and star network topologies. Besides, FedAvg is employed as the aggregation function across all nodes without additional defense mechanisms.

\paragraph{\textbf{\2 Attack Effectiveness}} \mbox{}\\

\figurename~\ref{fig:untargetedattacks} illustrates the impact, \ie average F1-Score, of three untargeted attacks on CFL and DFL with three datasets as the PNR increases. Overall, the effects of the Model Poisoning attack are much more destructive, with 10\% of the nodes attacked having a significant reduction in the modeling effects of both CFL and DFL. With more than 30\% of the nodes attacked, the model of the benign node is close to being completely destroyed and no longer able to provide meaningful inferences in all of these three datasets. The Label Flipping Attack shows a strong effect only after more than 30\% of malicious nodes, and the model of benign nodes is destroyed when the percentage of malicious nodes exceeds 50\%. In contrast, the effect of Sample Poisoning is weaker, and the effect of the attack is shown when the percentage of malicious nodes is greater than 50\%, but even if the number of nodes attacked reaches 90\%, for the MNIST dataset, the rest of the benign node model still has close to 80\% of the F1-Score, which suggests that Sample Poisoning in the attack strategy is not as effective as the other two in the untargeted attack. The reason is that since the noise injected by Sample Poisoning does not always cover all of the features, the model is able to learn some patterns from limited information.

\begin{figure*}[h]
\centering
\includegraphics[width=0.70\linewidth]{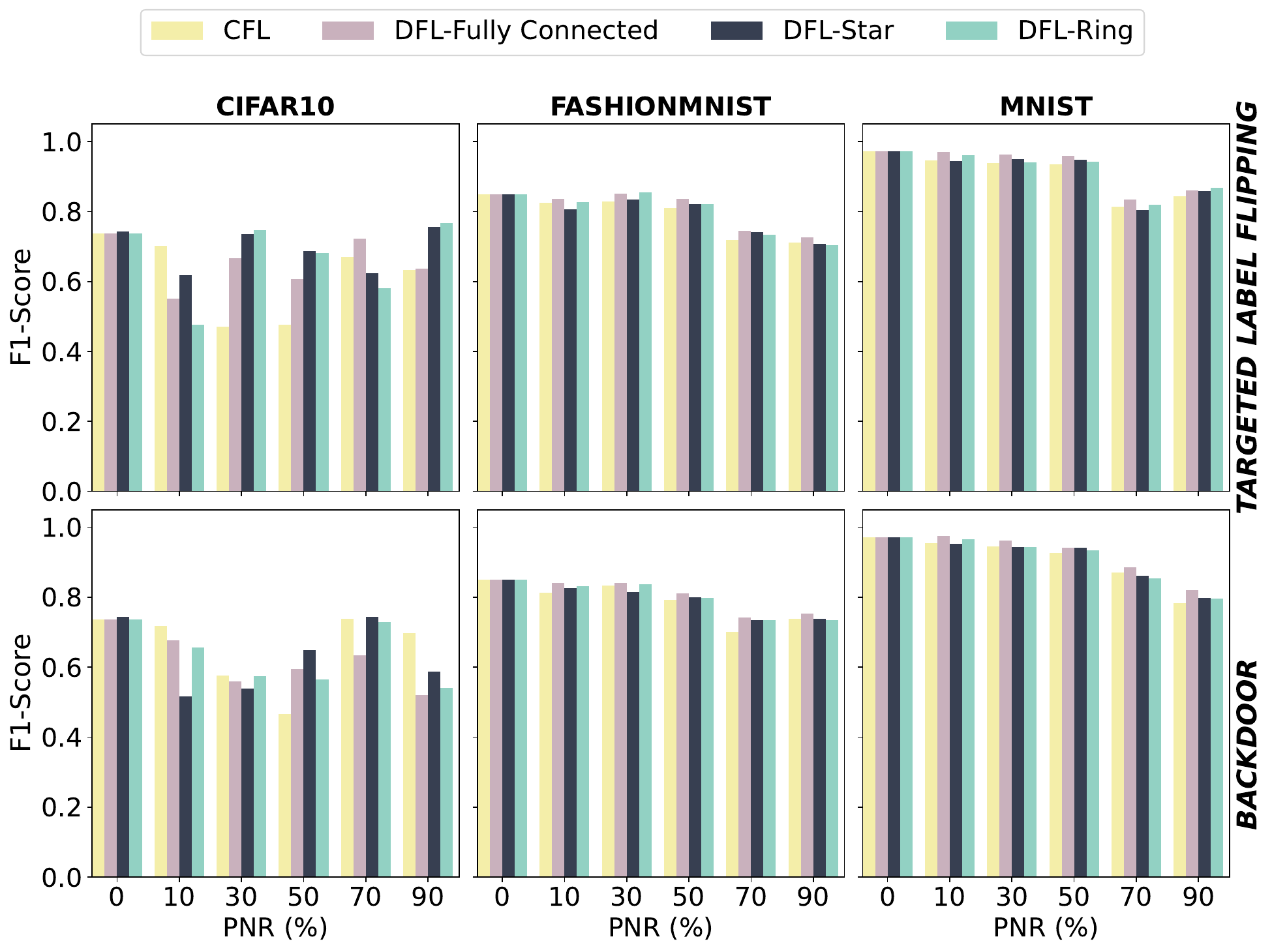}
\caption{Average F1-Score Results for Targeted Poisoning Attacks}
\label{fig:targetedattacks_f1}
\end{figure*}

\paragraph{\textbf{\3 Model Robustness in Different Paradigms and Topologies}} \mbox{}\\

For different FL paradigms, the CFL and DFL with fully connected topology exhibit a similar decrease in the model F1-Score across the attacks. However, the model robustness of DFL is affected by the manner in which participants are interconnected, \ie the topology. The results from \figurename~\ref{fig:untargetedattacks} indicate that DFL with the ring and star topology exhibit reduced levels of model robustness in counter when faced with untargeted poisoning attacks.

\cite{feng2023voyager} suggests that a benign node's connections to malicious nodes follow a hypergeometric distribution. Nodes with fewer average neighbors in a DFL network are more susceptible to poisoned attacks. Compared to fully connected, star and ring topologies are sparser, resulting in a lower decrease in F1-Score when facing untargeted attacks. To validate this hypothesis, this paper conducts an experiment on different DFL connected by random networks, using the Watts-Strogatz~\cite{watts1998collective} model to generate random small-world networks with average neighbor numbers of 2, 4, 6, and 8. 

\figurename~\ref{fig:untargeted-random} illustrates the average F1-Score of three untargeted attacks on the random topology of DFL with different average connected neighbors with three datasets as the PNR increases. The DFL model with 8 neighbors performs the best against the three attacks, followed by 6 neighbors, while 2 neighbors perform the worst. These results demonstrate that the network sparsity significantly influences the model robustness of DFL. Specifically, a denser network, characterized by a higher average number of connected neighbors, enhances the robustness of the model against poisoning attacks. This is because nodes need to aggregate fewer models when the average number of connected neighbors is small. Therefore, once aggregated with a malicious model, the benign node model will be greatly affected by the malicious ones. Whereas for a dense network, such as fully connected, although there is a high probability for a benign node to aggregate with a malicious node since the aggregation occurs in a large number of models, the maliciousness is thereby diluted and demonstrates better robustness.

In conclusion, when confronted with untargeted attacks, CFL and DFL exhibit similar performance. The model robustness of DFL is influenced by the network topology, with denser networks yield greater robustness. In terms of attacks, Model Poisoning Attacks prove to be the most efficient across various attack strategies. Label Flipping produces comparable outcomes but necessitates a higher proportion of malicious nodes. Conversely, Sample Poisoning Attacks are less effective than the aforementioned strategies regarding untargeted attacks.

\subsubsection{Experimental Analysis of Targeted Attacks}
\label{subsection:targetedattack}

In this section, two targeted attacks are designed and implemented: Targeted Label Flipping and Backdoor Attack. The detailed attack setups are listed below:

\begin{itemize}
 
 \item \textbf{Targeted Label Flipping}: Replacing only specific labels, specifically, replacing all the label 7 with 4 in the training dataset. Thus, for the model, the feature patterns of the training samples originally belonging to label 7 will be recognized as label 4. Consequently, all the samples that should have corresponded to 7 will be incorrectly classified as label 4, whereas for the other labels, their classification results are not affected.
 
 \item \textbf{Backdoor Attack}: A $10 \times 10$ pixels of an X shape watermark is added to the top right corner of the samples that belong to label 4 in the training dataset, such that the model learns the pattern corresponding to label 4 as the implanted watermark. In the testing phase, when this watermark does not appear, the model behaves normally, while the samples containing the watermark are misclassified as label 4.

\end{itemize}

\begin{figure*}[h]
     \centering
     \begin{subfigure}[b]{0.7\textwidth}
         \includegraphics[width=\textwidth]{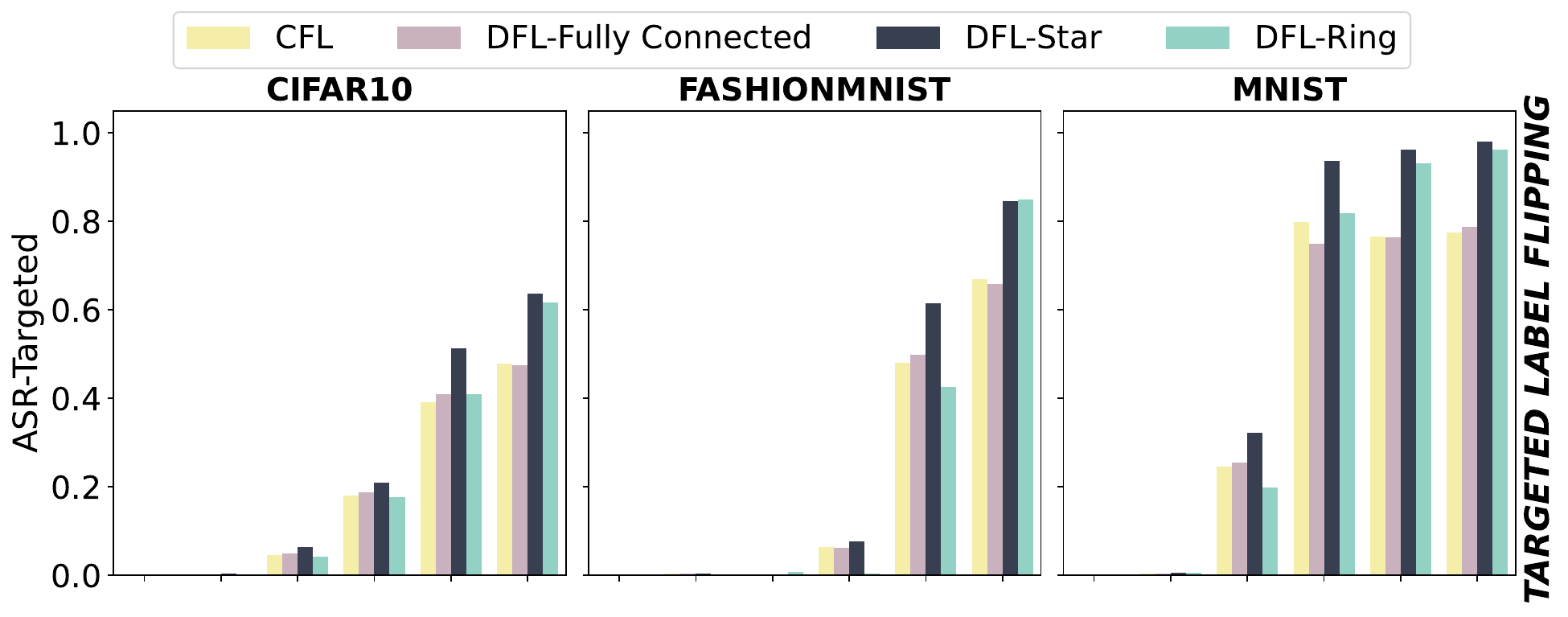} \hspace*{-0.9em}
         \label{fig:targetedlabelflipping-asr}
     \end{subfigure}
     \begin{subfigure}[b]{0.7\textwidth}
         \includegraphics[width=\textwidth]{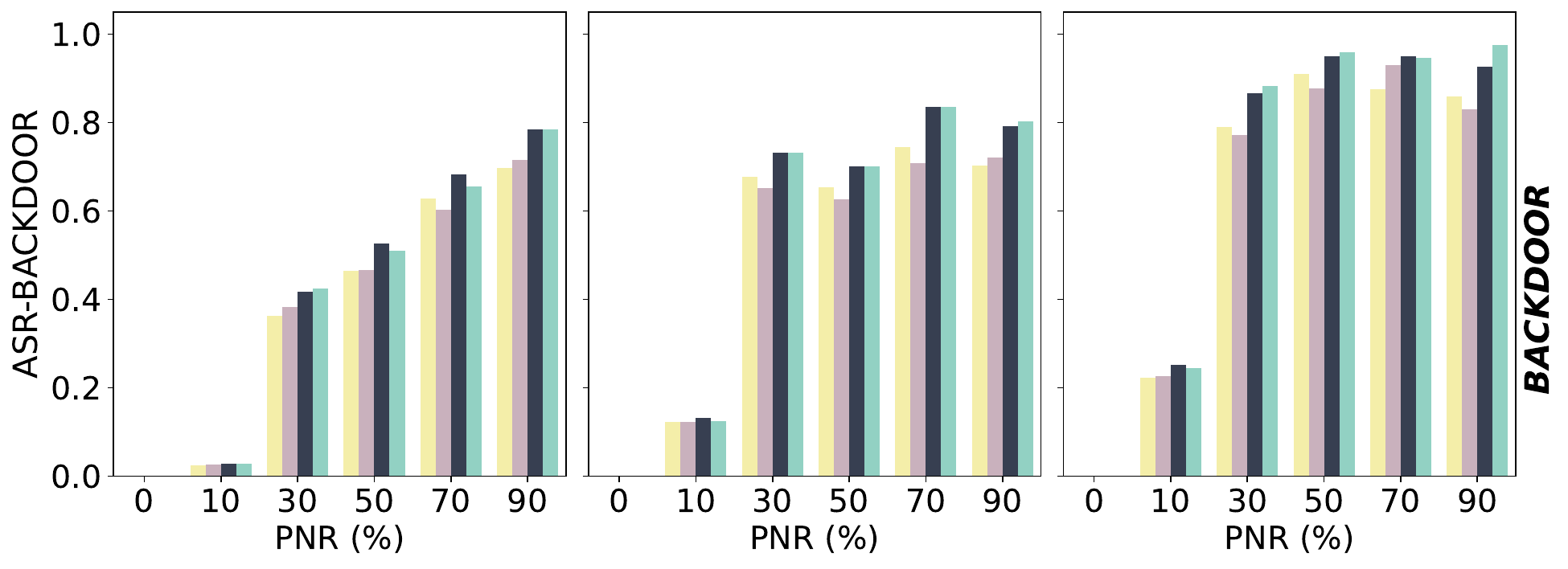}
         \label{fig:targetedbackdoor-asr}
     \end{subfigure}
    \caption{Average ASR Results for Targeted Poisoning Attacks}
    \label{fig:targeted-asr}
\end{figure*}

\paragraph{\textbf{\1 Evaluation Metrics for Targeted Attacks}} \mbox{}\\

This experiment first evaluates the overall performance of the model, \ie the F1-Score, when subjected to these targeted attacks, as shown in \figurename~\ref{fig:targetedattacks_f1}. The results show that even when 90\% of the nodes in both CFL and DFL models are suffering from targeted attacks, the overall performance on the model is not significantly dropped compared with no attack presence. This finding suggests that targeted attacks are more challenging to be detected through overall performance metrics. Therefore, this section adopts Attack Success Rate (ASR) to measure the effectiveness of targeted attacks.

The objective of Targeted Label Flipping is to cause misclassification of a source label $l_{src}$ to a desired target label $l_{t}$. The ASR-Targeted is determined by the number of samples in which the true label $y = l_{src}$ is incorrectly predicted as the target label $\hat{y} = l_{t}$, as described in Equation \eqref{eq:ASR_lf}. $c_{ij}$ represents the count of samples with the true label $y_i$ and predicted label $\hat{y}_j$, and the set $L$ represents the labels present in the dataset.

\begin{equation} \label{eq:ASR_lf}
    ASR-Targeted = \frac{c_{src, t}}{\sum_{j=0}^{|L|} c_{src, j}}
\end{equation}

The Backdoor attack involves the adversary inserting a trigger into local data samples that are associated with a particular target label $l_{t}$. The effectiveness of the Backdoor attack is evaluated using ASR-Backdoor, which is defined in Equation \eqref{eq:BA}. Here, $c_{ij}$ denotes the number of samples that have a true label $y_i$ but a predicted label $\hat{y}_j$, while $L$ represents the set of labels in the dataset under consideration.

\begin{equation} \label{eq:BA}
    ASR-Backdoor = \frac{\sum_{j=0}^{|L|} c_{j, t} - c_{t, t}}{|D| - c_{t, t}}
\end{equation}

The ASR metrics are calculated as the mean result across all benign participants, with higher values indicating greater effectiveness of the targeted attack.

\begin{figure*}[h]
     \centering
     \begin{subfigure}[b]{0.7\textwidth}
         \includegraphics[width=\textwidth]{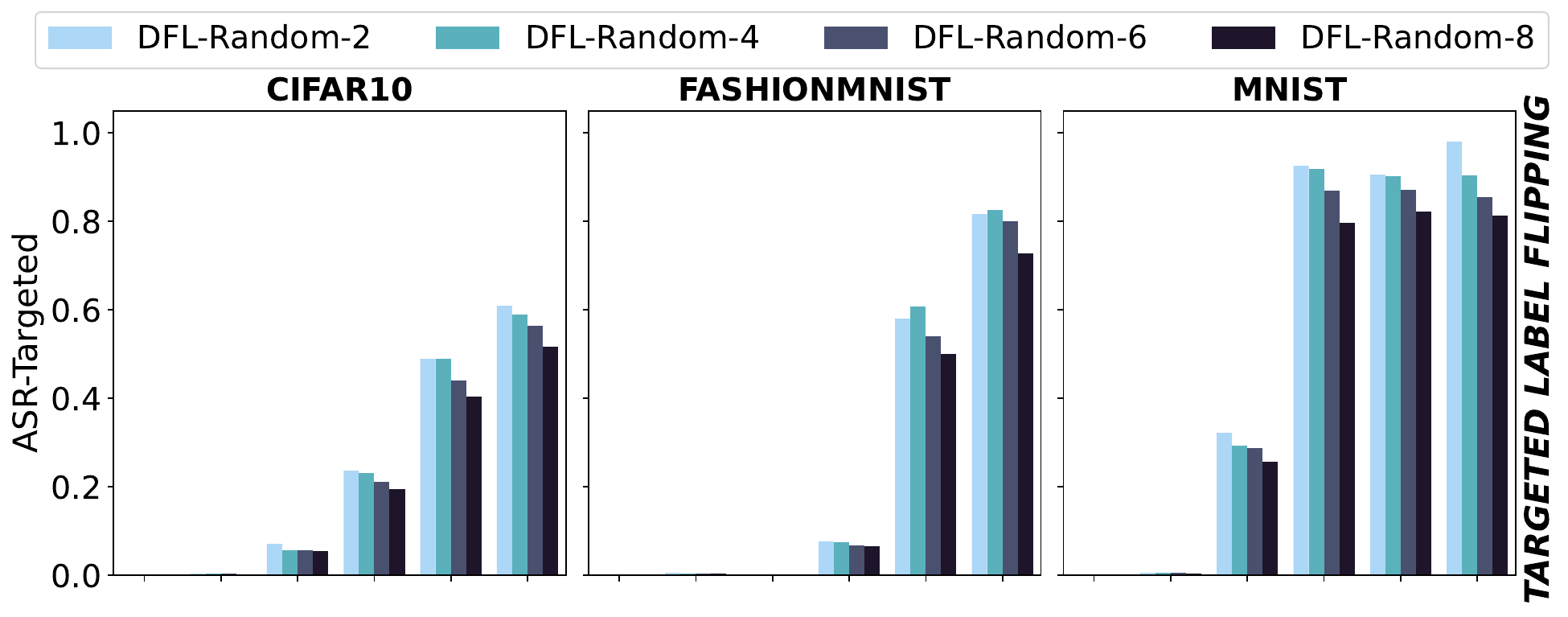} \hspace*{-0.9em}
         \label{fig:targetedlabelflipping-asr-random}
     \end{subfigure}
     \begin{subfigure}[b]{0.7\textwidth}
         \includegraphics[width=\textwidth]{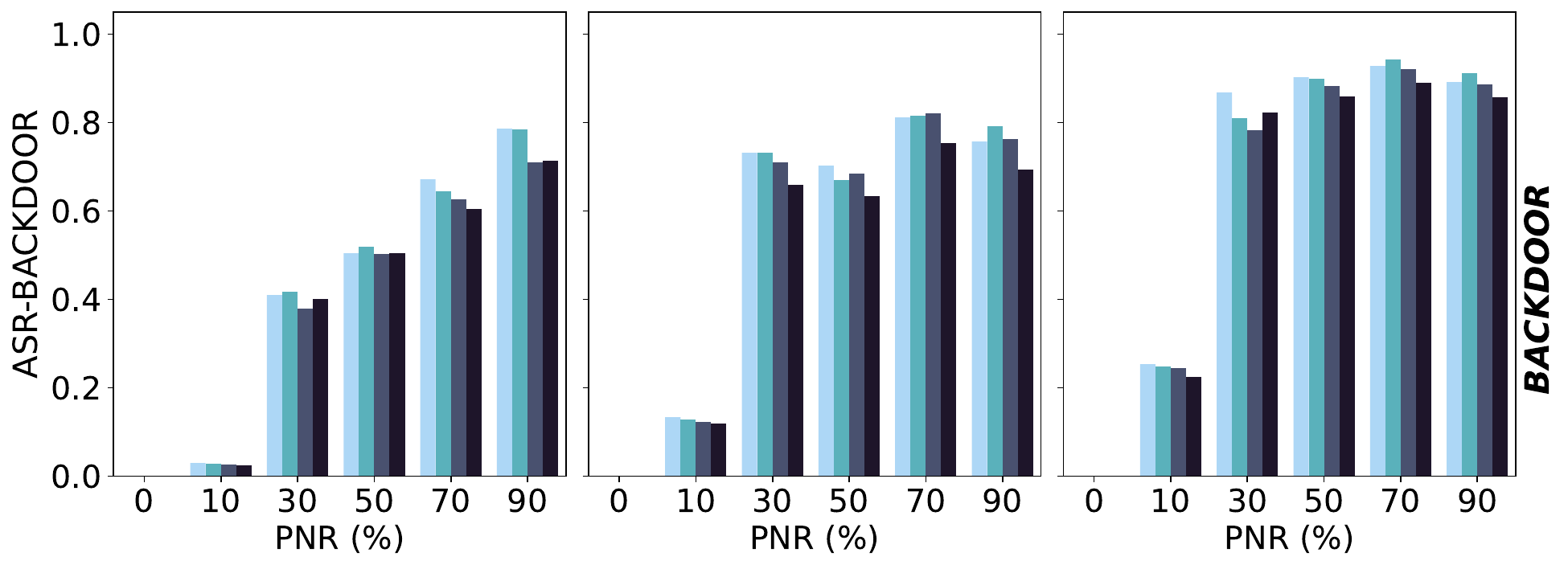}
         \label{fig:targetedbackdoor-asr-random}
     \end{subfigure}
    \caption{Average ASR Results for Targeted Poisoning Attacks with Random Topologies}
    \label{fig:targeted-asr-random}
\end{figure*}

\paragraph{\textbf{\2 Attack Effectiveness}} \mbox{}\\

\figurename~\ref{fig:targeted-asr} presents the average ASR metrics of Targeted Label Flipping and Backdoor attacks on CFL and DFL with three datasets as the PNR increases. Based on the results, in terms of attack effectiveness, the Backdoor Attack, which utilizes manipulated data samples, is found to be more effective than the Targeted Label Flipping attack, which uses manipulated labels. From the experimental results, it becomes evident that more than 30\% of nodes injected with a backdoor can lead to a backdoor injection success rate of over 50\% in all three datasets. Targeted Label Flipping requires at least 50\% of the nodes to be attacked to achieve similar results, and if the PNR is below 30\%, the influence on the models is considered negligible.

\paragraph{\textbf{\3 Model Robustness in Different Paradigms and Topologies}} \mbox{}\\

As in the case of untargeted attacks, CFL and DFL with fully connected topology demonstrate similar model robustness against targeted attacks. When the poisoned node proportion is low, the influence of this type of attack on the overall performance of the federation is insignificant, particularly in the case of Targeted Label Flipping.

This section evaluates the impact of different topologies on targeted poisoning attacks by implementing the same setups of random topology experiments of untargeted attacks. The results of this experiment are shown in  \figurename~\ref{fig:targeted-asr-random}. Similar to untargeted attacks, topology has an impact on the effectiveness of targeted attacks. A denser network facilitates the spread of targeted attacks, but the aggregation of more models also has a negative effect on the maliciousness. Thus, a higher ASR is observed in a low-density network than in a high-density network. However, for the Backdoor Attack, once the PNR exceeds 50\%, the impact of topology becomes less significant.

To summarize, Backdoor attacks by manipulating data are more effective than Target Label Flipping attacks by manipulating labels. Meanwhile, similar to untargeted attacks, the robustness of the model against targeted poisoning attacks for CFL and DFL with fully connected topology is roughly the same. Besides, the topology of the network plays a crucial role in the spread of maliciousness. When benign nodes are directly connected to malicious nodes, the more sparse the network, the more vulnerable they are to attacks.

% \usepackage{multirow}
% \usepackage{graphicx}
\begin{table*}[]
\centering
\caption{Benchmark of Average F1-Score Results for Defense Mechanisms in Mitigating Untargeted Poisoning Attacks}
\label{tab:untargeted}
\resizebox{\textwidth}{!}{%
\begin{tabular}{llllllll|lllll|lllll}
\toprule
 &  &  & \multicolumn{5}{c}{\textbf{Label Flipping }} & \multicolumn{5}{c}{\textbf{Model Poisoning }} &  \multicolumn{5}{c}{\textbf{Sample Poisoning }} \\ \midrule
 
 &  & \textbf{[PNR (\%)]} & \textbf{10} & \textbf{30} & \textbf{50} & \textbf{70} & \textbf{90} & \textbf{10} & \textbf{30} & \textbf{50} & \textbf{70} & \textbf{90} & \textbf{10} & \textbf{30} & \textbf{50} & \textbf{70} & \textbf{90} \\ \hline
 
 \multirow{28}{*}{\textbf{CIFAR10}}&  \multirow{7}{*}{\textbf{CFL}} & 
       \textit{FedAvg} & \textbf{69.2\%} & \textbf{67.8\%} & 1.7\% & 1.9\% & 1.9\% & 1.8\% & 1.8\% & 2.0\% & 1.7\% & 2.0\% & \textbf{71.2\%} & 58.7\% & 20.9\% & 8.0\% & 3.3\% \\
 &  & \textit{FLTrust} & 60.1\% & 61.0\% & 59.6\% & 56.2\% & 29.5\% & 41.4\% & 54.7\% & 50.2\% & 1.8\% & 2.9\% & 63.2\% & 59.0\% & 49.8\% & 39.6\% & 2.2\% \\
 &  & \textit{Krum} & 62.7\% & 63.6\% & \textbf{61.8\%} & \textbf{60.8\%} & 30.2\% & 52.9\% & \textbf{63.6\%} & \textbf{61.8\%} & 1.8\% & 1.9\% & 65.2\% & \textbf{63.3\% }& \textbf{61.8\%} &\textbf{43.8\% }& 1.9\% \\
 &  & \textit{Median} & 68.9\% & 52.7\% & 55.0\% & 47.3\% & \textbf{32.5\%} & \textbf{60.5\%} & 40.3\% & 18.2\% & \textbf{14.3\%} & 6.3\% & 62.7\% & 43.1\% & 57.4\% & 1.8\% & \textbf{27.1\%} \\
 &  & \sentinel{} & - & - & - & - & - & - & - & - & - & - & - & - & - & - & - \\
 &  & \textit{TrimmedMean} & 60.6\% & 54.9\% & 53.0\% & 47.3\% & 32.2\% & 2.6\% & 22.5\% & 7.3\% & 1.8\% & \textbf{8.0\%} & 68.1\% & 54.6\% & 13.7\% & 32.7\% & 4.2\% \\
 &  & \voyager{} & - & - & - & - & - & - & - & - & - & - & - & - & - & - & - \\  \cmidrule(l){2-18}
 & \multirow{7}{*}{\textbf{DFL-Fully}} 
 & \textit{FedAvg} & \textbf{72.9\%} & \textbf{70.4\% }& 1.8\% & 1.8\% & 1.9\% & 1.8\% & 1.8\% & 1.9\% & 1.8\% & 2.0\% & 69.9\% & 58.9\% & 21.9\% & 8.0\% & 3.2\% \\
 & & \textit{FLTrust} & 51.9\% & 52.2\% & 34.4\% & 1.8\% & 1.8\% & 52.3\% & 51.3\% & 1.9\% & 1.8\% & 1.8\% & 68.1\% & 61.9\% & 38.5\% & 2.5\% & 1.8\% \\
 &  & \textit{Krum} & 65.6\% & 65.8\% & 43.8\% & 2.0\% & 1.8\% & 66.0\% &\textbf{67.7\% }& 1.8\% & 1.8\% & 1.9\% & 66.5\% & 65.3\% & 47.8\% & 1.8\% & 1.8\% \\
 &  & \textit{Median} & 66.1\% & 68.4\% & 12.5\% & 1.8\% & 1.9\% & \textbf{70.1\% }& 18.7\% & 1.8\% & 1.7\% & 1.6\% & 70.3\% & 61.2\% & 39.0\% & 3.6\% & 2.3\% \\
 & \textbf{Connected} & \sentinel{} & 66.4\% & 66.0\% &\textbf{67.0\%} &\textbf{65.9\%} & \textbf{67.4\%} & 67.1\% & 65.5\% & \textbf{67.1\%} & \textbf{66.8\%} & \textbf{67.4\%} & 66.1\% & \textbf{67.4\%} & \textbf{65.4\% }&\textbf{66.4\% }& \textbf{67.4\%} \\
 &  & \textit{TrimmedMean} & 1.8\% & 1.8\% & 1.8\% & 1.8\% & 1.8\% & 1.8\% & 1.8\% & 1.8\% & 1.9\% & 1.7\% & \textbf{72.1\%} & 56.7\% & 11.9\% & 6.1\% & 1.8\% \\
 &  & \voyager{} & 63.8\% & 64.6\% & 65.0\% & 62.8\% & 65.5\% & 65.4\% & 64.0\% & 65.2\% & 64.3\% & 64.3\% & 65.0\% & 65.9\% & 63.3\% & 63.9\% & 65.2\% \\ \cmidrule(l){2-18}
 & \multirow{7}{*}{\textbf{DFL-Ring}} 
    & \textit{FedAvg} & 64.0\% & 62.6\% & 1.6\% & 1.6\% & 1.8\% & 1.7\% & 1.6\% & 1.6\% & 1.6\% & 1.7\% & 64.1\% & 54.2\% & 19.4\% & 7.4\% & 1.7\% \\
 &  & \textit{FLTrust} & 52.8\% & 51.3\% & 21.8\% & 1.8\% & 1.8\% & 53.1\% & 52.2\% & 3.5\% & 1.8\% & 1.7\% & 64.8\% & 62.2\% & 45.3\% & 5.1\% & 4.6\% \\
 &  & \textit{Krum} & 65.9\% & 65.3\% & 27.1\% & 1.9\% & 1.9\% & 67.0\% & \textbf{66.8\%} & 4.0\% & 1.8\% & 1.7\% & 67.2\% & 64.8\% & 51.2\% & 2.3\% & 5.4\% \\
 &  & \textit{Median} &\textbf{72.7\% }& \textbf{70.5\% }& 63.7\% & 1.9\% & 1.8\% & \textbf{71.0\%} & 4.6\% & 1.8\% & 2.0\% & 1.9\% & 69.6\% & 66.3\% & 59.3\% & 8.5\% & 4.3\% \\
 &  & \sentinel{} & 65.7\% & 65.6\% & \textbf{66.3\%} &\textbf{65.1\%} &\textbf{65.4\%} & 67.0\% & 66.1\% & \textbf{66.5\% }& \textbf{66.5\%} & \textbf{66.6\% }& 64.5\% & \textbf{67.0\%} & \textbf{66.3\% }& \textbf{65.5\% }& \textbf{67.0\%} \\
 &  & \textit{TrimmedMean} & 1.8\% & 1.8\% & 1.8\% & 1.8\% & 1.9\% & 2.1\% & 1.9\% & 1.8\% & 1.7\% & 1.7\% & \textbf{70.2\%} & 61.5\% & 27.4\% & 15.9\% & 1.8\% \\
 &  & \voyager{} & 63.2\% & 62.9\% & 64.2\% & 63.6\% & 65.2\% & 64.2\% & 63.0\% & 65.1\% & 65.0\% & 63.8\% & 62.6\% & 64.5\% & 64.1\% & 64.5\% & 64.6\% \\ \cmidrule(l){2-18}
 & \multirow{7}{*}{\textbf{DFL-Star}} & \textit{FedAvg} & 62.9\% & 1.7\% & 1.6\% & 1.6\% & 1.7\% & 1.6\% & 1.6\% & 1.7\% & 1.6\% & 1.7\% & 57.5\% & 52.2\% & 19.0\% & 3.1\% & 3.0\% \\
 &  & \textit{FLTrust} & 53.8\% & 49.3\% & 52.5\% & 1.7\% & 1.8\% & 50.1\% & 51.5\% & 1.8\% & 1.9\% & 2.0\% & 62.2\% & 59.3\% & 53.2\% & 4.6\% & 1.8\% \\
 &  & \textit{Krum} & 69.1\% & \textbf{63.9\%} & 65.4\% & 1.8\% & 1.8\% & 64.8\% & \textbf{67.2\%} & 1.9\% & 1.9\% & 2.0\% & 63.6\% & 62.7\% & 60.4\% & 1.7\% & 1.8\% \\
 &  & \textit{Median} & \textbf{71.2\%} & 24.5\% & 36.7\% & 1.8\% & 1.9\% & \textbf{71.0\%} & 1.9\% & 1.9\% & 1.8\% & 1.6\% & \textbf{71.1\%} & 36.1\% & 30.5\% & 4.9\% & 2.1\% \\
 &  & \sentinel{} & 67.5\% & 65.5\% & \textbf{66.4\%} & \textbf{65.5\% }& \textbf{65.6\%} & 66.8\% & 66.3\% & \textbf{67.0\%} & \textbf{67.0\%} & \textbf{67.6\%} & 66.2\% & \textbf{67.8\%} & \textbf{68.4\%} & \textbf{66.3\%} & \textbf{67.1\%} \\
 &  & \textit{TrimmedMean} & 2.0\% & 1.8\% & 1.8\% & 1.8\% & 1.9\% & 3.5\% & 1.9\% & 1.7\% & 1.8\% & 2.1\% & 70.5\% & 60.0\% & 34.6\% & 16.7\% & 2.2\% \\
 &  & \voyager{} & 63.7\% & 65.3\% & 65.8\% & 63.8\% & 64.6\% & 65.6\% & 64.2\% & 64.9\% & 64.3\% & 65.1\% & 65.0\% & 64.6\% & 64.9\% & 63.7\% & 64.0\% \\ \cmidrule(l){1-18}
\multirow{28}{*}{\textbf{FASHIONMNIST}} & \multirow{7}{*}{\textbf{CFL}} & \textit{FedAvg} & 80.5\% & 75.0\% & 48.5\% & 1.7\% & 0.5\% & 1.8\% & 1.9\% & 1.8\% & 1.8\% & 1.7\% & \textbf{82.1\%} & 71.4\% & 70.4\% & 64.5\% & 55.2\% \\
 &  & \textit{FLTrust} & 71.6\% & 69.0\% & 77.3\% & 59.5\% & 40.8\% & 76.6\% & 76.0\% & 72.5\% & 46.0\% & 28.6\% & 70.3\% & 68.8\% & 67.7\% & 61.9\% & 46.0\% \\
 &  & \textit{Krum} & 72.4\% & 68.9\% & \textbf{80.9\%} & 60.8\% & \textbf{42.0\%} & \textbf{81.0\%} & \textbf{81.8\%} & \textbf{80.8\%}& 44.4\% & 26.1\% & 72.6\% & 69.9\% & 67.9\% & 62.7\% & 42.4\% \\
 &  & \textit{Median} & \textbf{81.4\%} & 77.9\% & 80.0\% & 62.0\% & 0.3\% & 80.2\% & 70.7\% & 67.7\% & 23.0\% & 39.2\% & 80.9\% & \textbf{81.2\%} & 78.8\% & \textbf{68.9\%} & 40.5\% \\
 &  & \sentinel{} & - & - & - & - & - & - & - & - & - & - & - & - & - & - & - \\
 &  & \textit{TrimmedMean} & 81.3\% & \textbf{80.8\%} & 80.0\% & \textbf{65.7\%} & 41.2\% & 72.6\% & 65.8\% & 63.5\% & \textbf{59.1\%} & \textbf{40.7\%} & 80.3\% & 78.8\% & \textbf{80.2\%} & 65.7\% & \textbf{66.6\%} \\
 &  & \voyager{} & - & - & - & - & - & - & - & - & - & - & - & - & - & - & - \\  \cmidrule(l){2-18}
 & \multirow{7}{*}{\textbf{DFL-Fully}} & \textit{FedAvg} & 84.5\% & 79.4\% & 47.9\% & 1.8\% & 0.5\% & 1.8\% & 1.8\% & 1.8\% & 1.9\% & 1.8\% & 83.8\% & 71.2\% & 69.2\% & 64.1\% & 57.0\% \\
 &  & \textit{FLTrust} & 80.8\% & 67.0\% & 2.0\% & 0.8\% & 1.6\% & 67.8\% & 64.1\% & 63.9\% & 1.8\% & 1.8\% & 80.3\% & 78.6\% & 14.7\% & 10.6\% & 11.4\% \\
 &  & \textit{Krum} & 83.7\% & 69.4\% & 0.3\% & 0.4\% & 0.5\% & \textbf{83.2\%} & \textbf{83.0\%} & \textbf{82.5\%} & 1.8\% & 1.9\% & \textbf{83.7\%} & \textbf{81.6\%} & 2.9\% & 1.9\% & 1.3\% \\
 &  & \textit{Median} & 84.1\% & \textbf{79.6\%} & 55.9\% & 0.7\% & 0.7\% & 75.8\% & 20.6\% & 1.7\% & 1.8\% & 2.0\% & 83.4\% & 76.5\% & 64.1\% & 47.0\% & 9.4\% \\
 & \textbf{Connected}  & \sentinel{} & 77.9\% & 75.6\% & \textbf{79.3\%} & \textbf{74.3\%} & \textbf{75.2\%} & 77.2\% & 75.5\% & 76.0\% & \textbf{75.6\%} & \textbf{76.6\%} & 76.7\% & 76.3\% & \textbf{77.8\%} & \textbf{74.7\%} & \textbf{75.4\%} \\
 &  & \textit{TrimmedMean} & \textbf{84.7\%} & 78.5\% & 9.2\% & 2.3\% & 6.1\% & 13.9\% & 1.9\% & 1.8\% & 1.9\% & 1.8\% & 82.4\% & 76.3\% & 69.5\% & 48.1\% & 55.7\% \\
 &  & \voyager{} & 75.2\% & 73.7\% & 75.7\% & 73.3\% & 74.9\% & 75.3\% & 73.8\% & 73.3\% & 74.2\% & 73.7\% & 74.2\% & 72.6\% & 75.5\% & 72.9\% & 72.7\% \\ \cmidrule(l){2-18}
 & \multirow{7}{*}{\textbf{DFL-Ring}} & \textit{FedAvg} & 74.4\% & 55.0\% & 4.4\% & 1.6\% & 0.2\% & 1.7\% & 1.6\% & 1.6\% & 1.6\% & 1.5\% & 75.8\% & 63.6\% & 60.0\% & 47.6\% & 44.7\% \\
 &  & \textit{FLTrust} & 81.3\% & 69.8\% & 12.1\% & 0.7\% & 61.0\% & 64.2\% & 36.2\% & 1.8\% & 1.8\% & 1.8\% & 80.6\% & 45.4\% & 46.2\% & 27.1\% & 16.6\% \\
 &  & \textit{Krum} & 83.9\% & 71.1\% & 0.9\% & 0.5\% & 77.3\% & 84.2\% & 45.9\% & 1.9\% & 1.8\% & 1.9\% & 83.1\% & 40.3\% & 40.8\% & 21.4\% & 7.3\% \\
 &  & \textit{Median} & \textbf{84.7\%} & \textbf{83.5\%} & 55.0\% & 41.5\% & 0.0\% & \textbf{85.2\%} & 50.7\% & 1.9\% & 1.9\% & 1.9\% & \textbf{84.9\%} & 73.2\% & 69.6\% & 75.3\% & 34.9\% \\
 &  & \sentinel{} & 82.9\% & 76.7\% & \textbf{75.8\%} & \textbf{78.3\% }& \textbf{77.2\%} & 81.3\% & \textbf{75.1\%} & \textbf{76.3\%} & \textbf{75.8\% }& \textbf{75.6\% }& 76.2\% & \textbf{77.6\% }& \textbf{77.3\%} & \textbf{76.2\%} & \textbf{77.2\%} \\
 &  & \textit{TrimmedMean} & 84.6\% & 79.8\% & 59.0\% & 1.5\% & 0.8\% & 1.8\% & 5.3\% & 1.9\% & 1.9\% & 1.9\% & 84.3\% & 74.6\% & 72.1\% & 53.9\% & 56.4\% \\
 &  & \voyager{} & 74.2\% & 75.0\% & 75.0\% & 75.4\% & 74.7\% & 72.7\% & 73.4\% & 74.7\% & 75.6\% & 72.8\% & 73.0\% & 74.3\% & 74.8\% & 73.3\% & 75.5\% \\ \cmidrule(l){2-18}
 & \multirow{7}{*}{\textbf{DFL-Star}} & \textit{FedAvg} & 73.1\% & 47.6\% & 44.1\% & 1.6\% & 0.5\% & 1.6\% & 1.6\% & 1.6\% & 1.7\% & 1.5\% & 74.6\% & 64.9\% & 53.1\% & 22.8\% & 45.3\% \\
 &  & \textit{FLTrust} & 81.8\% & 14.8\% & 68.0\% & 26.5\% & 0.4\% & 65.3\% & 66.0\% & 23.4\% & 1.7\% & 1.7\% & 81.8\% & 76.7\% & 77.6\% & 11.6\% & 12.1\% \\
 &  & \textit{Krum} & 80.7\% & 0.6\% & \textbf{80.2\%} & 27.6\% & 0.4\% & \textbf{83.5\% }& 83.3\% & 30.0\% & 1.8\% & 1.7\% & \textbf{84.4\%} & \textbf{80.9\%} & \textbf{81.6\%} & 2.2\% & 3.3\% \\
 &  & \textit{Median} & 84.2\% & \textbf{84.4\%} & 77.6\% & 0.2\% & 0.3\% & \textbf{85.0\%} & 31.5\% & 1.8\% & 1.9\% & 1.8\% & 72.8\% & 72.1\% & 62.9\% & 27.3\% & 73.4\% \\
 &  & \sentinel{} & 78.1\% & 75.0\% & 76.9\% & \textbf{76.0\%} & \textbf{76.7\%} & 76.8\% & 76.9\% & \textbf{76.6\%} & \textbf{76.7\%} & \textbf{76.9\% }& 77.3\% & 75.8\% & 77.9\% & \textbf{76.4\%} & \textbf{75.7\%} \\
 &  & \textit{TrimmedMean} & \textbf{84.6\%} & 73.0\% & 40.5\% & 29.7\% & 0.2\% & 1.8\% & 3.8\% & 1.9\% & 1.8\% & 1.9\% & 83.4\% & 79.2\% & 61.6\% & 49.9\% & 52.8\% \\
 &  & \voyager{} & 74.5\% & 73.6\% & 74.1\% & 72.2\% & 73.5\% & 73.6\% & 74.6\% & 73.9\% & 74.5\% & 74.0\% & 74.4\% & 73.5\% & 76.6\% & 75.1\% & 73.7\% \\ \cmidrule(l){1-18}
\multirow{28}{*}{\textbf{MNIST}} & \multirow{7}{*}{\textbf{CFL}} & 
      \textit{FedAvg} & \textbf{95.3\%} & \textbf{94.7\% }& 80.5\% & 35.3\% & 0.8\% & 18.8\% & 15.0\% & 1.8\% & 1.6\% & 1.5\% & \textbf{99.8\%} & 93.0\% & 90.9\% & \textbf{88.0\%} & \textbf{73.6\%} \\
 &  & \textit{FLTrust} & 88.6\% & 89.9\% & 87.2\% & 67.6\% & 43.9\% & 88.3\% & 86.2\% & 88.8\% & 62.6\% & 24.8\% & 92.0\% & 81.0\% & 77.2\% & 71.7\% & 46.8\% \\
 &  & \textit{Krum} & 93.5\% & 92.8\% & 93.2\% & 69.5\% & 44.7\% & 93.3\% & \textbf{93.2\%} & \textbf{93.9\%} & 64.6\% & 21.8\% & 92.7\% & 79.5\% & 77.2\% & 70.2\% & 47.4\% \\
 &  & \textit{Median} & 94.1\% & 93.5\% & \textbf{95.4\%} & 70.0\% & 44.4\% & \textbf{93.9\%} & 85.2\% & 83.7\% & \textbf{68.4\%} & \textbf{48.3\%} & 94.4\% & \textbf{93.3\%} & 92.9\% & 71.4\% & 48.0\% \\
 &  & \sentinel{} & - & - & - & - & - & - & - & - & - & - & - & - & - & - & - \\
 &  & \textit{TrimmedMean} & 93.7\% & 80.7\% & 91.1\% & \textbf{88.0\%} & \textbf{46.0\%} & 90.6\% & 78.0\% & 77.1\% & 66.3\% & 40.9\% & \textbf{93.3\%} & 92.7\% & 93.3\% & 83.1\% & 53.9\% \\
 &  & \voyager{} & - & - & - & - & - & - & - & - & - & - & - & - & - & - & - \\ \cmidrule(l){2-18}
 & \multirow{7}{*}{\textbf{DFL-Fully}} & \textit{FedAvg} & 94.1\% & 95.9\% & 82.7\% & 35.2\% & 0.9\% & 19.5\% & 15.4\% & 1.8\% & 1.8\% & 1.6\% & \textbf{97.3\%} & 96.3\% & 93.9\% & 91.3\% & 75.8\% \\
 &  & \textit{FLTrust} & 93.0\% & 90.0\% & 48.7\% & 32.8\% & 0.3\% & 78.6\% & 77.2\% & 57.2\% & 1.7\% & 1.7\% & 88.3\% & 90.0\% & 19.9\% & 21.9\% & 18.7\% \\
 &  & \textit{Krum} & 95.8\% & 95.6\% & 47.4\% & 30.5\% & 0.3\% & 95.6\% & \textbf{95.8\%} & 72.7\% & 1.7\% & 1.8\% & 95.6\% & 93.2\% & 2.1\% & 4.6\% & 2.2\% \\
 &  & \textit{Median} & 97.1\% & \textbf{96.0\%} & 75.3\% & 18.4\% & 0.1\% & 96.7\% & 76.0\% & 1.6\% & 1.7\% & 1.8\% & 95.3\% & 94.6\% & \textbf{93.2\%} & 60.4\% & 2.2\% \\
 & \textbf{Connected}  & \sentinel{} & 95.1\% & 85.9\% & \textbf{88.9\%} & \textbf{88.4\%} & \textbf{89.2\%} & \textbf{97.8\%} & 89.2\% & \textbf{88.3\%} & \textbf{87.7\%} & \textbf{87.6\%} & 95.6\% & 87.4\% & \textbf{86.0\%} & 86.9\% & \textbf{87.9\%} \\
 &  & \textit{TrimmedMean} & \textbf{97.3\%} & 88.1\% & 67.4\% & 45.2\% & 0.4\% & 23.2\% & 1.7\% & 1.8\% & 1.8\% & 1.9\% & 96.0\% & \textbf{96.8\%} & 92.7\% & \textbf{93.2\%} & 80.3\% \\
 &  & \voyager{} & 93.8\% & 84.8\% & 86.2\% & 84.7\% & 85.6\% & 94.4\% & 86.1\% & 86.0\% & 84.8\% & 84.7\% & 94.1\% & 85.0\% & 83.2\% & 84.4\% & 87.2\% \\ \cmidrule(l){2-18}
 & \multirow{7}{*}{\textbf{DFL-Ring}} & \textit{FedAvg} & 88.0\% & 87.5\% & 42.3\% & 15.5\% & 0.4\% & 8.2\% & 1.7\% & 1.7\% & 1.5\% & 1.4\% & 88.0\% & 85.9\% & 84.0\% & 81.3\% & 64.6\% \\
 &  & \textit{FLTrust} & 94.9\% & 19.1\% & 18.2\% & 75.2\% & 0.4\% & 27.8\% & 74.5\% & 17.9\% & 1.8\% & 1.8\% & 94.3\% & 88.0\% & 78.2\% & 85.6\% & 16.5\% \\
 &  & \textit{Krum} & 96.1\% & 1.6\% & 1.2\% & \textbf{94.3\%} & 0.3\% & 36.0\% & \textbf{93.8\%} & 23.0\% & 1.9\% & 1.9\% & 95.6\% & 90.4\% & 80.7\% & 86.3\% & 2.0\% \\
 &  & \textit{Median} & 96.9\% & \textbf{95.7\%} & 86.8\% & 72.1\% & 0.4\% & 94.8\% & 67.6\% & 1.8\% & 1.8\% & 1.9\% & \textbf{97.0\%} & 95.7\% & 86.4\% & 74.9\% & 2.0\% \\
 &  & \sentinel{} & 96.5\% & 87.4\% & 87.1\% & 88.4\% & \textbf{86.5\% }& \textbf{96.7\%} & 88.4\% & \textbf{87.2\%} & \textbf{88.3\%} & \textbf{88.2\%} & 90.7\% & 84.5\% & 87.3\% & 87.8\% & \textbf{90.3\%} \\
 &  & \textit{TrimmedMean} & \textbf{97.4\%} & 95.5\% & \textbf{88.3\%} & 5.9\% & 1.1\% & 1.9\% & 1.8\% & 1.8\% & 1.8\% & 1.9\% & 96.9\% & \textbf{96.8\%} & \textbf{95.3\% }& \textbf{90.5\%} & 73.0\% \\
 &  & \voyager{} & 95.2\% & 84.0\% & 83.9\% & 85.9\% & 84.9\% & 83.7\% & 85.3\% & 83.9\% & 85.0\% & 86.7\% & 86.5\% & 83.8\% & 86.0\% & 86.6\% & 86.2\% \\ \cmidrule(l){2-18}
 & \multirow{7}{*}{\textbf{DFL-Star}} & \textit{FedAvg} & 86.5\% & 85.7\% & 46.8\% & 9.9\% & 0.8\% & 1.7\% & 3.4\% & 1.6\% & 1.5\% & 1.4\% & 88.0\% & 84.5\% & 82.6\% & 77.4\% & 66.8\% \\
 &  & \textit{FLTrust} & 89.2\% & 93.4\% & 89.6\% & 10.2\% & 0.7\% & 59.9\% & 55.4\% & 43.3\% & 1.7\% & 1.7\% & 91.8\% & 93.4\% & 87.1\% & 18.2\% & 17.4\% \\
 &  & \textit{Krum} & 95.2\% & 95.0\% & \textbf{94.5\%} & 0.3\% & 0.6\% & 74.8\% & 69.6\% & 55.6\% & 1.8\% & 1.8\% & 94.4\% & 94.2\% & \textbf{93.5\%} & 2.0\% & 2.3\% \\
 &  & \textit{Median} & \textbf{97.2\%} & \textbf{96.5\%} & 67.9\% & 50.0\% & 1.1\% & \textbf{97.2\%} & 1.7\% & 1.7\% & 1.9\% & 1.7\% & \textbf{97.1\%} & 93.8\% & 79.4\% & 70.7\% & 2.1\% \\
 &  & \sentinel{} & 86.9\% & 85.6\% & 87.0\% & \textbf{88.5\%} & \textbf{90.0\% }& 86.7\% & \textbf{88.0\%} & \textbf{88.3\%} & \textbf{88.2\%} & \textbf{87.7\%} & 87.3\% & 87.9\% & 87.5\% & 87.7\% & 84.4\% \\
 &  & \textit{TrimmedMean} & 91.8\% & 95.5\% & 85.2\% & 51.2\% & 1.0\% & 6.7\% & 6.9\% & 1.8\% & 1.8\% & 1.7\% & \textbf{97.1\% }& \textbf{96.0\%} & 91.0\% & \textbf{89.0\%} & 82.2\% \\
 &  & \voyager{} & 85.0\% & 84.2\% & 86.8\% & 85.4\% & 87.4\% & 85.3\% & 86.2\% & 86.4\% & 85.0\% & 84.4\% & 84.5\% & 84.4\% & 83.0\% & 86.6\% & \textbf{84.7\%} \\\bottomrule
\end{tabular}%
}
\end{table*}
% \usepackage{graphicx}
\begin{table*}[]
\centering
\caption{Benchmark of Average ASR-Targeted Results for Defense Mechanisms in Mitigating Targeted Label Flipping Attack}
\label{tab:targeted_label_flipping}
\resizebox{\textwidth}{!}{%
\begin{tabular}{lllllll|lllll|lllll}

\toprule
 &  & \multicolumn{5}{c}{\textbf{CIFAR10 }} & \multicolumn{5}{c}{\textbf{FASHIONMNIST }} &  \multicolumn{5}{c}{\textbf{MNIST }} \\ \midrule
 
 & \textbf{[PNR (\%)]}  & \textbf{10} & \textbf{30} & \textbf{50} & \textbf{70} & \textbf{90} & \textbf{10} & \textbf{30} & \textbf{50} & \textbf{70} & \textbf{90} & \textbf{10} & \textbf{30} & \textbf{50} & \textbf{70} & \textbf{90} \\ \hline
\multirow{7}{*}{\textbf{CFL}} & \textit{FedAvg} & 0.3\% & 4.5\% & 18.0\% & 39.0\% & 47.9\% & 0.4\% & 0.0\% & 6.4\% & 48.0\% & 66.9\% & 0.5\% & 24.6\% & 79.9\% & 76.6\% & 77.5\% \\
 & \textit{FLTrust} & 0.2\% & \textbf{4.2\%} & \textbf{16.8\%} & 34.6\% & \textbf{41.7\%} & 0.0\% & 1.6\% & 11.1\% & 17.6\% & 39.8\% & 1.0\% & \textbf{0.8\% }& 15.6\% & \textbf{21.6\%} & 48.4\% \\
 & \textit{Krum} & 0.3\% & 4.4\% & 17.1\% & 38.6\% & 41.8\% & 0.0\% & 0.0\% & 11.5\% & 19.2\% & 40.7\% & 1.0\% & 0.7\% & 16.5\% & 22.5\% & 49.6\% \\
 & \textit{Median} & 0.3\% & 4.3\% & 21.1\% & 46.7\% & 44.2\% & 0.0\% & 0.0\% & \textbf{2.7\%} & 15.9\% & \textbf{39.0\%} & 0.8\% & 1.2\% & \textbf{3.0\%} & 22.9\% & 48.3\% \\
 & \sentinel{} & - & - & - & - & - & - & - & - & - & - & - & - & - & - & - \\
 & \textit{TrimmedMean} & 0.3\% & 4.8\% & 25.5\% & \textbf{31.7\%} & 44.8\% & 0.0\% & 8.8\% & 9.9\% & \textbf{14.8\%} & 41.2\% & 1.1\% & 1.2\% & 15.5\% & 22.9\% & \textbf{47.3\%} \\
 & \voyager{} & - & - & - & - & - & - & - & - & - & - & - & - & - & - & - \\\hline
\multirow{7}{*}{\textbf{DFL-Fully}}  & \textit{FedAvg} & 0.3\% & 4.9\% & 18.7\% & 41.0\% & 47.5\% & 0.4\% & 0.0\% & 6.2\% & 49.8\% & 65.9\% & 0.4\% & 25.5\% & 74.9\% & 76.5\% & 78.8\% \\
 & \textit{FLTrust} & 0.2\% & 4.3\% & 16.2\% & 35.6\% & 41.8\% & 0.0\% & 19.7\% & 12.8\% & 82.9\% & 82.1\% & 0.7\% & 0.9\% & 91.7\% & 91.5\% & 90.6\% \\
 & \textit{Krum} & 0.2\% & 4.4\% & 16.6\% & 42.7\% & 52.6\% & 0.0\% & 25.0\% & 0.0\% & 89.4\% & 84.1\% & 0.7\% & 0.4\% & 97.3\% & 96.3\% & 97.0\% \\
 & \textit{Median} & 0.2\% & 4.5\% & 15.3\% & 38.3\% & 60.0\% & 0.0\% & 0.6\% & 25.7\% & 76.3\% & 86.4\% & 0.7\% & 5.1\% & 22.2\% & 89.3\% & 97.8\% \\
 \textbf{Connected}& \sentinel{} & 0.3\% & \textbf{0.2\%} & \textbf{0.2\%} & \textbf{0.2\%} & \textbf{0.2\%} & 0.0\% & \textbf{0.0\% }& \textbf{0.0\%} & \textbf{0.0\%} & \textbf{0.0\%} & 0.1\% & \textbf{1.3\%} & \textbf{1.2\%} & \textbf{0.8\%} & \textbf{1.6\%} \\
 & \textit{TrimmedMean} & 0.2\% & 4.4\% & 17.1\% & 47.1\% & 45.6\% & 0.0\% & 0.8\% & 67.1\% & 75.9\% & 86.9\% & 0.7\% & 3.0\% & 75.8\% & 96.8\% & 95.3\% \\
 & \voyager{} & 0.2\% & 4.3\% & 16.6\% & 36.7\% & 42.3\% & 0.3\% & 0.0\% & 5.4\% & 44.0\% & 57.6\% & 0.4\% & 22.4\% & 64.7\% & 65.0\% & 68.7\% \\\hline
\multirow{7}{*}{\textbf{DFL-Ring}} & \textit{FedAvg} & 0.1\% & 4.3\% & 17.6\% & 40.9\% & 61.6\% & 0.0\% & 0.8\% & 0.4\% & 42.7\% & 84.9\% & 0.6\% & 19.9\% & 81.9\% & 93.1\% & 96.3\% \\
 & \textit{FLTrust} & 0.2\% & 4.3\% & 16.5\% & 35.8\% & 43.3\% & 0.0\% & 0.1\% & 41.2\% & 57.8\% & 82.1\% & 0.9\% & 12.3\% & 78.8\% & 17.1\% & 91.9\% \\
 & \textit{Krum} & 0.3\% & 6.5\% & 20.0\% & 32.5\% & 48.8\% & 0.0\% & 0.0\% & 47.9\% & 56.6\% & 84.5\% & 0.9\% & 15.3\% & 97.7\% & 1.0\% & 95.2\% \\
 & \textit{Median} & 0.2\% & 5.1\% & 15.3\% & 42.1\% & 42.6\% & 0.0\% & 0.0\% & 62.1\% & 59.7\% & 72.6\% & 0.8\% & 4.3\% & 56.3\% & 96.0\% & 94.5\% \\
 & \sentinel{} & 0.3\% & \textbf{0.2\%} & \textbf{0.2\%} & \textbf{0.2\%} & \textbf{0.2\%} & 0.0\% & \textbf{0.0\%} & \textbf{0.0\%} & \textbf{0.0\%} & \textbf{0.0\%} & \textbf{0.5\%} & \textbf{1.0\%} & \textbf{1.0\%} & \textbf{0.8\%} & \textbf{1.1\%} \\
 & \textit{TrimmedMean} & 0.2\% & 5.0\% & 17.9\% & 47.4\% & 41.4\% & 0.0\% & 0.3\% & 19.9\% & 70.1\% & 83.5\% & 0.9\% & 1.8\% & 9.8\% & 96.8\% & 95.3\% \\
 & \voyager{} & 0.2\% & 4.4\% & 16.2\% & 35.1\% & 41.8\% & 0.3\% & 0.0\% & 5.5\% & 44.3\% & 57.9\% & 0.4\% & 22.1\% & 65.0\% & 66.1\% & 68.5\% \\\hline
\multirow{7}{*}{\textbf{DFL-Star}} & \textit{FedAvg} & 0.3\% & 6.3\% & 20.9\% & 51.3\% & 63.7\% & 0.5\% & 0.0\% & 7.6\% & 61.5\% & 84.5\% & 0.6\% & 32.2\% & 93.7\% & 96.3\% & 98.0\% \\
 & \textit{FLTrust} & 0.2\% & 4.3\% & 16.1\% & 35.8\% & 41.4\% & 15.9\% & 0.0\% & 79.7\% & 88.1\% & 85.2\% & 0.4\% & 1.8\% & 17.9\% & 17.4\% & 94.1\% \\
 & \textit{Krum} & 0.3\% & 5.3\% & 19.0\% & 45.5\% & 39.1\% & 20.3\% & 0.0\% & 89.1\% & 91.6\% & 92.1\% & 0.4\% & 1.6\% & 1.4\% & \textbf{0.7\%} & 99.1\% \\
 & \textit{Median} & 0.3\% & 5.7\% & 20.8\% & 46.4\% & 48.0\% & 4.1\% & 64.5\% & 56.4\% & 78.3\% & 0.0\% & 0.8\% & 0.8\% & 98.2\% & 93.5\% & 97.8\% \\
 & \sentinel{} & 0.3\% & \textbf{0.2\%} & \textbf{0.2\% }& \textbf{0.2\%} & \textbf{0.2\%} & 0.0\% & \textbf{0.0\%} & \textbf{0.0\%} & \textbf{0.0\% }& \textbf{0.0\%} & 0.5\% & \textbf{1.0\%} & \textbf{1.0\%} & 0.9\% & \textbf{1.2\%} \\
 & \textit{TrimmedMean} & 0.3\% & 5.6\% & 18.2\% & 42.8\% & 42.0\% & 0.0\% & 0.2\% & 60.9\% & 79.8\% & 78.6\% & 0.6\% & 3.1\% & 95.0\% & 95.4\% & 98.0\% \\
 & \voyager{} & 0.2\% & 4.2\% & 16.0\% & 36.2\% & 42.3\% & 0.3\% & 0.0\% & 5.5\% & 44.0\% & 56.2\% & 0.4\% & 22.5\% & 64.6\% & 65.4\% & 65.1\%
 \\\bottomrule
\end{tabular}%
}
\end{table*}
% \usepackage{graphicx}
\begin{table*}[]
\centering
\caption{Benchmark of Average ASR-Backdoor Results for Defense Mechanisms in Mitigating Backdoor Attack}
\label{tab:targeted_backdoor}
\resizebox{\textwidth}{!}{%
\begin{tabular}{lllllll|lllll|lllll}

\toprule
 &  & \multicolumn{5}{c}{\textbf{CIFAR10 }} & \multicolumn{5}{c}{\textbf{FASHIONMNIST }} &  \multicolumn{5}{c}{\textbf{MNIST }} \\ \midrule
 
 & \textbf{[PNR (\%)]} & \textbf{10} & \textbf{30} & \textbf{50} & \textbf{70} & \textbf{90} & \textbf{10} & \textbf{30} & \textbf{50} & \textbf{70} & \textbf{90} & \textbf{10} & \textbf{30} & \textbf{50} & \textbf{70} & \textbf{90} \\ \hline
\multirow{7}{*}{\textbf{CFL}} 
& \textit{FedAvg} & \textbf{2.5\%} & 36.3\% & 46.5\% & 62.7\% & 69.8\% & 12.3\% & 67.7\% & 65.4\% & 74.4\% & 70.3\% & 22.2\% & 78.9\% & 91.1\% & 87.6\% & 85.9\% \\
 & \textit{FLTrust} & 3.9\% & \textbf{8.3\%} & \textbf{6.4\%} & \textbf{33.2\%} & \textbf{49.6\%} & \textbf{1.0\%} & 21.4\% & 29.1\% & 39.5\% & 70.5\% & 3.2\% & 53.1\% & 60.2\% & 57.9\% & 63.7\% \\
 & \textit{Krum} & 2.5\% & 34.3\% & 48.8\% & 51.0\% & 68.3\% & \textbf{1.0\%} & 25.9\% & 34.6\% & 46.4\% & 73.5\% & \textbf{2.5\%} & 59.4\% & 65.3\% & \textbf{57.6\%} & \textbf{57.1\%} \\
 & \textit{Median} & 2.7\% & 33.6\% & 34.6\% & 59.6\% & 73.2\% & 1.2\% & 4.9\% & 20.8\% & \textbf{14.9\%} & \textbf{42.6\%} & 5.4\% & \textbf{11.3\%} & \textbf{24.1\%} & 59.0\% & 97.1\% \\
 & \sentinel{} & - & - & - & - & - & - & - & - & - & - & - & - & - & - & - \\
 & \textit{TrimmedMean} & 2.5\% & 35.3\% & 39.9\% & 65.5\% & 70.1\% & \textbf{1.0\%} & \textbf{4.7\%} & \textbf{13.0\%} & 23.4\% & 66.5\% & 6.9\% & 31.1\% & 36.9\% & 70.0\% & 97.0\% \\
 & \voyager{} & - & - & - & - & - & - & - & - & - & - & - & - & - & - & - \\\hline
\multirow{7}{*}{\textbf{DFL-Fully}} & \textit{FedAvg} & 2.6\% & 38.3\% & 46.7\% & 60.2\% & 71.6\% & 12.3\% & 65.2\% & 62.7\% & 70.8\% & 72.1\% & 22.7\% & 77.2\% & 87.8\% & 93.1\% & 83.1\% \\
 & \textit{FLTrust} & 3.5\% & \textbf{18.3\%} & 61.2\% & 60.3\% & 91.8\% & 10.8\% & 35.9\% & 74.6\% & 63.7\% & 75.6\% & 36.1\% & 77.6\% & 94.1\% & 68.8\% & 96.1\% \\
 & \textit{Krum} & \textbf{2.3\%} & 36.9\% & \textbf{37.0\%} & 58.5\% & 61.5\% & 9.9\% & 32.9\% & 78.2\% & 60.5\% & 78.1\% & 37.5\% & 78.5\% & 93.3\% & 65.4\% & 99.6\% \\
 & \textit{Median} & 2.6\% & 31.3\% & 45.2\% & 60.5\% & 71.0\% & 3.1\% & 61.0\% & 74.4\% & 89.1\% & 84.5\% & 12.0\% & 70.5\% & 97.8\% & 96.8\% & 98.5\% \\
\textbf{Connected} & \sentinel{} & \textbf{2.3\%} & 38.5\% & 36.8\% & 53.5\% & 67.5\% & \textbf{1.0\%} & \textbf{1.0\%} & \textbf{1.2\%} & \textbf{1.0\%} & \textbf{1.0\%} & \textbf{3.2\%} & \textbf{5.4\%} & \textbf{2.5\%} & \textbf{3.2\%} & \textbf{6.9\%} \\
 & \textit{TrimmedMean} & 2.6\% & 36.9\% & 44.5\% & \textbf{48.4\%} & 65.5\% & 17.1\% & 53.4\% & 72.5\% & 85.3\% & 74.6\% & 39.8\% & 92.5\% & 98.8\% & 97.9\% & 99.0\% \\
 & \voyager{} & \textbf{2.3\%} & 32.8\% & 40.9\% & 52.1\% & \textbf{62.4\%} & 10.7\% & 57.2\% & 54.6\% & 62.2\% & 62.7\% & 19.2\% & 68.1\% & 78.8\% & 80.7\% & 70.3\% \\ \hline
\multirow{7}{*}{\textbf{DFL-Ring}} & \textit{FedAvg} & 2.9\% & 42.4\% & 51.1\% & 65.6\% & 78.5\% & 12.5\% & 73.1\% & 70.1\% & 83.6\% & 80.2\% & 24.5\% & 88.2\% & 95.8\% & 94.6\% & 97.5\% \\
 & \textit{FLTrust} & 3.0\% & 32.3\% & 64.5\% & 80.8\% & 85.9\% & 3.0\% & 28.2\% & 64.0\% & 84.9\% & 80.7\% & 11.7\% & 93.0\% & 92.3\% & 96.4\% & 90.3\% \\
 & \textit{Krum} & 3.0\% & \textbf{31.1\%} & 62.0\% & 53.6\% & 69.1\% & 1.4\% & 23.5\% & 65.7\% & 90.6\% & 83.6\% & 5.6\% & 98.6\% & 97.7\% & 99.3\% & 90.8\% \\
 & \textit{Median} & 2.7\% & 38.9\% & 42.7\% & \textbf{44.1\%} & 70.3\% & 6.3\% & 54.1\% & 72.2\% & 86.6\% & 86.7\% & 18.8\% & 67.0\% & 78.5\% & 98.7\% & 99.1\% \\
 & \sentinel{} & 2.5\% & 34.5\% & 50.9\% & 76.3\% & 81.3\% & \textbf{1.2\% }& \textbf{1.0\%} & \textbf{1.2\%} & \textbf{1.0\%} & \textbf{1.0\%} & \textbf{2.5\%} & \textbf{7.5\%} & \textbf{5.3\%} & \textbf{11.5\%} & \textbf{5.8\%} \\
 & \textit{TrimmedMean} & 2.8\% & 41.4\% & 53.9\% & 52.5\% & 61.8\% & 9.6\% & 50.6\% & 60.9\% & 79.1\% & 81.9\% & 38.7\% & 89.5\% & 98.0\% & 99.7\% & 98.5\% \\
 & \voyager{} & \textbf{2.2\%} & 32.5\% & \textbf{40.9\%} & 52.2\% & \textbf{61.1\%} & 11.2\% & 58.6\% & 54.5\% & 61.4\% & 63.1\% & 19.6\% & 65.9\% & 76.4\% & 80.9\% & 72.1\% \\ \hline
\multirow{7}{*}{\textbf{DFL-Star}} & \textit{FedAvg} & 2.8\% & 41.8\% & 52.6\% & 68.3\% & 78.5\% & 13.2\% & 73.1\% & 70.1\% & 83.6\% & 79.1\% & 25.2\% & 86.7\% & 95.0\% & 95.0\% & 92.6\% \\
 & \textit{FLTrust} & 4.7\% & 44.6\% & 61.9\% & 72.9\% & 87.5\% & 1.7\% & 55.8\% & 70.4\% & 72.5\% & 78.7\% & 7.5\% & 76.0\% & 91.1\% & 96.0\% & 84.6\% \\
 & \textit{Krum} & 2.6\% & 35.4\% & 42.3\% & 56.2\% & 80.8\% & \textbf{0.9\%} & 59.4\% & 70.8\% & 73.6\% & 81.3\% & \textbf{2.4\%} & 72.8\% & 92.1\% & 97.4\% & 88.0\% \\
 & \textit{Median} & \textbf{2.1\%} & 40.2\% & \textbf{38.6\%} & 61.8\% & 66.1\% & 2.8\% & 58.1\% & 73.0\% & 85.8\% & 68.0\% & 41.9\% & 88.2\% & 97.7\% & 98.5\% & 97.6\% \\
 & \sentinel{} & 2.9\% & \textbf{31.7\%} & 44.9\% & 56.4\% & 76.9\% & 2.7\% & \textbf{4.9\%} & \textbf{5.4\% }& \textbf{6.4\%} & \textbf{7.1\%} & 13.0\% & \textbf{6.0\%} & \textbf{11.7\%} & \textbf{13.0\%} & \textbf{5.9\%} \\
 & \textit{TrimmedMean} & 2.2\% & 46.1\% & 56.2\% & 52.6\% & 76.8\% & 5.3\% & 59.3\% & 74.0\% & 80.0\% & 80.0\% & 28.9\% & 94.6\% & 99.1\% & 98.7\% & 98.6\% \\
 & \voyager{} & 2.3\% & 33.0\% & 40.1\% & \textbf{51.8\%} & \textbf{61.0\%} & 10.5\% & 57.6\% & 54.7\% & 61.6\% & 62.6\% & 19.6\% & 68.2\% & 77.9\% & 79.8\% & 72.6\%
 \\\bottomrule
\end{tabular}%
}
\end{table*}

\subsubsection{Benchmark of Defense Mechanisms for Untargeted Poisoning Attacks}
\label{benchmark_untarget}
This section conducts extensive experiments to assess the effectiveness of defense mechanisms against poisoning attacks. It specifically looks at how well these mechanisms, initially designed for CFL, can adapt to DFL. The experiments benchmark these defense mechanisms across diverse datasets against various attack strategies.

This benchmark includes a variety of defense mechanisms, including:
\begin{itemize}
\item \textbf{\textit{FedAvg}}: Averaging over the received models, designed for CFL, without providing additional protection.
\item \textbf{\textit{FLTrust}}: Filtering anomalous models using ReLU-clipped cosine similarity, designed for CFL.
\item \textbf{\textit{Krum}}: Finding the model with the shortest distance to all other models as the aggregated model, designed for CFL.
\item \textbf{\textit{Median}}: Finding the median model of all models as the aggregated model, designed for CFL.
\item \textbf{\sentinel{}}: Combining similarity and loss-based methods to filter and regularize models, designed for DFL.
\item \textbf{\textit{TrimmedMean}}: Filtering the extreme values in the received models and then averaging them to get the aggregated model, designed for CFL.
\item \textbf{\voyager{}}: An MTD-based defense that isolates malicious nodes by means of dynamic topology, designed for DFL.
\end{itemize}

The first experiment seeks to evaluate the effectiveness of these defense mechanisms in mitigating Untargeted Label Flipping, Untargeted Sample Poisoning, and Random Model Poisoning attacks. The effectiveness of the defense mechanisms is assessed using the F1-Score, with higher values indicating better protection against adversarial attacks. 

For those defense mechanisms originally designed for CFL, such as \textit{FedAvg}, \textit{FLTrust}, \textit{Krum}, \textit{Median}, and \textit{TrimmedMean}, they are tested on both CFL and DFL. For the defense mechanisms designed for DFL, \sentinel{} and \voyager{}, since they do not work on CFL, the testing is done on DFL only.

\paragraph{\textbf{\1 Defense Effectiveness}} \mbox{}\\ 

The average F1-Score of the defense mechanisms under three datasets, three attack strategies, and different PNRs is shown in \tablename~\ref{tab:untargeted}. Overall, \sentinel{} achieves the best defense against untargeted attacks and is able to maintain the maximum model robustness, with \voyager{} coming in second. However, when the percentage of poisoned nodes exceeds 50\%, most defenses prove to be ineffective. This is due to Byzantine-robust aggregation techniques, such as\textit{Krum}, \textit{Median}, and \textit{TrimmedMean}, being more likely to select malicious models over benign ones in such scenarios. Although \textit{FLTrust} demonstrates a good result at low PNR, the results show that it does not exhibit an advantage for Byzantine-robust aggregation techniques. These defense strategies exhibit similar outcomes to \textit{FedAvg} across different network topologies and datasets. Meanwhile, the defense mechanism is more effective against Sample Poisoning and Label Flipping than Model Poisoning. Except for \sentinel{} and \voyager{}, most defense mechanisms are ineffective in mitigating Model Poisoning attacks.

\paragraph{\textbf{\2 Compatibility in DFL}} \mbox{}\\

In contrast to CFL, the network topologies influence the effectiveness of Byzantine defense strategies and \textit{FLTrust} in DFL. A smaller average number of connected neighbors in the network results in malicious nodes posing a greater threat to their directly connected benign nodes. In such cases, benign nodes are more likely to aggregate with more than 50\% of malicious neighbors, rendering Byzantine-robust aggregation techniques ineffective. It can be seen from the results in  \tablename~\ref{tab:untargeted} that the results in the ring and star topologies are worse than the fully connected ones. Therefore, those defense mechanisms designed for CFL have some difficulties adapting to DFL, and they are more effective in dense topologies.

To conclude, when the FL system contains a small number of malicious nodes, all defense mechanisms prove to be effective. However, if the percentage of malicious nodes exceeds 50\%, the Byzantine-robust defenses and \textit{FLTrust} become ineffective. In contrast, the DFL-focused defense mechanisms, Voyager and Sentinel, are capable of effectively countering attacks from a high percentage of malicious nodes. Furthermore, the network's topology also plays a role in the effectiveness of the defenses, with Byzantine-robust defenses and \textit{FLTrust} being less effective in sparse networks but performing better in dense networks.

\subsubsection{Benchmark of Defense Mechanisms for Targeted Poisoning Attacks}

As analyzed in Section~\ref{subsection:targetedattack}, targeted attacks are more difficult to detect. Therefore, this section experimentally benchmarks the performance of defense mechanisms listed in section~\ref{benchmark_untarget} under targeted attacks, including Targeted Label Flipping and Backdoor Attacks.

\paragraph{\textbf{\1 Defense Effectiveness for Targeted Label Flipping}} \mbox{}\\

Overall, \textit{Krum}, \textit{TrimmedMean}, \textit{Median}, \voyager{}, and \textit{FLTrust} are not effective against Targeted Label Flipping attacks, as shown in the \tablename~\ref{tab:targeted_label_flipping}. The reason is that since the Targeted Label Flipping attacks have a small impact on the overall model, it is difficult for distance-based or extreme value filtering defenses to work. In contrast, \sentinel{}'s loss-based layer-wise normalization mechanism is much more effective against Targeted Label Flipping, and the experimental results show that \sentinel{} can effectively prevent the spread of Targeted Label Flipping attacks.

\paragraph{\textbf{\2 Defense Effectiveness for Backdoor Attack}} \mbox{}\\

According to the data presented in the \tablename~\ref{tab:targeted_backdoor}, the defense mechanisms exhibit weaknesses when confronted with the Backdoor Attack. In general, \sentinel{} successfully counters the Backdoor Attack on the MNIST and FashionMNIST datasets. However, it fails to effectively address this attack on the Cifar10 dataset, which task is more complex and has a larger number of model parameters. The reason behind this failure is that in more complex models, the implanted backdoor has a limited effect on both the similarity and the loss of the model. Therefore, neither the Byzantine-robust defense mechanisms nor the hybrid \sentinel{} and \textit{FLTrust} approaches are able to mitigate this type of attack effectively.

In conclusion, compared to untargeted attacks, targeted attacks are more difficult to defend, and the Byzantine-robust defense is often ineffective. On the contrary, the combination of model similarity and loss-based layer-wise normalization, \ie \sentinel{}, can effectively defend against targeted attacks.

\section{Lessons Learned, Open Challenges, and Research Opportunities}

\label{sec:challenges}
Drawing on the analysis of the diverse model robustness of DFL in previous sections, this section endeavors to address the lessons learned and challenges that have surfaced in the research of DFL model robustness. Additionally, it suggests potential avenues for further research.

% \paragraph{\textbf{\1 Lessons Learned}} \mbox{}\\ 
\subsection{Lessons Learned}

Through the literature research and experimental analysis on both attack and defense sides, the following lessons were summarized:

\begin{itemize}
    \item There is limited research on model robustness for DFL. The current research on model robustness is mainly focused on CFL. In addition, there is a lack of research on designing attacks optimized for DFL from an attack perspective or designing defense mechanisms to enhance the robustness of DFL models from a defense perspective.
    
    \item The model robustness degradation caused by untargeted attacks is more tangible than with targeted attacks. The latter have an insignificant impact on the overall effectiveness of the model and are more challenging to detect by common metrics. When considering attack strategy, sample poisoning is found to be less effective in untargeted attacks, but it exhibits higher efficacy in targeted attacks. In terms of the extent of harm inflicted by the attacks, untargeted attacks, particularly those involving model poisoning, result in significant damage.
    
    \item Topology plays a crucial role in determining the success of an attack in DFL. A densely connected network is more vulnerable to the spread of malicious attacks, while a sparse network can result in more severe consequences when under attack.
    
    \item The majority of defense mechanisms are ineffective in a high percentage (\ie more than 50\%) of malicious nodes. Meanwhile, defense mechanisms designed for CFL encounter challenges when applied to DFL. Byzantine-robust countermeasures, such as \textit{Krum} and \textit{TrimmedMean}, have limited effectiveness in sparse DFL networks.
    
    \item Most defense mechanisms are insufficient when mitigating targeted attacks, either in CFL or DFL. The limited effect of targeted attacks on the model's parameters poses challenges for distance- or similarity-based defense mechanisms.
\end{itemize}

% \paragraph{\textbf{\2 Open Challenges and Opportunities}} \mbox{}\\ 

\subsection{Application in Practical Scenarios}
% how the experimental findings can be applied in practical scenarios, like IoT security, healthcare, finance.
\label{sec:practical}
In general, current research on model robustness for FL, especially DFL, still tends to focus on the general problem rather than practical scenarios, i.e., most of the studies are validated on commonly used benchmark datasets, such as MNIST and FashionMNIST. There is a limited amount of research exploring the impacts of poisoning attacks on the robustness of FL models in practical scenarios, particularly  in the fields of IoT, finance, and healthcare.

{\cite{nguyen2020poisoning}} discussed the feasibility of implementing poisoning attacks on FL systems in IoT environments and evaluated the hazards caused by poisoning attacks in an IoT Intrusion Detection System. {\cite{feng2023cyberforce}} proposed a framework for Federated Reinforcement Learning (FRL) to recognize malware attacks and automatically implement mitigation strategies in IoT environments. They verified the robustness of their proposed FRL framework to data poisoning and model poisoning attacks. {\cite{yoo2021federated}} and {\cite{ali2022federated}} provided an overview of the various problems associated with FL applications in medical and healthcare scenarios, including the problems associated with heterogeneities, and also the security problems associated with poisoning attacks. {\cite{kuo2022detecting}} proposed a framework for data poisoning attacks against healthcare FL, by comparing local models of different rounds and models from other nodes. In {\cite{xie2019dba}}, a distributed backdoor attack approach was proposed and validated on a dataset in the financial domain. The research on FL in practical applications has garnered significant attention, yet there remains a noticeable absence of discourse surrounding the robustness of its models to both offensive and defensive sides.

Through the experimental validation in this paper, there are the following insights for the deployment of FL in practice scenarios, especially in the fields of IoT, healthcare, and finance:
\begin{itemize}
    \item The decentralized nature of DFL results in a lack of model auditing and verification mechanisms for the federation, leading to potential security concerns that participants should be mindful of. While the FedAvg algorithm demonstrates strong performance in non-attacked scenarios, the presence of a malicious node within the federation can compromise the security of all nodes. To mitigate these risks,  aggregation methods specifically tailored for DFL are advised, especially in scenarios where heightened security is required.

    \item Since overlay topology greatly impacts the model robustness of DFL, nodes should carefully consider which neighbors to connect and aggregate with. The remarkable performance of the \textit{Voyager} algorithm, which relies on altering the connections with neighbors in DFL, indicates that employing a dynamic topology can effectively reduce the impact of poisoning attacks.

    \item Due to the minimal alterations made by targeted attacks, particularly backdoor attacks, conventional methods of model auditing that focus on performance or similarity between nodes have shown limited efficacy. Nevertheless, these attacks can have significant consequences in critical sectors like healthcare and finance. Therefore, additional audit analysis should be counted for targeted attacks, such as analyzing the stability of model results across categories. 
    
\end{itemize}

\subsection{Real-World Applicability}
\label{sec:realtime}
% the deployment challenges in real work application, for both attack side and defense side.
When deploying the DART method for evaluating the robustness of DFL models in a real-world environment, the following difficulties pose limitations and difficulties:
\begin{itemize}
    \item \textbf{Limited Datasets.} As mentioned in Section {\ref{sec:practical}}, current evaluations of DFL model robustness focus on commonly used datasets and lack domain datasets, especially real-world datasets. However, datasets may contain sensitive information, and sharing them directly may bring privacy and security concerns. Hence, the process of desensitizing and restructuring authentic data using techniques like data augmentation and synthetic data generation presents viable options {\cite{gan2018}}. These methods serve to eliminate private information within the raw data, while maintaining the raw data's characteristics and patterns. This, in turn, allows for a more thorough examination of the potential challenges in deploying DFL models in practical settings.
    
    \item \textbf{Realistic Attacks.} In the research on poisoning attacks, it is commonly assumed that malicious nodes can coordinate with each other to target benign nodes. However, in real-world scenarios of DFL, individual nodes often have different affiliations that complement each other. This poses a challenge for attackers, as the lack of synergy between malicious nodes can diminish the impact of their attacks or even cancel out the effect brought by the attack. Combining poisoning attacks with Sybil attacks can have a more significant impact, particularly when a large number of manipulated nodes are present in a DFL. Experimental findings indicate that a high proportion of malicious nodes within a federation can render defense mechanisms ineffective. Thus, if the Sybil attack has more than half of the manipulated nodes in the federation, defenses, including Krum and TrimmedMean, fail. Besides, a combination of poisoning attacks and Eclipse attacks can isolate a target benign node from other benign nodes. In such a case, the target node is surrounded by malicious nodes, and most of the defenses are ineffective. The same attacker can execute these two types of attacks and does not require collaboration with other attackers. However, they require attackers to know the overlay topology of the DFL in advance. Consequently, a topology manipulation defense mechanism can effectively defend against such attacks.

    \item \textbf{Collaboration of Defenses.} Decentralization not only poses a synchronization difficulty for the attacker but also a collaboration difficulty for the defender. Ideally, if a benign node in a DFL detects an attack, it should notify the other benign nodes to take appropriate defense measures. However, these collaborative messages can be intercepted and tampered with by the attacker, making collaboration among defenders difficult in real-world scenarios.DFL participants can adopt a more "selfish" defense, i.e., proactive defenses, to protect their models from attacks and not rely on notifications from other nodes. Possible proactive strategies include dynamically changing the aggregation algorithm, or proactively changing connected neighbors.
\end{itemize}

\subsection{Ethical Considerations}
\label{sec:ethical}
% compliance with GDPR, federated unlearning
Although DFL offers advantages regarding privacy and effectiveness, it still has some ethical concerns:
\begin{itemize}
    \item \textbf{Data Privacy.} Although in DFL, only the parameters or gradients of the model are transmitted through the network, studies have shown that this can still cause privacy leakage {\cite{benmalek:hal-03620400}}, {\cite{lyu_PrivacyRobustness_2022}}, {\cite{Kumar10274102}}. An attacker can infer privacy information from model parameters, model gradients, or model outputs, including the hyper-parameters of model training, or restore the original training data. This type of attack is called an inference attack, bringing privacy concerns to DFL. Since the successful implementation of inference attacks relies on model overfitting, pruning and data augmentation techniques can be used to mitigate inference attacks. Besides, differential privacy or homomorphic encryption techniques can be used to protect the privacy-preserving capabilities of DFL models.

    \item \textbf{Fairness and Bias.} As a data-driven process, the bias of the data brings about the bias of the DFL model. Therefore, the fairness and bias of the DFL model are also important aspects of the ethical consideration of DFL. {\cite{SANCHEZSANCHEZ202483}} proposed a qualitative analysis method for FL fairness, which can effectively measure the fairness of FL models.

    \item \textbf{Sustainability.} The training of DFL models consumes energy and produces greenhouse gases. Thus, it poses concerns in terms of sustainability. {\cite{celdran2023sustainability}} proposed a framework for measuring the sustainability of FL, which calculates the sustainability of FL models by considering multiple aspects such as grid efficiency, hardware efficiency, and computational efficiency.

    \item \textbf{Compliance with Regulations.} DFL is required to comply to a range of data protection regulations, such as General Data Protection Regulation (GDPR) {\cite{gdpr2016}} and Health Insurance Portability and Accountability Act (HIPAA) {\cite{hipaa1996}}. These regulations were not specifically tailored to address the challenges faced by DFL, making compliance a complex undertaking. The concept of the right to be forgotten, as outlined in GDPR, presents particular challenges for DFL, as the deletion of user data from one node may not guarantee its removal from other nodes through model aggregation. Research on machine unlearning {\cite{machineunlearning}} is gaining attention as a potential avenue for DFL to align with data protection requirements.
    
\end{itemize}

\subsection{Open Challenges and Opportunities}
\label{sec:challenge}
Regarding the lessons learned, application in practical scenarios, and ethical considerations, the following challenges and opportunities for model robustness research are identified:

\begin{itemize}
    \item  \textbf{Increased Attack Surfaces.} In CFL, only the central aggregator has a global view of the whole federation and has model information of all nodes. However, in DFL, depending on the topology, any node may have model information about all other nodes in the federation, which actually increases the security vulnerabilities. These increased attack surfaces pose a persistent challenge in preserving the robustness of DFL models. Therefore, it is crucial to customize defense mechanisms specifically for DFL systems. The robustness of DFL heavily relies on topology, suggesting that incorporating dynamic topology could serve as a viable defense mechanism. Additionally, leveraging nodes' local data and models within DFL systems can enhance the effectiveness of defense mechanisms against attacks.    
    \item \textbf{Optimized Attacks.} In DFL, nodes receive models from their neighbors. Therefore, it is possible for malicious nodes to exploit the models received from their neighboring nodes as knowledge to optimize their attack strategies. By incorporating their own knowledge of defense mechanisms, these malicious nodes can manipulate the attack direction and increase the likelihood of achieving their objectives. This increases the difficulty and challenge for defense. In benign nodes, dynamic changes in aggregation functions can increase the difficulty of the attacker learning about the defense mechanism and increase the cost of executing the optimized attack. Besides, techniques such as homomorphic encryption can effectively increase the cost of such attacks.
    \item \textbf{Balance between Overhead and Robustness.} Incorporating defense mechanisms can enhance the robustness of DFL models, but it also presents challenges like increased network overhead, high computational resource usage, and scalability limitations. For example, when employing filtering-based strategies as a defense approach, setting a threshold too low may lead to ineffective filtering of malicious models, while setting the threshold too high may result in the exclusion of valuable models and the loss of benefits from information sharing. In terms of network traffic, defenses with inter-node connection operations may introduce communication overhead since they may change the overlay topology. Besides, more complex defenses may necessitate a higher number of computations, resulting in computation overhead and increased communication latency. This can lead to prolonged convergence times for the federation. It is crucial to balance overhead and robustness when designing defenses. MTD is a dynamic defense framework that enables defenders to adjust security levels based on specific requirements rather than striving for absolute security. As such, MTD is suggested as a design framework for DFL defenses to manage the trade-off between security and overhead effectively. Viable MTD strategies include model backup and rollback, dynamic topology, and dynamic aggregation. Furthermore, conducting data auditing and validation for each node before starting federated training can effectively protect against data poisoning attacks.
    \item \textbf{Data Heterogeneity.} As a data-driven process, performance of DFL relies on how the data is distributed across the nodes. It is often assumed that the data follows the IID settings when investigating the robustness of DFL models. Algorithms like Krum are based on the assumption that models trained on similar data should be highly similar, but anomalies are not. However, this assumption is invalidated in non-IID scenarios where data distribution across nodes is highly varsion. In non-IID settings, model similarity decreases, making it challenging to differentiate between benign and malicious models using similarity or clustering. In these cases, data-independent metrics can be utilized to detect malicious models. Rather than solely depending on model similarity, utilizing intrinsic model characteristics, such as eXplainable AI (XAI) metrics, can improve the performance of defense mechanisms while preserving the knowledge exchanged by non-IID nodes.

\end{itemize}

\subsection{Roadmap for Future Work}
\label{sec:roadmap}
%MTD
Based on the preceding discussion, the following aspects can serve as a guiding framework for future studies aimed at strengthening the robustness of DFL models:
\begin{itemize}
    \item \textbf{Topology is the Key.} As mentioned several times above, overlay topology is the key factor affecting model robustness in DFL. Therefore, adopting the strategy of dynamic topology, either proactive or reactive, can mitigate the effects of different kinds of poisoning attacks while optimizing the model performance of DFL.

    \item \textbf{Dynamic Aggregation.} Advanced attacks optimize attack strategies for specific aggregation functions. Therefore, using dynamic aggregation functions, such as changing the aggregation algorithm in every round, can effectively prevent the knowledge of aggregation strategies from being identified by the attackers.

    \item \textbf{Endogenous Properties of Models.} Metrics such as model similarity strongly depend on the distribution of the data and thus are inefficient in the non-IID environment. The endogenous properties of the model, such as convergence and entropy and other metrics used in XAI, are a viable way to enhance the robustness of the model in the non-IID environment.

    \item \textbf{Data Augmentation and Synthetic Data.} Utilizing techniques such as data augmentation and synthetic data generation can maintain the integrity of the original data while also ensuring data privacy. Consequently, these approaches can be employed in DFL to generate synthetic data that balances data distribution among nodes. By leveraging the synthetic data generated, it is possible to approximate non-IID as IID, thereby enabling the detection and mitigation of poisoning attacks.
    
\end{itemize}

\section{Conclusion}
\label{sec:conclusion}
DFL is gaining prominence as a novel paradigm for achieving stable, collaborative, and privacy-preserving ML. Its distinctive architecture renders it vulnerable to various forms of malicious attacks, particularly poisoning attacks. There is a growing interest in researching model robustness in FL. However, existing research, which explores both offensive and defensive approaches, mainly concentrates on investigating the CFL paradigm. Therefore, there is a great need for studies on model robustness for DFL paradigm.

Therefore, this paper provides a comprehensive literature review of poisoning attacks that target model robustness in the DFL paradigm, along with the corresponding defense mechanisms. Additionally, a novel module named \solution{} is introduced to assess the robustness of DFL models, which is then implemented and integrated into a DFL platform. Through a series of thorough experiments, this study contrasts the behavior of CFL and DFL when subjected to various poisoning attacks, identifying the critical factors that influence the spread and effectiveness of attacks within DFL. Furthermore, it benchmarks the efficacy of different defense mechanisms and explores the compatibility of CFL-based defense strategies with DFL.

The empirical findings offer valuable insights into the challenges of research and propose potential directions to enhance the robustness of DFL models for future research.

\section*{Declaration of Competing Interest}
The authors declare that they have no known competing financial interests or personal relationships that could have appeared to influence the work reported in this paper. 

\section*{CRediT authorship contribution statement}
\textbf{Chao Feng.}  Methodology, Conceptualization, Writing, Review \& Editing.
\textbf{Alberto Huertas Celdran.} Methodology, Writing, Review. 
\textbf{Jan von der Assen.} Writing, Review. 
\textbf{Enrique Tom\'as Mart\'inez Beltr\'an.} Writing, Review.
\textbf{Gérôme Bovet.} Project administration, Funding acquisition.
\textbf{Burkhard Stiller.} Supervision, Funding acquisition.

\section*{Acknowledgment}

This work has been partially supported by \textit{(a)} the Swiss Federal Office for Defense Procurement (armasuisse) with the CyberMind and DATRIS (CYD-C-2020003) projects and \textit{(b)} the University of Zürich UZH.

\balance
\bibliographystyle{cas-model2-names}
\bibliography{main}

\begin{thebibliography}{73}
\expandafter\ifx\csname natexlab\endcsname\relax\def\natexlab#1{#1}\fi
\providecommand{\url}[1]{\texttt{#1}}
\providecommand{\href}[2]{#2}
\providecommand{\path}[1]{#1}
\providecommand{\DOIprefix}{doi:}
\providecommand{\ArXivprefix}{arXiv:}
\providecommand{\URLprefix}{URL: }
\providecommand{\Pubmedprefix}{pmid:}
\providecommand{\doi}[1]{\href{http://dx.doi.org/#1}{\path{#1}}}
\providecommand{\Pubmed}[1]{\href{pmid:#1}{\path{#1}}}
\providecommand{\bibinfo}[2]{#2}
\ifx\xfnm\relax \def\xfnm[#1]{\unskip,\space#1}\fi
%Type = Misc
\bibitem[{Abadi et~al.(2016)Abadi, Barham, Chen, Chen, Davis, Dean, Devin, Ghemawat, Irving, Isard, Kudlur, Levenberg, Monga, Moore, Murray, Steiner, Tucker, Vasudevan, Warden, Wicke, Yu and Zheng}]{tensorboard}
\bibinfo{author}{Abadi, M.}, \bibinfo{author}{Barham, P.}, \bibinfo{author}{Chen, J.}, \bibinfo{author}{Chen, Z.}, \bibinfo{author}{Davis, A.}, \bibinfo{author}{Dean, J.}, \bibinfo{author}{Devin, M.}, \bibinfo{author}{Ghemawat, S.}, \bibinfo{author}{Irving, G.}, \bibinfo{author}{Isard, M.}, \bibinfo{author}{Kudlur, M.}, \bibinfo{author}{Levenberg, J.}, \bibinfo{author}{Monga, R.}, \bibinfo{author}{Moore, S.}, \bibinfo{author}{Murray, D.G.}, \bibinfo{author}{Steiner, B.}, \bibinfo{author}{Tucker, P.}, \bibinfo{author}{Vasudevan, V.}, \bibinfo{author}{Warden, P.}, \bibinfo{author}{Wicke, M.}, \bibinfo{author}{Yu, Y.}, \bibinfo{author}{Zheng, X.}, \bibinfo{year}{2016}.
\newblock \bibinfo{title}{Tensorflow: A system for large-scale machine learning}.
\newblock \URLprefix \url{https://www.tensorflow.org/tensorboard}, \href{http://arxiv.org/abs/1605.08695}{\tt arXiv:1605.08695}. \bibinfo{note}{{Last} Visit July 2024}.
%Type = Article
\bibitem[{Ali et~al.(2022)Ali, Naeem, Tariq and Kaddoum}]{ali2022federated}
\bibinfo{author}{Ali, M.}, \bibinfo{author}{Naeem, F.}, \bibinfo{author}{Tariq, M.}, \bibinfo{author}{Kaddoum, G.}, \bibinfo{year}{2022}.
\newblock \bibinfo{title}{Federated learning for privacy preservation in smart healthcare systems: A comprehensive survey}.
\newblock \bibinfo{journal}{IEEE journal of biomedical and health informatics} \bibinfo{volume}{27}, \bibinfo{pages}{778--789}.
%Type = Inproceedings
\bibitem[{Bagdasaryan et~al.(2019)Bagdasaryan, Veit, Hua, Estrin and Shmatikov}]{bagdasaryan_HowBackdoor_2019}
\bibinfo{author}{Bagdasaryan, E.}, \bibinfo{author}{Veit, A.}, \bibinfo{author}{Hua, Y.}, \bibinfo{author}{Estrin, D.}, \bibinfo{author}{Shmatikov, V.}, \bibinfo{year}{2019}.
\newblock \bibinfo{title}{How {To} {Backdoor} {Federated} {Learning}}, in: \bibinfo{booktitle}{arXiv:1807.00459 [cs]}, pp. \bibinfo{pages}{1--15}.
%Type = Article
\bibitem[{Beltran et~al.(2024)Beltran, Gómez, Feng, Sanchez, Bernal, Bovet, Perez, Perez and Celdran}]{beltran_FedstellarPlatform_2023}
\bibinfo{author}{Beltran, E.T.M.}, \bibinfo{author}{Gómez, a.L.P.}, \bibinfo{author}{Feng, C.}, \bibinfo{author}{Sanchez, P.M.S.}, \bibinfo{author}{Bernal, S.L.}, \bibinfo{author}{Bovet, G.}, \bibinfo{author}{Perez, M.G.}, \bibinfo{author}{Perez, G.M.}, \bibinfo{author}{Celdran, A.H.}, \bibinfo{year}{2024}.
\newblock \bibinfo{title}{Fedstellar: A platform for decentralized federated learning}.
\newblock \bibinfo{journal}{Expert Systems with Applications} \bibinfo{volume}{242}, \bibinfo{pages}{122861}.
%Type = Article
\bibitem[{Beltran et~al.(2023)Beltran, Perez, Sanchez, Bernal, Bovet, Perez, Perez and Celdran}]{beltran_DecentralizedFederated_2022}
\bibinfo{author}{Beltran, E.T.M.}, \bibinfo{author}{Perez, M.Q.}, \bibinfo{author}{Sanchez, P.M.S.}, \bibinfo{author}{Bernal, S.L.}, \bibinfo{author}{Bovet, G.}, \bibinfo{author}{Perez, M.G.}, \bibinfo{author}{Perez, G.M.}, \bibinfo{author}{Celdran, A.H.}, \bibinfo{year}{2023}.
\newblock \bibinfo{title}{Decentralized {Federated} {Learning}: {Fundamentals}, {State}-of-the-art, {Frameworks}, {Trends}, and {Challenges}}.
\newblock \bibinfo{journal}{IEEE Communications Surveys and Tutorials} \bibinfo{volume}{25}, \bibinfo{pages}{2983--3013}.
%Type = Article
\bibitem[{Benmalek et~al.(2022)Benmalek, Benrekia and Challal}]{benmalek:hal-03620400}
\bibinfo{author}{Benmalek, M.}, \bibinfo{author}{Benrekia, M.A.}, \bibinfo{author}{Challal, Y.}, \bibinfo{year}{2022}.
\newblock \bibinfo{title}{{Security of Federated Learning: Attacks, Defensive Mechanisms, and Challenges}}.
\newblock \bibinfo{journal}{{Revue des Sciences et Technologies de l'Information - S{\'e}rie RIA : Revue d'Intelligence Artificielle}} \bibinfo{volume}{36}, \bibinfo{pages}{49--59}.
%Type = Inproceedings
\bibitem[{Blanchard et~al.(2017)Blanchard, El~Mhamdi, Guerraoui and Stainer}]{blanchard_MachineLearning_2017}
\bibinfo{author}{Blanchard, P.}, \bibinfo{author}{El~Mhamdi, E.M.}, \bibinfo{author}{Guerraoui, R.}, \bibinfo{author}{Stainer, J.}, \bibinfo{year}{2017}.
\newblock \bibinfo{title}{Machine {Learning} with {Adversaries}: {Byzantine} {Tolerant} {Gradient} {Descent}}, in: \bibinfo{booktitle}{Proceedings of the 31st International Conference on Neural Information Processing Systems}, \bibinfo{publisher}{{California}, USA}. pp. \bibinfo{pages}{118--128}.
%Type = Article
\bibitem[{Blanco-Justicia et~al.(2021)Blanco-Justicia, Domingo-Ferrer, Martínez, Sánchez, Flanagan and Tan}]{BLANCOJUSTICIA2021104468}
\bibinfo{author}{Blanco-Justicia, A.}, \bibinfo{author}{Domingo-Ferrer, J.}, \bibinfo{author}{Martínez, S.}, \bibinfo{author}{Sánchez, D.}, \bibinfo{author}{Flanagan, A.}, \bibinfo{author}{Tan, K.E.}, \bibinfo{year}{2021}.
\newblock \bibinfo{title}{Achieving security and privacy in federated learning systems: Survey, research challenges and future directions}.
\newblock \bibinfo{journal}{Engineering Applications of Artificial Intelligence} \bibinfo{volume}{106}, \bibinfo{pages}{104468}.
%Type = Inproceedings
\bibitem[{Bourtoule et~al.(2021)Bourtoule, Chandrasekaran, Choquette-Choo, Jia, Travers, Zhang, Lie and Papernot}]{machineunlearning}
\bibinfo{author}{Bourtoule, L.}, \bibinfo{author}{Chandrasekaran, V.}, \bibinfo{author}{Choquette-Choo, C.A.}, \bibinfo{author}{Jia, H.}, \bibinfo{author}{Travers, A.}, \bibinfo{author}{Zhang, B.}, \bibinfo{author}{Lie, D.}, \bibinfo{author}{Papernot, N.}, \bibinfo{year}{2021}.
\newblock \bibinfo{title}{Machine unlearning}, in: \bibinfo{booktitle}{2021 IEEE Symposium on Security and Privacy (SP)}, pp. \bibinfo{pages}{141--159}.
\newblock \DOIprefix\doi{10.1109/SP40001.2021.00019}.
%Type = Article
\bibitem[{Cai et~al.(2016)Cai, Wang, Hu and Wang}]{cai2016moving}
\bibinfo{author}{Cai, G.}, \bibinfo{author}{Wang, B.}, \bibinfo{author}{Hu, W.}, \bibinfo{author}{Wang, T.}, \bibinfo{year}{2016}.
\newblock \bibinfo{title}{{Moving Target Defense: State of the Art and Characteristics}}.
\newblock \bibinfo{journal}{Frontiers of Information Technology \& Electronic Engineering} \bibinfo{volume}{17}, \bibinfo{pages}{1122--1153}.
%Type = Inproceedings
\bibitem[{Cao et~al.(2019)Cao, Chang, Lin, Liu and Sun}]{cao_UnderstandingDistributed_2019}
\bibinfo{author}{Cao, D.}, \bibinfo{author}{Chang, S.}, \bibinfo{author}{Lin, Z.}, \bibinfo{author}{Liu, G.}, \bibinfo{author}{Sun, D.}, \bibinfo{year}{2019}.
\newblock \bibinfo{title}{Understanding {Distributed} {Poisoning} {Attack} in {Federated} {Learning}}, in: \bibinfo{booktitle}{{IEEE} 25th {International} {Conference} on {Parallel} and {Distributed} {Systems}}.
%Type = Inproceedings
\bibitem[{Cao et~al.(2021)Cao, Fang, Liu and Gong}]{cao_FLTrustByzantinerobust_2021}
\bibinfo{author}{Cao, X.}, \bibinfo{author}{Fang, M.}, \bibinfo{author}{Liu, J.}, \bibinfo{author}{Gong, N.Z.}, \bibinfo{year}{2021}.
\newblock \bibinfo{title}{{FLTrust}: {Byzantine}-robust {Federated} {Learning} via {Trust} {Bootstrapping}}, in: \bibinfo{booktitle}{Proceedings 2021 {Network} and {Distributed} {System} {Security} {Symposium}}.
%Type = Misc
\bibitem[{Celdran et~al.(2023)Celdran, Feng, Sanchez, Zumtaugwald, Bovet and Stiller}]{celdran2023sustainability}
\bibinfo{author}{Celdran, A.H.}, \bibinfo{author}{Feng, C.}, \bibinfo{author}{Sanchez, P.M.S.}, \bibinfo{author}{Zumtaugwald, L.}, \bibinfo{author}{Bovet, G.}, \bibinfo{author}{Stiller, B.}, \bibinfo{year}{2023}.
\newblock \bibinfo{title}{Assessing the sustainability and trustworthiness of federated learning models}.
\newblock \URLprefix \url{https://arxiv.org/abs/2310.20435}, \href{http://arxiv.org/abs/2310.20435}{\tt arXiv:2310.20435}.
%Type = Inproceedings
\bibitem[{Chen et~al.(2022)Chen, Gui, Lin, Gan and Wu}]{Chen10020431}
\bibinfo{author}{Chen, Y.}, \bibinfo{author}{Gui, Y.}, \bibinfo{author}{Lin, H.}, \bibinfo{author}{Gan, W.}, \bibinfo{author}{Wu, Y.}, \bibinfo{year}{2022}.
\newblock \bibinfo{title}{Federated learning attacks and defenses: A survey}, in: \bibinfo{booktitle}{2022 IEEE International Conference on Big Data (Big Data)}, pp. \bibinfo{pages}{4256--4265}.
%Type = Misc
\bibitem[{Duarte(2023)}]{Duarte2023}
\bibinfo{author}{Duarte, F.}, \bibinfo{year}{2023}.
\newblock \bibinfo{title}{{Number of IOT devices (2023-2030)}}.
\newblock \URLprefix \url{https://explodingtopics.com/blog/number-of-iot-devices}. \bibinfo{note}{{Last} Visit December 2023}.
%Type = Misc
\bibitem[{{European Parliament and Council of the European Union}(2016)}]{gdpr2016}
\bibinfo{author}{{European Parliament and Council of the European Union}}, \bibinfo{year}{2016}.
\newblock \bibinfo{title}{Regulation (eu) 2016/679 of the european parliament and of the council of 27 april 2016 on the protection of natural persons with regard to the processing of personal data and on the free movement of such data, and repealing directive 95/46/ec (general data protection regulation)}.
\newblock \URLprefix \url{https://eur-lex.europa.eu/eli/reg/2016/679/oj}. \bibinfo{note}{{Last} Visit July 2024}.
%Type = Misc
\bibitem[{Falcon and team(2019)}]{pytorch_lightning}
\bibinfo{author}{Falcon, W.}, \bibinfo{author}{team, T.P.L.}, \bibinfo{year}{2019}.
\newblock \bibinfo{title}{Pytorch lightning}.
\newblock \URLprefix \url{https://www.pytorchlightning.ai}. \bibinfo{note}{{Last} Visit July 2024}.
%Type = Misc
\bibitem[{Fang et~al.(2021)Fang, Cao, Jia and Gong}]{fang_LocalModel_2021}
\bibinfo{author}{Fang, M.}, \bibinfo{author}{Cao, X.}, \bibinfo{author}{Jia, J.}, \bibinfo{author}{Gong, N.Z.}, \bibinfo{year}{2021}.
\newblock \bibinfo{title}{Local {Model} {Poisoning} {Attacks} to {Byzantine}-{Robust} {Federated} {Learning}}.
\newblock \bibinfo{note}{ArXiv:1911.11815 [cs]}.
%Type = Misc
\bibitem[{Feng et~al.(2023a)Feng, Celdran, Baltensperger, Beltran, Bovet and Stiller}]{feng2023sentinel}
\bibinfo{author}{Feng, C.}, \bibinfo{author}{Celdran, A.H.}, \bibinfo{author}{Baltensperger, J.}, \bibinfo{author}{Beltran, E.T.M.}, \bibinfo{author}{Bovet, G.}, \bibinfo{author}{Stiller, B.}, \bibinfo{year}{2023}a.
\newblock \bibinfo{title}{Sentinel: An aggregation function to secure decentralized federated learning}.
%Type = Misc
\bibitem[{Feng et~al.(2023b)Feng, Celdran, Sanchez, Kreischer, von~der Assen, Bovet, Perez and Stiller}]{feng2023cyberforce}
\bibinfo{author}{Feng, C.}, \bibinfo{author}{Celdran, A.H.}, \bibinfo{author}{Sanchez, P.M.S.}, \bibinfo{author}{Kreischer, J.}, \bibinfo{author}{von~der Assen, J.}, \bibinfo{author}{Bovet, G.}, \bibinfo{author}{Perez, G.M.}, \bibinfo{author}{Stiller, B.}, \bibinfo{year}{2023}b.
\newblock \bibinfo{title}{Cyberforce: A federated reinforcement learning framework for malware mitigation}.
\newblock \URLprefix \url{https://arxiv.org/abs/2308.05978}, \href{http://arxiv.org/abs/2308.05978}{\tt arXiv:2308.05978}.
%Type = Article
\bibitem[{Feng et~al.(2024)Feng, Celdran, Vuong, Bovet and Stiller}]{feng2023voyager}
\bibinfo{author}{Feng, C.}, \bibinfo{author}{Celdran, A.H.}, \bibinfo{author}{Vuong, M.}, \bibinfo{author}{Bovet, G.}, \bibinfo{author}{Stiller, B.}, \bibinfo{year}{2024}.
\newblock \bibinfo{title}{Voyager: Mtd-based aggregation protocol for mitigating poisoning attacks on dfl}.
\newblock \bibinfo{journal}{IEEE/IFIP Network Operations and Management Symposium} .
%Type = Misc
\bibitem[{Flask(2024)}]{Flask}
\bibinfo{author}{Flask}, \bibinfo{year}{2024}.
\newblock \bibinfo{title}{{Flask}}.
\newblock \URLprefix \url{https://flask.palletsprojects.com/en/3.0.x/}. \bibinfo{note}{{Last} Visit July 2024}.
%Type = Inproceedings
\bibitem[{Frid-Adar et~al.(2018)Frid-Adar, Klang, Amitai, Goldberger and Greenspan}]{gan2018}
\bibinfo{author}{Frid-Adar, M.}, \bibinfo{author}{Klang, E.}, \bibinfo{author}{Amitai, M.}, \bibinfo{author}{Goldberger, J.}, \bibinfo{author}{Greenspan, H.}, \bibinfo{year}{2018}.
\newblock \bibinfo{title}{Synthetic data augmentation using gan for improved liver lesion classification}, in: \bibinfo{booktitle}{2018 IEEE 15th International Symposium on Biomedical Imaging (ISBI 2018)}, pp. \bibinfo{pages}{289--293}.
\newblock \DOIprefix\doi{10.1109/ISBI.2018.8363576}.
%Type = Misc
\bibitem[{Fung et~al.(2020)Fung, Yoon and Beschastnikh}]{fung_MitigatingSybils_2020}
\bibinfo{author}{Fung, C.}, \bibinfo{author}{Yoon, C.J.M.}, \bibinfo{author}{Beschastnikh, I.}, \bibinfo{year}{2020}.
\newblock \bibinfo{title}{Mitigating {Sybils} in {Federated} {Learning} {Poisoning}}.
%Type = Inproceedings
\bibitem[{Gholami et~al.(2022)Gholami, Torkzaban and Baras}]{gholami_TrustedDecentralized_2022}
\bibinfo{author}{Gholami, A.}, \bibinfo{author}{Torkzaban, N.}, \bibinfo{author}{Baras, J.S.}, \bibinfo{year}{2022}.
\newblock \bibinfo{title}{Trusted {Decentralized} {Federated} {Learning}}, in: \bibinfo{booktitle}{2022 {IEEE} 19th {Annual} {Consumer} {Communications} \& {Networking} {Conference} ({CCNC})}, \bibinfo{publisher}{IEEE}, \bibinfo{address}{Las Vegas, NV, USA}. pp. \bibinfo{pages}{1--6}.
%Type = Misc
\bibitem[{Guo et~al.(2021)Guo, Zhang, Yu, Xie, Ma, Xiang and Liu}]{guo_ByzantineresilientDecentralized_2021}
\bibinfo{author}{Guo, S.}, \bibinfo{author}{Zhang, T.}, \bibinfo{author}{Yu, H.}, \bibinfo{author}{Xie, X.}, \bibinfo{author}{Ma, L.}, \bibinfo{author}{Xiang, T.}, \bibinfo{author}{Liu, Y.}, \bibinfo{year}{2021}.
\newblock \bibinfo{title}{Byzantine-resilient {Decentralized} {Stochastic} {Gradient} {Descent}}.
\newblock \bibinfo{note}{ArXiv:2002.08569}.
%Type = Article
\bibitem[{Howard et~al.(2017)Howard, Zhu, Chen, Kalenichenko, Wang, Weyand, Andreetto and Adam}]{howard_MobileNetsEfficient_2017}
\bibinfo{author}{Howard, A.G.}, \bibinfo{author}{Zhu, M.}, \bibinfo{author}{Chen, B.}, \bibinfo{author}{Kalenichenko, D.}, \bibinfo{author}{Wang, W.}, \bibinfo{author}{Weyand, T.}, \bibinfo{author}{Andreetto, M.}, \bibinfo{author}{Adam, H.}, \bibinfo{year}{2017}.
\newblock \bibinfo{title}{Mobilenets: Efficient convolutional neural networks for mobile vision applications}.
\newblock \bibinfo{journal}{arXiv preprint arXiv:1704.04861} .
%Type = Article
\bibitem[{Huertas~Celdran et~al.(2023)Huertas~Celdran, Sanchez~Sanchez, Feng, Bovet, Perez and Stiller}]{huertas2022iotj}
\bibinfo{author}{Huertas~Celdran, A.}, \bibinfo{author}{Sanchez~Sanchez, P.M.}, \bibinfo{author}{Feng, C.}, \bibinfo{author}{Bovet, G.}, \bibinfo{author}{Perez, G.M.}, \bibinfo{author}{Stiller, B.}, \bibinfo{year}{2023}.
\newblock \bibinfo{title}{{Privacy-Preserving and Syscall-Based Intrusion Detection System for IoT Spectrum Sensors Affected by Data Falsification Attacks}}.
\newblock \bibinfo{journal}{IEEE Internet of Things Journal} \bibinfo{volume}{10}, \bibinfo{pages}{8408--8415}.
%Type = Article
\bibitem[{Jere et~al.(2021)Jere, Farnan and Koushanfar}]{Jere9308910}
\bibinfo{author}{Jere, M.S.}, \bibinfo{author}{Farnan, T.}, \bibinfo{author}{Koushanfar, F.}, \bibinfo{year}{2021}.
\newblock \bibinfo{title}{A taxonomy of attacks on federated learning}.
\newblock \bibinfo{journal}{IEEE Security \& Privacy} \bibinfo{volume}{19}, \bibinfo{pages}{20--28}.
%Type = Techreport
\bibitem[{Krizhevsky(2009)}]{krizhevsky_LearningMultiple_2009}
\bibinfo{author}{Krizhevsky, A.}, \bibinfo{year}{2009}.
\newblock \bibinfo{title}{Learning multiple layers of features from tiny images}.
\newblock \bibinfo{type}{Technical Report}.
%Type = Article
\bibitem[{Kumar et~al.(2023)Kumar, Mohan and Cenkeramaddi}]{Kumar10274102}
\bibinfo{author}{Kumar, K.N.}, \bibinfo{author}{Mohan, C.K.}, \bibinfo{author}{Cenkeramaddi, L.R.}, \bibinfo{year}{2023}.
\newblock \bibinfo{title}{The impact of adversarial attacks on federated learning: A survey}.
\newblock \bibinfo{journal}{IEEE Transactions on Pattern Analysis and Machine Intelligence} , \bibinfo{pages}{1--20}.
%Type = Article
\bibitem[{Kuo and Pham(2022)}]{kuo2022detecting}
\bibinfo{author}{Kuo, T.T.}, \bibinfo{author}{Pham, A.}, \bibinfo{year}{2022}.
\newblock \bibinfo{title}{Detecting model misconducts in decentralized healthcare federated learning}.
\newblock \bibinfo{journal}{International journal of medical informatics} \bibinfo{volume}{158}, \bibinfo{pages}{104658}.
%Type = Article
\bibitem[{LeCun and Cortes(2010)}]{lecun_MNISTHandwritten_2010}
\bibinfo{author}{LeCun, Y.}, \bibinfo{author}{Cortes, C.}, \bibinfo{year}{2010}.
\newblock \bibinfo{title}{{MNIST} handwritten digit database} .
%Type = Article
\bibitem[{Li et~al.(2019)Li, Xu, Chen, Giannakis and Ling}]{li_RSAByzantineRobust_2019}
\bibinfo{author}{Li, L.}, \bibinfo{author}{Xu, W.}, \bibinfo{author}{Chen, T.}, \bibinfo{author}{Giannakis, G.B.}, \bibinfo{author}{Ling, Q.}, \bibinfo{year}{2019}.
\newblock \bibinfo{title}{{RSA}: {Byzantine}-{Robust} {Stochastic} {Aggregation} {Methods} for {Distributed} {Learning} from {Heterogeneous} {Datasets}}.
\newblock \bibinfo{journal}{Proceedings of the AAAI Conference on Artificial Intelligence} \bibinfo{volume}{33}, \bibinfo{pages}{1544--1551}.
%Type = Misc
\bibitem[{Li et~al.(2020)Li, Cheng, Wang, Liu and Chen}]{li_LearningDetect_2020}
\bibinfo{author}{Li, S.}, \bibinfo{author}{Cheng, Y.}, \bibinfo{author}{Wang, W.}, \bibinfo{author}{Liu, Y.}, \bibinfo{author}{Chen, T.}, \bibinfo{year}{2020}.
\newblock \bibinfo{title}{Learning to {Detect} {Malicious} {Clients} for {Robust} {Federated} {Learning}}.
\newblock \bibinfo{note}{ArXiv:2002.00211}.
%Type = Article
\bibitem[{Liu et~al.(2022)Liu, Xu and Wang}]{liu2022threats}
\bibinfo{author}{Liu, P.}, \bibinfo{author}{Xu, X.}, \bibinfo{author}{Wang, W.}, \bibinfo{year}{2022}.
\newblock \bibinfo{title}{Threats, attacks and defenses to federated learning: issues, taxonomy and perspectives}.
\newblock \bibinfo{journal}{Cybersecurity} \bibinfo{volume}{5}, \bibinfo{pages}{1--19}.
%Type = Article
\bibitem[{Lyu et~al.(2022)Lyu, Yu, Ma, Chen, Sun, Zhao, Yang and Yu}]{lyu_PrivacyRobustness_2022}
\bibinfo{author}{Lyu, L.}, \bibinfo{author}{Yu, H.}, \bibinfo{author}{Ma, X.}, \bibinfo{author}{Chen, C.}, \bibinfo{author}{Sun, L.}, \bibinfo{author}{Zhao, J.}, \bibinfo{author}{Yang, Q.}, \bibinfo{author}{Yu, P.S.}, \bibinfo{year}{2022}.
\newblock \bibinfo{title}{Privacy and {Robustness} in {Federated} {Learning}: {Attacks} and {Defenses}}.
\newblock \bibinfo{journal}{IEEE Transactions on Neural Networks and Learning Systems} , \bibinfo{pages}{1--21}.
%Type = Misc
\bibitem[{Merkel(2014)}]{docker}
\bibinfo{author}{Merkel, D.}, \bibinfo{year}{2014}.
\newblock \bibinfo{title}{Docker: Lightweight linux containers for consistent development and deployment}.
\newblock \URLprefix \url{https://doi.org/10.1016/B978-0-12-800729-7.00004-2}. \bibinfo{note}{accessed: 2024-07-09}.
%Type = Article
\bibitem[{Mhamdi et~al.(2018)Mhamdi, Guerraoui and Rouault}]{mhamdi_HiddenVulnerability_2018}
\bibinfo{author}{Mhamdi, E.M.E.}, \bibinfo{author}{Guerraoui, R.}, \bibinfo{author}{Rouault, S.}, \bibinfo{year}{2018}.
\newblock \bibinfo{title}{The {H}idden {V}ulnerability of {D}istributed{ L}earning in {B}yzantium}.
\newblock \bibinfo{journal}{arXiv preprint arXiv:1802.07927} .
%Type = Article
\bibitem[{Mothukuri et~al.(2021)Mothukuri, Parizi, Pouriyeh, Huang, Dehghantanha and Srivastava}]{MOTHUKURI2021619}
\bibinfo{author}{Mothukuri, V.}, \bibinfo{author}{Parizi, R.M.}, \bibinfo{author}{Pouriyeh, S.}, \bibinfo{author}{Huang, Y.}, \bibinfo{author}{Dehghantanha, A.}, \bibinfo{author}{Srivastava, G.}, \bibinfo{year}{2021}.
\newblock \bibinfo{title}{A survey on security and privacy of federated learning}.
\newblock \bibinfo{journal}{Future Generation Computer Systems} \bibinfo{volume}{115}, \bibinfo{pages}{619--640}.
%Type = Misc
\bibitem[{Muñoz-Gonzalez et~al.(2019)Muñoz-Gonzalez, Co and Lupu}]{munoz-gonzalez_ByzantineRobustFederated_2019}
\bibinfo{author}{Muñoz-Gonzalez, L.}, \bibinfo{author}{Co, K.T.}, \bibinfo{author}{Lupu, E.C.}, \bibinfo{year}{2019}.
\newblock \bibinfo{title}{Byzantine-{Robust} {Federated} {Machine} {Learning} through {Adaptive} {Model} {Averaging}}.
\newblock \bibinfo{note}{ArXiv:1909.05125}.
%Type = Article
\bibitem[{Nair et~al.(2023)Nair, Raj and Sahoo}]{NAIR2023103723}
\bibinfo{author}{Nair, A.K.}, \bibinfo{author}{Raj, E.D.}, \bibinfo{author}{Sahoo, J.}, \bibinfo{year}{2023}.
\newblock \bibinfo{title}{A robust analysis of adversarial attacks on federated learning environments}.
\newblock \bibinfo{journal}{Computer Standards \& Interfaces} \bibinfo{volume}{86}, \bibinfo{pages}{103723}.
%Type = Misc
\bibitem[{Nguyen et~al.(2021)Nguyen, Rieger, Chen, Yalame, Möllering, Fereidooni, Marchal, Miettinen, Mirhoseini, Zeitouni, Koushanfar, Sadeghi and Schneider}]{nguyen_FLAMETaming_2021}
\bibinfo{author}{Nguyen, T.}, \bibinfo{author}{Rieger, P.}, \bibinfo{author}{Chen, H.}, \bibinfo{author}{Yalame, H.}, \bibinfo{author}{Möllering, H.}, \bibinfo{author}{Fereidooni, H.}, \bibinfo{author}{Marchal, S.}, \bibinfo{author}{Miettinen, M.}, \bibinfo{author}{Mirhoseini, A.}, \bibinfo{author}{Zeitouni, S.}, \bibinfo{author}{Koushanfar, F.}, \bibinfo{author}{Sadeghi, A.}, \bibinfo{author}{Schneider, T.}, \bibinfo{year}{2021}.
\newblock \bibinfo{title}{{FLAME}: {Taming} {Backdoors} in {Federated} {Learning}}.
%Type = Misc
\bibitem[{Nguyen et~al.(2023)Nguyen, Nguyen, Nguyen, Pham, Doan and Wong}]{nguyen_BackdoorAttacks_2023}
\bibinfo{author}{Nguyen, T.D.}, \bibinfo{author}{Nguyen, T.}, \bibinfo{author}{Nguyen, P.L.}, \bibinfo{author}{Pham, H.H.}, \bibinfo{author}{Doan, K.}, \bibinfo{author}{Wong, K.S.}, \bibinfo{year}{2023}.
\newblock \bibinfo{title}{Backdoor {Attacks} and {Defenses} in {Federated} {Learning}: {Survey}, {Challenges} and {Future} {Research} {Directions}}.
\newblock \bibinfo{note}{ArXiv:2303.02213}.
%Type = Inproceedings
\bibitem[{Nguyen et~al.(2020)Nguyen, Rieger, Miettinen and Sadeghi}]{nguyen2020poisoning}
\bibinfo{author}{Nguyen, T.D.}, \bibinfo{author}{Rieger, P.}, \bibinfo{author}{Miettinen, M.}, \bibinfo{author}{Sadeghi, A.R.}, \bibinfo{year}{2020}.
\newblock \bibinfo{title}{Poisoning attacks on federated learning-based iot intrusion detection system}, in: \bibinfo{booktitle}{Proc. Workshop Decentralized IoT Syst. Secur.(DISS)}.
%Type = Article
\bibitem[{Ozdayi et~al.(2021)Ozdayi, Kantarcioglu and Gel}]{ozdayi_DefendingBackdoors_2021}
\bibinfo{author}{Ozdayi, M.S.}, \bibinfo{author}{Kantarcioglu, M.}, \bibinfo{author}{Gel, Y.R.}, \bibinfo{year}{2021}.
\newblock \bibinfo{title}{Defending against {Backdoors} in {Federated} {Learning} with {Robust} {Learning} {Rate}}.
\newblock \bibinfo{journal}{Proceedings of the AAAI Conference on Artificial Intelligence} \bibinfo{volume}{35}, \bibinfo{pages}{9268--9276}.
%Type = Misc
\bibitem[{Paszke et~al.(2019)Paszke, Gross, Massa, Lerer, Bradbury, Chanan, Killeen, Lin, Gimelshein, Antiga, Desmaison, Köpf, Yang, DeVito, Raison, Tejani, Chilamkurthy, Steiner, Fang, Bai and Chintala}]{paszke2019pytorchimperativestylehighperformance}
\bibinfo{author}{Paszke, A.}, \bibinfo{author}{Gross, S.}, \bibinfo{author}{Massa, F.}, \bibinfo{author}{Lerer, A.}, \bibinfo{author}{Bradbury, J.}, \bibinfo{author}{Chanan, G.}, \bibinfo{author}{Killeen, T.}, \bibinfo{author}{Lin, Z.}, \bibinfo{author}{Gimelshein, N.}, \bibinfo{author}{Antiga, L.}, \bibinfo{author}{Desmaison, A.}, \bibinfo{author}{Köpf, A.}, \bibinfo{author}{Yang, E.}, \bibinfo{author}{DeVito, Z.}, \bibinfo{author}{Raison, M.}, \bibinfo{author}{Tejani, A.}, \bibinfo{author}{Chilamkurthy, S.}, \bibinfo{author}{Steiner, B.}, \bibinfo{author}{Fang, L.}, \bibinfo{author}{Bai, J.}, \bibinfo{author}{Chintala, S.}, \bibinfo{year}{2019}.
\newblock \bibinfo{title}{Pytorch: An imperative style, high-performance deep learning library}.
\newblock \URLprefix \url{https://arxiv.org/abs/1912.01703}, \href{http://arxiv.org/abs/1912.01703}{\tt arXiv:1912.01703}.
%Type = Misc
\bibitem[{Pillutla et~al.(2022)Pillutla, Kakade and Harchaoui}]{pillutla_RobustAggregation_2022}
\bibinfo{author}{Pillutla, K.}, \bibinfo{author}{Kakade, S.M.}, \bibinfo{author}{Harchaoui, Z.}, \bibinfo{year}{2022}.
\newblock \bibinfo{title}{Robust {Aggregation} for {Federated} {Learning}}.
\newblock \bibinfo{note}{ArXiv:1912.13445}.
%Type = Article
\bibitem[{Qammar et~al.(2022)Qammar, Ding and Ning}]{qammar2022federated}
\bibinfo{author}{Qammar, A.}, \bibinfo{author}{Ding, J.}, \bibinfo{author}{Ning, H.}, \bibinfo{year}{2022}.
\newblock \bibinfo{title}{Federated learning attack surface: taxonomy, cyber defences, challenges, and future directions}.
\newblock \bibinfo{journal}{Artificial Intelligence Review} , \bibinfo{pages}{1--38}.
%Type = Misc
\bibitem[{Research(2017)}]{googleresearch_FederatedLearning_2017}
\bibinfo{author}{Research, G.}, \bibinfo{year}{2017}.
\newblock \bibinfo{title}{Federated {Learning}: {Collaborative} {Machine} {Learning} without {Centralized} {Training} {Data}}.
%Type = Inproceedings
\bibitem[{Rieger et~al.(2022)Rieger, Nguyen, Miettinen and Sadeghi}]{rieger_DeepSightMitigating_2022}
\bibinfo{author}{Rieger, P.}, \bibinfo{author}{Nguyen, T.}, \bibinfo{author}{Miettinen, M.}, \bibinfo{author}{Sadeghi, A.}, \bibinfo{year}{2022}.
\newblock \bibinfo{title}{{DeepSight}: {Mitigating} {Backdoor} {Attacks} in {Federated} {Learning} {Through} {Deep} {Model} {Inspection}}, in: \bibinfo{booktitle}{Proceedings {Network} and {Distributed} {System} {Security} {Symposium}}.
%Type = Article
\bibitem[{Rodríguez-Barroso et~al.(2023)Rodríguez-Barroso, Jimenez-López, Luzón, Herrera and Martínez-Camara}]{rodriguez-barroso_SurveyFederated_2023}
\bibinfo{author}{Rodríguez-Barroso, N.}, \bibinfo{author}{Jimenez-López, D.}, \bibinfo{author}{Luzón, M.V.}, \bibinfo{author}{Herrera, F.}, \bibinfo{author}{Martínez-Camara, E.}, \bibinfo{year}{2023}.
\newblock \bibinfo{title}{{Survey on Federated Learning Threats: {Concepts}, Taxonomy on Attacks and Defences, Experimental Study and Challenges}}.
\newblock \bibinfo{journal}{Information Fusion} \bibinfo{volume}{90}, \bibinfo{pages}{148--173}.
%Type = Inproceedings
\bibitem[{Shejwalkar and Houmansadr(2021)}]{shejwalkar_ManipulatingByzantine_2021}
\bibinfo{author}{Shejwalkar, V.}, \bibinfo{author}{Houmansadr, A.}, \bibinfo{year}{2021}.
\newblock \bibinfo{title}{Manipulating the {Byzantine}: {Optimizing} {Model} {Poisoning} {Attacks} and {Defenses} for {Federated} {Learning}}, in: \bibinfo{booktitle}{Proceedings 2021 {Network} and {Distributed} {System} {Security} {Symposium}}.
%Type = Article
\bibitem[{Silva et~al.(2023)Silva, Vinagre and Gama}]{silva_FederatedLearning_2023}
\bibinfo{author}{Silva, P.R.}, \bibinfo{author}{Vinagre, J.}, \bibinfo{author}{Gama, J.}, \bibinfo{year}{2023}.
\newblock \bibinfo{title}{{Towards Federated Learning: {An} Overview of Methods and Applications}}.
\newblock \bibinfo{journal}{WIREs Data Mining and Knowledge Discovery} .
%Type = Misc
\bibitem[{Sun et~al.(2019)Sun, Kairouz, Suresh and McMahan}]{sun_CanYou_2019}
\bibinfo{author}{Sun, Z.}, \bibinfo{author}{Kairouz, P.}, \bibinfo{author}{Suresh, A.T.}, \bibinfo{author}{McMahan, H.B.}, \bibinfo{year}{2019}.
\newblock \bibinfo{title}{Can {You} {Really} {Backdoor} {Federated} {Learning}?}
\newblock \bibinfo{note}{ArXiv:1911.07963}.
%Type = Article
\bibitem[{{Sánchez Sánchez} et~al.(2024){Sánchez Sánchez}, {Huertas Celdrán}, Xie, Bovet, {Martínez Pérez} and Stiller}]{SANCHEZSANCHEZ202483}
\bibinfo{author}{{Sánchez Sánchez}, P.M.}, \bibinfo{author}{{Huertas Celdrán}, A.}, \bibinfo{author}{Xie, N.}, \bibinfo{author}{Bovet, G.}, \bibinfo{author}{{Martínez Pérez}, G.}, \bibinfo{author}{Stiller, B.}, \bibinfo{year}{2024}.
\newblock \bibinfo{title}{Federatedtrust: A solution for trustworthy federated learning}.
\newblock \bibinfo{journal}{Future Generation Computer Systems} \bibinfo{volume}{152}, \bibinfo{pages}{83--98}.
\newblock \URLprefix \url{https://www.sciencedirect.com/science/article/pii/S0167739X23003886}, \DOIprefix\doi{https://doi.org/10.1016/j.future.2023.10.013}.
%Type = Article
\bibitem[{Tian et~al.(2023)Tian, Cui, Liang and Yu}]{tian_ComprehensiveSurvey_2023}
\bibinfo{author}{Tian, Z.}, \bibinfo{author}{Cui, L.}, \bibinfo{author}{Liang, J.}, \bibinfo{author}{Yu, S.}, \bibinfo{year}{2023}.
\newblock \bibinfo{title}{A {Comprehensive} {Survey} on {Poisoning} {Attacks} and {Countermeasures} in {Machine} {Learning}}.
\newblock \bibinfo{journal}{ACM Computing Surveys} \bibinfo{volume}{55}, \bibinfo{pages}{1--35}.
%Type = Misc
\bibitem[{{U.S. Department of Health and Human Services}(1996)}]{hipaa1996}
\bibinfo{author}{{U.S. Department of Health and Human Services}}, \bibinfo{year}{1996}.
\newblock \bibinfo{title}{Health insurance portability and accountability act of 1996 (hipaa)}.
\newblock \URLprefix \url{https://www.hhs.gov/hipaa/for-professionals/privacy/index.html}. \bibinfo{note}{{Last} Visit July 2024}.
%Type = Inproceedings
\bibitem[{Wang et~al.(2022)Wang, Kang, Zhang and Hu}]{Wang9771619}
\bibinfo{author}{Wang, Z.}, \bibinfo{author}{Kang, Q.}, \bibinfo{author}{Zhang, X.}, \bibinfo{author}{Hu, Q.}, \bibinfo{year}{2022}.
\newblock \bibinfo{title}{Defense strategies toward model poisoning attacks in federated learning: A survey}, in: \bibinfo{booktitle}{2022 IEEE Wireless Communications and Networking Conference (WCNC)}, pp. \bibinfo{pages}{548--553}.
%Type = Article
\bibitem[{Watts and Strogatz(1998)}]{watts1998collective}
\bibinfo{author}{Watts, D.J.}, \bibinfo{author}{Strogatz, S.H.}, \bibinfo{year}{1998}.
\newblock \bibinfo{title}{Collective dynamics of ‘small-world’networks}.
\newblock \bibinfo{journal}{nature} \bibinfo{volume}{393}, \bibinfo{pages}{440--442}.
%Type = Misc
\bibitem[{Wu et~al.(2021)Wu, Yang, Zhu and Mitra}]{wu_MitigatingBackdoor_2021}
\bibinfo{author}{Wu, C.}, \bibinfo{author}{Yang, X.}, \bibinfo{author}{Zhu, S.}, \bibinfo{author}{Mitra, P.}, \bibinfo{year}{2021}.
\newblock \bibinfo{title}{Mitigating {Backdoor} {Attacks} in {Federated} {Learning}}.
\newblock \bibinfo{note}{ArXiv:2011.01767}.
%Type = Article
\bibitem[{Xia et~al.(2023)Xia, Chen, Yu and Ma}]{xia_PoisoningAttacks_2023}
\bibinfo{author}{Xia, G.}, \bibinfo{author}{Chen, J.}, \bibinfo{author}{Yu, C.}, \bibinfo{author}{Ma, J.}, \bibinfo{year}{2023}.
\newblock \bibinfo{title}{Poisoning {Attacks} in {Federated} {Learning}: {A} {Survey}}.
\newblock \bibinfo{journal}{IEEE Access} \bibinfo{volume}{11}, \bibinfo{pages}{10708--10722}.
%Type = Misc
\bibitem[{Xiao et~al.(2017)Xiao, Rasul and Vollgraf}]{xiao_FashionMNISTNovel_2017}
\bibinfo{author}{Xiao, H.}, \bibinfo{author}{Rasul, K.}, \bibinfo{author}{Vollgraf, R.}, \bibinfo{year}{2017}.
\newblock \bibinfo{title}{Fashion-{MNIST}: a {Novel} {Image} {Dataset} for {Benchmarking} {Machine} {Learning} {Algorithms}}.
\newblock \bibinfo{note}{ArXiv:1708.07747 [cs, stat]}.
%Type = Inproceedings
\bibitem[{Xie et~al.(2019)Xie, Huang, Chen and Li}]{xie2019dba}
\bibinfo{author}{Xie, C.}, \bibinfo{author}{Huang, K.}, \bibinfo{author}{Chen, P.Y.}, \bibinfo{author}{Li, B.}, \bibinfo{year}{2019}.
\newblock \bibinfo{title}{Dba: Distributed backdoor attacks against federated learning}, in: \bibinfo{booktitle}{International conference on learning representations}.
%Type = Misc
\bibitem[{Xie et~al.(2021)Xie, Koyejo and Gupta}]{xie_ZenoRobust_2021}
\bibinfo{author}{Xie, C.}, \bibinfo{author}{Koyejo, S.}, \bibinfo{author}{Gupta, I.}, \bibinfo{year}{2021}.
\newblock \bibinfo{title}{Zeno++: {Robust} {Fully} {Asynchronous} {SGD}}.
\newblock \bibinfo{note}{ArXiv:1903.07020 [cs, stat]}.
%Type = Inproceedings
\bibitem[{Yin et~al.(2018)Yin, Chen, Kannan and Bartlett}]{yin_ByzantineRobustDistributed_2018}
\bibinfo{author}{Yin, D.}, \bibinfo{author}{Chen, Y.}, \bibinfo{author}{Kannan, R.}, \bibinfo{author}{Bartlett, P.}, \bibinfo{year}{2018}.
\newblock \bibinfo{title}{Byzantine-{Robust} {Distributed Learning}: {Towards} {Optimal Statistical Rates}}, in: \bibinfo{booktitle}{Proceedings of the 35th international conference on machine learning}, pp. \bibinfo{pages}{5650--5659}.
%Type = Inproceedings
\bibitem[{Yin et~al.(2017)Yin, Cui and Li}]{yin_ReputationBasedResilient_2017}
\bibinfo{author}{Yin, J.}, \bibinfo{author}{Cui, X.}, \bibinfo{author}{Li, K.}, \bibinfo{year}{2017}.
\newblock \bibinfo{title}{A {Reputation}-{Based} {Resilient} and {Recoverable} {P2P} {Botnet}}, in: \bibinfo{booktitle}{2017 {IEEE} {Second} {International} {Conference} on {Data} {Science} in {Cyberspace} ({DSC})}, pp. \bibinfo{pages}{275--282}.
%Type = Inproceedings
\bibitem[{Yoo et~al.(2021)Yoo, Jeong, Lee and Chung}]{yoo2021federated}
\bibinfo{author}{Yoo, J.H.}, \bibinfo{author}{Jeong, H.}, \bibinfo{author}{Lee, J.}, \bibinfo{author}{Chung, T.M.}, \bibinfo{year}{2021}.
\newblock \bibinfo{title}{Federated learning: Issues in medical application}, in: \bibinfo{booktitle}{Future Data and Security Engineering: 8th International Conference, FDSE 2021, Virtual Event, November 24--26, 2021, Proceedings 8}, \bibinfo{organization}{Springer}. pp. \bibinfo{pages}{3--22}.
%Type = Inproceedings
\bibitem[{Zhang et~al.(2021)Zhang, Li, Zeng, Xie and Zhao}]{Zhang9634881}
\bibinfo{author}{Zhang, J.}, \bibinfo{author}{Li, M.}, \bibinfo{author}{Zeng, S.}, \bibinfo{author}{Xie, B.}, \bibinfo{author}{Zhao, D.}, \bibinfo{year}{2021}.
\newblock \bibinfo{title}{A survey on security and privacy threats to federated learning}, in: \bibinfo{booktitle}{2021 International Conference on Networking and Network Applications (NaNA)}, pp. \bibinfo{pages}{319--326}.
%Type = Misc
\bibitem[{Zhang et~al.(2023)Zhang, Zeng, Luo, Xu and King}]{zhang_SurveyTrustworthy_2023}
\bibinfo{author}{Zhang, Y.}, \bibinfo{author}{Zeng, D.}, \bibinfo{author}{Luo, J.}, \bibinfo{author}{Xu, Z.}, \bibinfo{author}{King, I.}, \bibinfo{year}{2023}.
\newblock \bibinfo{title}{A {Survey} of {Trustworthy} {Federated} {Learning} with {Perspectives} on {Security}, {Robustness}, and {Privacy}}.
\newblock \bibinfo{note}{ArXiv:2302.10637 [cs]}.
%Type = Misc
\bibitem[{Zhang et~al.(2022)Zhang, Cao, Jia and Gong}]{zhang_FLDetectorDefending_2022}
\bibinfo{author}{Zhang, Z.}, \bibinfo{author}{Cao, X.}, \bibinfo{author}{Jia, J.}, \bibinfo{author}{Gong, N.Z.}, \bibinfo{year}{2022}.
\newblock \bibinfo{title}{{FLDetector}: {Defending} {Federated} {Learning} {Against} {Model} {Poisoning} {Attacks} via {Detecting} {Malicious} {Clients}}.
\newblock \bibinfo{note}{ArXiv:2207.09209 [cs]}.
%Type = Article
\bibitem[{Zhao et~al.(2022)Zhao, Sun, Wang and Jiang}]{zhao_FedInvByzantineRobust_2022}
\bibinfo{author}{Zhao, B.}, \bibinfo{author}{Sun, P.}, \bibinfo{author}{Wang, T.}, \bibinfo{author}{Jiang, K.}, \bibinfo{year}{2022}.
\newblock \bibinfo{title}{{FedInv}: {Byzantine}-{Robust} {Federated} {Learning} by {Inversing} {Local} {Model} {Updates}}.
\newblock \bibinfo{journal}{Proceedings of the AAAI Conference on Artificial Intelligence} \bibinfo{volume}{36}, \bibinfo{pages}{9171--9179}.
\newblock \bibinfo{note}{Number: 8}.
%Type = Inproceedings
\bibitem[{Zhao et~al.(2020)Zhao, Chen, Zhang, Wu, Teng and Yu}]{zhao_PDGANNovel_2020}
\bibinfo{author}{Zhao, Y.}, \bibinfo{author}{Chen, J.}, \bibinfo{author}{Zhang, J.}, \bibinfo{author}{Wu, D.}, \bibinfo{author}{Teng, J.}, \bibinfo{author}{Yu, S.}, \bibinfo{year}{2020}.
\newblock \bibinfo{title}{{PDGAN}: {A} {Novel} {Poisoning} {Defense} {Method} in {Federated} {Learning} {Using} {Generative} {Adversarial} {Network}}, in: \bibinfo{booktitle}{Algorithms and {Architectures} for {Parallel} {Processing}}.

\end{thebibliography}

\end{document}